%% file: apssamp.tex
\documentclass[%
 preprint,
 superscriptaddress,
nofootinbib,
amsmath,amssymb,
aps,
]{revtex4-2}

\usepackage{mathtools}
\usepackage{booktabs}
\usepackage{pgfplots}
\pgfplotsset{compat=1.18}
\usepackage{tikz}
\usetikzlibrary{arrows.meta,positioning,shapes.geometric,shapes.misc,fit,calc}
\usetikzlibrary{shapes.geometric, arrows.meta, positioning}
\tikzstyle{process} = [rectangle, minimum width=2.8cm, minimum height=1.2cm, text centered, draw=black, fill=gray!10]
\tikzstyle{arrow} = [thick, -{Latex[length=3mm, width=2mm]}]
\usepackage[most]{tcolorbox}
\usepackage{graphicx}
\usepackage{dcolumn}
\usepackage{bm}
\usepackage{physics}
\usepackage{comment}
\pgfplotsset{compat=1.18}
\usetikzlibrary{positioning}
\usepackage{graphicx} 
\usepackage{dcolumn} 
\usepackage[table,xcdraw]{xcolor}
\usepackage{subcaption}
\usepackage{caption}
\usepackage{placeins}
\usepackage{graphicx}
\usepackage{textcomp}
\usepackage{subcaption}
\usepackage{braket}
\usepackage{bm}
\usepackage{multirow}
\usepackage{booktabs}
\usepackage{hyperref}
\usepackage[nameinlink,noabbrev]{cleveref} 
\usepackage[normalem]{ulem}
\usepackage{mathtools}
\usepackage{amsthm}
\theoremstyle{definition}
 
\theoremstyle{plain}

\makeatletter

\def\frontmatter@preabstractspace{0.5cm}

\def\frontmatter@postabstractspace{0.5cm}
\makeatother

\makeatletter
\renewcommand{\p@subsection}{}
\makeatother

\begin{document}

\preprint{}

\title{Transformer-Based Neural Quantum Digital Twins for Many-Body Spectral Reconstruction and Adaptive Quantum-Annealing Schedule Design}

\author{Jianlong Lu}

\affiliation{Department of Mathematics, Faculty of Science, National University of Singapore}

\author{Hanqiu Peng}

\affiliation{Department of Mathematics, Faculty of Science, National University of Singapore}

\author{Hongrui Zhang}
\affiliation{Department of Mathematics, Faculty of Science, National University of Singapore}

\author{Ying Chen}
\email[Corresponding author: ]{matcheny@nus.edu.sg}
\affiliation{Centre for Quantitative Finance, Department of Mathematics, Faculty of Science, National University of Singapore}
\affiliation{SIA-NUS Digital Aviation Lab, Institute of Operations Research \& Analytics, National University of Singapore}
\affiliation{Risk Management Institute, National University of Singapore}

\date{\today}

\begin{abstract}

\vspace{1em}

We introduce Transformer-based Neural Quantum Digital Twins (Tx-NQDTs) to reconstruct the low-energy spectral evolution of many-body quantum systems along quantum-annealing paths, including ground- and first-excited-state energies, spectral gaps, and transition matrix elements, at efficient computational cost. Tx-NQDTs employ a graph-informed Transformer neural network trained to estimate the spectral information needed for annealing-schedule design. We integrate these estimates with an adaptive schedule-construction procedure guided by first-order adiabatic perturbation theory (FOAPT), which is used as a closed-system spectral diagnostic to allocate annealing time near predicted spectral bottlenecks. Experiments on a D-Wave quantum annealer ($N=10,15,20$ logical variables, with schedules represented by up to 12 control points) show that Tx-NQDT-informed schedules can improve empirical ground-state success probabilities relative to the default $20\,\mu\mathrm{s}$ linear schedule under the tested hardware conditions. The proposed schedules achieve success probabilities $2.2$--$11.7$ percentage points higher across the reported easy and hard subsets and outperform the default baseline in 44 of 60 cases. 
The results demonstrate the feasibility of using learned logical spectral information to automatically generate adaptive quantum-annealing schedules for practical hardware experiments.

\end{abstract}

\maketitle

\input{Intro.tex}
\input{method.tex}

\input{sim}

\input{dwave}

\input{conclusion}

\begin{acknowledgments}
J. Lu respectfully acknowledges the late Associate Professor Stéphane Bressan for his valuable discussions.\\
This research project is supported by the National Research Foundation Singapore under its Industry Alignment Fund – Pre-positioning (IAF-PP) Funding Initiative. Any opinions, findings and conclusions or recommendations expressed in this material are those of the authors and do not reflect the views of National Research Foundation, Singapore.

\end{acknowledgments}

\section*{Author Contributions}

J. Lu conceived the study, developed the methodology and the Tx-NQDT
framework, implemented the computational methods, performed the numerical
simulations and quantum-annealing experiments, analyzed and interpreted the
results, prepared the figures, and co-wrote the original draft. H. Peng implemented the computational methods, performed the numerical simulations and quantum-annealing experiments, analyzed and interpreted the results, prepared the figures, and co-wrote the original draft.
H. Zhang contributed to the numerical simulations and preparation of the figures. 
Y. Chen supervised the research, contributed
to the interpretation of the results, and co-wrote the original draft. All
authors reviewed and edited the manuscript and approved the final version.

\appendix

\section{D-Wave Instance Construction and Hardware Schedule Implementation}
\label{app:dwave_implementation}

This appendix records the logical instances, embedding procedure, and
programming conditions used in the D-Wave experiments of
Sec.~\ref{sec:application}. The continuous Tx-NQDT schedule is defined in
Sec.~\ref{sec:foapt_schedule}, and its projection onto
hardware-compatible control points is specified in
Appendix~\ref{app:schedule_projection}. Tx-NQDT reconstructs spectral
quantities for the unembedded logical Hamiltonian, whereas the processor
executes a minor-embedded, autoscaled, calibrated, finite-temperature
physical implementation. Hardware diagnostics that were unavailable
uniformly, their interpretive consequences, and recommended future
measurements are discussed separately in
Appendix~\ref{app:dwave_hardware_diagnostics}.

\subsection{Logical instances, embedding, and programming conditions}
\label{app:dwave_instances}

We use dense random Ising instances, equivalently expressible as QUBOs, with
$N=10,15,$ and $20$. For each size, $20$ independent logical
instances are tested, giving $60$ problems in total. The local fields
$h_i$ and couplings $J_{ij}$, $i<j$, are sampled independently from
the uniform distribution on $[-5,5]$. The logical problem and annealing
Hamiltonians are
\begin{align}
H_{\rm P}^{\rm log}
&=
\frac{1}{2}
\left(
\sum_i h_i\hat{\sigma}_z^{(i)}
+
\sum_{i<j}J_{ij}
\hat{\sigma}_z^{(i)}
\hat{\sigma}_z^{(j)}
\right),
\label{eq:logical_problem_hardware_section}
\\
H_{\rm QA}^{\rm log}(s)
&=
-\frac{A(s)}{2}
\sum_i\hat{\sigma}_x^{(i)}
+
\frac{B(s)}{2}
\left(
\sum_i h_i\hat{\sigma}_z^{(i)}
+
\sum_{i<j}J_{ij}
\hat{\sigma}_z^{(i)}
\hat{\sigma}_z^{(j)}
\right).
\label{eq:logical_qa_hardware_section}
\end{align}
All Tx-NQDT energies, gaps, transition matrix elements, and FOAPT scores
are evaluated for Eq.~\eqref{eq:logical_qa_hardware_section} before
hardware embedding and coefficient rescaling.

The random logical graphs are complete with probability one and are
therefore effectively $K_N$ graphs. They are not native Pegasus
subgraphs and require minor embedding. Logical variables may consequently
be represented by connected chains of physical qubits coupled
ferromagnetically. Schematically, the programmed problem contains
\begin{equation}
H_{\rm P}^{\rm phys}
=
H_{\rm P}^{\rm emb}
+
H_{\rm chain},
\label{eq:physical_embedded_problem}
\end{equation}
where
\begin{equation}
H_{\rm chain}
=
-\frac{J_{\rm chain}}{2}
\sum_i
\sum_{(q,r)\in C_i}
\hat{\sigma}_z^{(q)}
\hat{\sigma}_z^{(r)}
\label{eq:chain_hamiltonian}
\end{equation}
represents the chain couplers associated with the physical chain $C_i$
for logical variable $i$. Equation~\eqref{eq:chain_hamiltonian} is
schematic; it displays the additional degrees of freedom and energy
scale introduced by embedding.

The embeddings were generated through the standard D-Wave Ocean
minor-embedding pipeline rather than through an analytical native
Pegasus mapping
\cite{cai2014practical,dwaveMinorEmbeddingDocs,dwaveEmbeddingGuidance}.
The resulting physical implementation can contain chain excitations and
chain breaks and depends on the selected chain strength.

The reported submissions used \texttt{auto\_scale=True}. If the submitted
coefficients are $(h_i,J_{ij})$, the solver programs coefficients
proportional to
\begin{equation}
\left(\frac{h_i}{\alpha},\frac{J_{ij}}{\alpha}\right),
\qquad
\alpha>0,
\label{eq:autoscaling_schematic}
\end{equation}
where $\alpha$ is selected to satisfy the hardware field, coupler, and
coupling-sum constraints. For each logical instance, the default and
Tx-NQDT schedules used the same submitted coefficients and
\texttt{auto\_scale=True} convention. Autoscaling was therefore held
fixed within each paired schedule comparison.

The experiments relied on the standard D-Wave calibration. We did not
perform an additional user-level shimming or calibration-refinement
procedure, infer instance-specific persistent field or coupler offsets,
or apply manually calibrated \texttt{flux\_biases}.

Complete job-level embedding, autoscaling, and calibration metadata were
not retained uniformly for all historical submissions.

At each $N$, the $20$ instances are divided into $10$ easy and
$10$ hard cases according to the logical minimum gap
\begin{equation}
\Delta_{\min}
=
\min_{s\in[0,1]}
\left[E_1(s)-E_0(s)\right].
\label{eq:hardware_delta_min}
\end{equation}
Relatively large-gap instances are labeled easy and small-gap instances
hard. These labels provide a closed-system logical stratification and
are not interpreted as a complete measure of physical hardware
difficulty.

\subsection{Hardware implementation of the Tx-NQDT schedule}
\label{app:dwave_schedule}

For each logical instance, Tx-NQDT supplies the low-energy spectral
quantities of Eq.~\eqref{eq:logical_qa_hardware_section}. These quantities
are converted into the FOAPT score, normalized time density, cumulative
time allocation, and continuous schedule according to the definitions
in Sec.~\ref{sec:foapt_schedule}. Those definitions are not repeated
here.

The continuous schedule is evaluated on the discrete anneal grid and
converted into an uploaded piecewise-linear schedule using the numerical
quadrature, cumulative-map inversion, control-point placement, and
ramp-rate enforcement described in
Appendix~\ref{app:schedule_projection}. The reported schedules contain
at most $K_{\rm cp}=12$ monotone control points and satisfy the maximum
segment ramp rate
$\gamma_{\max}=2.0~(\mu{\rm s})^{-1}$. 

The processor executes the resulting uploaded schedule
$s_{\rm upload}(t)$. The corresponding programmed control trajectories
are $A(s_{\rm upload}(t))$ and $B(s_{\rm upload}(t))$, as defined in
Appendix~\ref{app:schedule_projection}. These trajectories change the
rate at which the fixed annealing path is traversed; they do not measure
the instantaneous spectrum of the embedded physical Hamiltonian.

No instance-specific freeze-out correction was applied in the reported
schedule construction.

\section{Hardware Diagnostics and Unresolved Experimental Systematics}
\label{app:dwave_hardware_diagnostics}

Appendix~\ref{app:dwave_implementation} records the instance
construction, embedding route, programming settings, and schedule
implementation used in the reported D-Wave experiments. The present
appendix instead collects the hardware-level diagnostics that were not
available uniformly for the historical data, explains how their absence
limits interpretation, and specifies measurements required for more
controlled future benchmarks. These effects were not deliberately varied
between the default and Tx-NQDT-informed schedules for a given logical
instance, but they prevent a complete reconstruction of the physical
hardware dynamics.

\subsection{Embedding and chain diagnostics}

As described in Appendix~\ref{app:dwave_implementation}, the dense
logical instances require minor embedding on the Pegasus hardware graph.
A logical variable may therefore be represented by a connected chain of
physical qubits with an additional ferromagnetic coupling scale. Chain
strength competes with the logical problem scale, transverse-field scale,
effective temperature, analog-control errors, and available coefficient
ranges. Chains that are too weak may break, whereas excessively strong
chains can increase coefficient compression and reduce the programmed
logical energy scale relative to temperature and control errors
\cite{dwaveEmbeddingGuidance,gilbert2024chainstrength}.

Chain excitations, chain breaks, and the rule used to resolve broken
chains can change the returned logical distribution. The embedded
Hamiltonian also contains physical degrees of freedom and energy scales
that are absent from the unembedded logical Hamiltonian used to train
Tx-NQDT. Consequently, the learned logical gap, transition matrix
element, and FOAPT score cannot be identified directly with the
corresponding quantities of the embedded physical system.

Complete job-level embedding metadata were not retained uniformly for
all historical submissions. We therefore cannot report for every run the
exact embedding map, number of physical qubits, chain-length distribution,
chain strength, chain-break fraction, chain-break resolution rule, or
unembedding method.

\subsection{Freeze-out diagnostics}

During a forward anneal, the transverse-field scale $A(s)$ decreases
while the problem scale $B(s)$ increases. Spin-flip,
tunneling-assisted, relaxation-mediated, and thermally assisted
transitions may eventually become slow relative to the remaining anneal
time, causing the output distribution to become approximately frozen at
an instance-, embedding-, and schedule-dependent fraction $s_f$
\cite{dwaveFreezeoutDocs,dwaveFreezeoutEffectiveTemperature}.

The reported experiments did not include pause--quench, anneal-slicing,
reverse-annealing, or effective-temperature measurements from which
$s_f$ could be inferred for each embedded instance
\cite{dwaveAnnealingControls,marshall2019power,chen2020pausing,
pelofske2019peering,pelofske2020inferring}. Consequently, the programmed
slowdown location cannot be identified with the physical freeze-out
region. If an independent estimate of $s_f$ were available, a
freeze-out-aware schedule could suppress time reallocation after that
point, for example through the weighted density
\begin{equation}
\rho_{\rm fr}(s)
=
\frac{
[\Lambda(s)+\epsilon_\Lambda]w_f(s)
}{
\displaystyle
\int_0^1
[\Lambda(u)+\epsilon_\Lambda]w_f(u)\,\mathrm{d}u
},
\label{eq:appendix_freezeout_density}
\end{equation}
where $w_f(s)$ decreases across the estimated freeze-out region.

\subsection{Calibration and shimming}

Residual analog-control errors may be represented schematically by
\begin{equation}
h_i
\longrightarrow
h_i+\delta h_i,
\qquad
J_{ij}
\longrightarrow
J_{ij}+\delta J_{ij}.
\label{eq:ice_schematic}
\end{equation}
The offsets $\delta h_i$ and $\delta J_{ij}$ can arise from persistent
flux biases, coupler-induced field shifts, crosstalk, magnetic-environment
variation, device nonuniformity, and calibration drift
\cite{perdomoortiz2016persistentbiases,chern2023shimming,
dwaveErrorsDocs}.

The reported experiments used the standard D-Wave calibration and did
not include an additional user-level shimming procedure. We did not
perform symmetry-based offset estimation, gauge-averaged calibration
loops, instance-specific bias inference, or manually calibrated
\texttt{flux\_biases}. Residual biases may therefore affect the absolute
success probabilities and prevent reconstruction of the exact calibrated
physical Hamiltonian.

\subsection{Effective temperature}

The nominal refrigerator temperature need not equal the effective
temperature relevant to sampling or freeze-out. A schematic
dimensionless thermal scale is
\begin{equation}
\beta_{\rm eff}(s)
\sim
\frac{B(s)}
{k_{\rm B}T_{\rm QPU}},
\label{eq:effective_temperature_scale}
\end{equation}
subject to coefficient normalization, embedding, autoscaling, and the
location at which the distribution freezes.

No run-resolved effective-temperature calibration or energy-scale
compensation was performed in the reported experiments. Temperature
drift may therefore change the ratio between the programmed problem
energy and the effective thermal energy and may contribute to
batch-to-batch variation
\cite{finiteTemperatureSampling2025}.

\subsection{Autoscaling and coefficient normalization}

As recorded in Appendix~\ref{app:dwave_implementation}, the reported
hardware submissions used \texttt{auto\_scale=True}. Positive autoscaling
preserves the minimizing bitstrings of the final classical Ising objective
but changes the physical problem scale relative to $A(s)$, chain
couplings, temperature, analog-control errors, and freeze-out.

Because both schedules for a given logical instance used the same
submitted coefficients and autoscaling convention, autoscaling was
controlled within the paired comparison. It nevertheless complicates
cross-instance comparisons and mechanistic interpretation. Complete
solver-selected scaling factors were not retained for all historical
jobs, so the exact programmed problem scale cannot be reconstructed
uniformly.

\subsection{Inter-sample correlations}

Let the ordered ground-state-success indicator be
\begin{equation}
X_r
=
\begin{cases}
1,
&
\text{if read $r$ returns a ground-state sample},
\\
0,
&
\text{otherwise}.
\end{cases}
\label{eq:ordered_success_indicator}
\end{equation}
A lag-$k$ autocorrelation diagnostic is
\begin{equation}
C_X(k)
=
\frac{
\displaystyle
\sum_{r=1}^{R-k}
(X_r-\bar X)(X_{r+k}-\bar X)
}{
\displaystyle
\sum_{r=1}^{R}
(X_r-\bar X)^2
}.
\label{eq:readout_autocorrelation}
\end{equation}
A complementary bitstring-level statistic is
\begin{equation}
D_r^{(k)}
=
\frac{1}{N}
d_{\rm H}
\!\left(
z^{(r)},z^{(r+k)}
\right),
\label{eq:hamming_lag_distance}
\end{equation}
where $d_{\rm H}$ is the Hamming distance between returned logical
bitstrings.

Complete ordered readout streams were not retained for all historical
runs, and additional QPU access was unavailable for controlled
single-read, delayed-readout, or interleaved tests. The quantities in
Eqs.~\eqref{eq:readout_autocorrelation} and
\eqref{eq:hamming_lag_distance} therefore cannot be evaluated uniformly
for the full data set.

\subsection{Statistical scope}

Repeated reads refine the empirical returned-sample frequency for a fixed
submission but do not substitute for independent logical instances,
embeddings, gauges, jobs, or calibration periods. The reported win
counts, means, and medians are therefore descriptive statistics for the
tested ensemble.

\subsection{Future hardware benchmarks}

A more controlled hardware benchmark should implement the following
protocol:

\begin{enumerate}
\item \textbf{Embeddings.}
Use a fixed embedding for each paired schedule comparison, retain the
complete logical-to-physical embedding map and physical-qubit count, and
repeat the benchmark over multiple recorded embeddings to quantify
embedding-to-embedding variability.

\item \textbf{Chain diagnostics.}
Record the chain-length distribution, chain strength, per-chain and
aggregate chain-break fractions, chain-break resolution rule, and
unembedding method for every submission.

\item \textbf{Coefficient normalization.}
Either set \texttt{auto\_scale=False} and apply an explicit recorded
normalization or retain the solver-selected scaling factor, programmed
field and coupler ranges, coupling-sum constraints, and chain-strength
normalization for every job.

\item \textbf{Calibration and shimming.}
Use multiple gauges, repeated calibration measurements, symmetry-based
offset estimation, and calibrated flux-bias corrections where
appropriate. Record all inferred field and coupler corrections together
with the coefficient-normalization convention.

\item \textbf{Effective temperature.}
Record the available temperature and calibration metadata for every job,
preserve their chronological order, and use auxiliary thermalization or
effective-temperature problems to estimate the relevant thermal scale
before each experimental batch.

\item \textbf{Freeze-out.}
Use pause--quench, anneal-slicing, reverse-annealing, or complementary
effective-temperature protocols to estimate the instance-, embedding-,
and schedule-dependent freeze-out fraction. Use the inferred $s_f$ to
test whether programmed slowdowns occur within the dynamically active
region and, where appropriate, to construct freeze-out-aware schedules.

\item \textbf{Read order and correlations.}
Preserve the complete ordered readout stream, report lag-dependent
success and bitstring correlations, estimate effective sample sizes using
blocking or block-bootstrap methods, and compare multi-read submissions
with single-read or delayed-readout protocols where available.

\item \textbf{Statistical design and tests.}
Use substantially larger independent instance ensembles, randomized or
interleaved schedule execution, multiple gauges and embeddings, and
repeated jobs across calibration periods. Report confidence intervals
based on independent submission blocks, paired nonparametric tests, and
hierarchical models that separate instance variation, temporal drift,
embedding effects, and within-job correlations.
\end{enumerate}

\section{Support and Variance Limitations of the VMC Gradient Estimator}
\label{app:vmc_support_variance}

This appendix gives the detailed support, nodal, and variance analysis
for the score-function VMC gradient used in
Eq.~\eqref{eq:vmc_gradient}. For a variational state
$\Psi_\theta$, the normalized energy is
\begin{equation}
E_\theta
=
\frac{
\langle\Psi_\theta|H|\Psi_\theta\rangle
}{
\langle\Psi_\theta|\Psi_\theta\rangle
}.
\label{eq:appendix_vmc_energy}
\end{equation}
In a discrete basis $\{|x\rangle\}$, define
\begin{equation}
p_\theta(x)
=
\frac{|\Psi_\theta(x)|^2}
{\sum_{x'}|\Psi_\theta(x')|^2},
\qquad
E_{\rm loc}(x)
=
\frac{\langle x|H|\Psi_\theta\rangle}
{\Psi_\theta(x)},
\label{eq:appendix_vmc_distribution_local_energy}
\end{equation}
and the logarithmic derivative
\begin{equation}
O_a(x)
=
\frac{
\partial_{\theta_a}\Psi_\theta(x)
}{
\Psi_\theta(x)
}
=
\partial_{\theta_a}\log\Psi_\theta(x),
\label{eq:appendix_log_derivative}
\end{equation}
where the expression is defined only when
$\Psi_\theta(x)\neq0$. Provided that these quantities are regular on the
sampled support and differentiation does not generate unaccounted
support-boundary terms, the energy gradient is
\begin{equation}
\partial_{\theta_a}E_\theta
=
2\,\mathrm{Re}
\left[
\left\langle
\left(E_{\rm loc}(x)-E_\theta\right)
O_a^*(x)
\right\rangle_{p_\theta}
\right].
\label{eq:appendix_vmc_gradient}
\end{equation}
For real wavefunctions away from their nodes, this reduces to the
score-function form involving
$\partial_{\theta_a}\log|\Psi_\theta(x)|$ used in the main text.

The support qualification is important. If
$\Psi_\theta(x)=0$, then $p_\theta(x)=0$, so the configuration is never
sampled, while both $O_a(x)$ and $E_{\rm loc}(x)$ may be undefined. If
a parameter variation changes the support by making
$\partial_{\theta_a}\Psi_\theta(x)\neq0$ at such a configuration, the
sampled expectation in
Eq.~\eqref{eq:appendix_vmc_gradient} does not automatically include the
corresponding contribution. The estimator should therefore be understood
as valid on a fixed, regular sampled support rather than as a globally
regular identity across arbitrary parameter-dependent nodal changes.

Even without exact zeros, very small amplitudes can produce large
variance. Expanding the local energy gives
\begin{equation}
E_{\rm loc}(x)
=
H_{xx}
+
\sum_{x'\neq x}
H_{xx'}
\frac{\Psi_\theta(x')}{\Psi_\theta(x)}.
\label{eq:appendix_local_energy_ratios}
\end{equation}
A small denominator can amplify off-diagonal amplitude ratios. Such
configurations may be sampled rarely, but when encountered they can
produce heavy-tailed local-energy and gradient estimates. For the
transverse-field Ising Hamiltonian, the off-diagonal terms connect
single-spin-flip configurations, giving contributions proportional to
\begin{equation}
\frac{\Psi_\theta(x^{(i)})}{\Psi_\theta(x)}.
\label{eq:appendix_spin_flip_ratio}
\end{equation}
The local energy contains a sum of these ratios over the $N$ spin-flip
neighbors. Errors or rare large values in individual ratios can therefore
accumulate across the transverse-field contribution.

This effect is relevant in two regimes encountered in the present work.
Near the driver-dominated beginning of the anneal, the exact low-energy
states are delocalized in the computational $z$ basis, and accurate
estimation depends on many neighboring amplitude ratios. The
first-excited level of
$H_0=-\frac{1}{2}\sum_i\hat{\sigma}_x^{(i)}$ is also $N$-fold
degenerate, so variational tracking of a particular excited-state
direction is sensitive to small optimization and sampling errors. Near
the classical endpoint, the target state can become sharply concentrated
on a small subset of configurations, making many other amplitudes
extremely small. Local-energy and gradient variances can therefore become
large in either regime even when the physical instantaneous gap is not
small.

For the reported ground-state calculations, the parameterization
\begin{equation}
\Psi_{0,\theta}(x)
=
\exp\!\left[g_{0,\theta}(x)\right]
\label{eq:appendix_positive_ground_state}
\end{equation}
is strictly positive for finite network outputs and therefore avoids
exact neural zeros. It does not prevent exponentially small amplitudes,
large amplitude ratios, or numerical underflow at finite precision. For
excited states, sign changes and nodal sets are physically required, so
support, phase, and near-node variance problems are more severe. These
features help explain the localized excited-state and early-$s$ errors
observed under the runtime-controlled training protocol.

Several improvements are possible. Longer optimization, larger Monte
Carlo batches, better proposal distributions, autoregressive or direct
sampling, and stochastic-reconfiguration or natural-gradient
preconditioning may reduce estimator noise or improve convergence.
Regularization or clipping of extreme local-energy contributions can
improve numerical stability, although such procedures may introduce bias
and must be reported explicitly. Useful diagnostics include the
local-energy variance, gradient variance, amplitude-ratio tail
statistics, Markov-chain autocorrelation time, and effective sample size.

A basis or Hamiltonian rotation can change the support and nodal
structure without changing the spectrum. For example, a local rotation
about the $x$ axis,
\begin{equation}
U_x
=
\bigotimes_{i=1}^{N}
\exp\!\left(
-\frac{i\pi}{4}\hat{\sigma}_x^{(i)}
\right),
\qquad
U_x\hat{\sigma}_z^{(i)}U_x^\dagger
=
\pm\hat{\sigma}_y^{(i)},
\label{eq:appendix_basis_rotation}
\end{equation}
with the sign determined by the rotation convention, transforms
computational-basis localization into phase structure. In the rotated
representation, endpoint eigenstates that are localized in the original
$z$ basis can have nonvanishing magnitudes over all computational-basis
configurations. This can remove some exact-support pathologies.

The rotation does not eliminate the computational problem: it generally
produces more complicated off-diagonal matrix elements and may require a
complex-valued NQS capable of representing the induced phase structure.
The Hamiltonian, variational state, local-energy estimator, and
transition matrix elements must all be transformed consistently.
Whether the reduction in nodal or support problems outweighs the more
general complex sampling problem is therefore model and
implementation dependent.

The reported calculations use neither basis rotations nor an explicit
nodal regularization scheme. The resulting errors should consequently be
interpreted as limitations of the present computational-basis,
finite-sample, AdamW-optimized VMC implementation, especially for
sign-changing excited states, rather than as a general limitation of
neural quantum states.

\section{Hermitian Projector-Shift Identity for Deflation}
\label{brauer_proof}

This appendix records the Hermitian projector-shift identity used in
Sec.~\ref{sec:brauer}. Although related to Brauer-type rank-one
eigenvalue-shifting results~\cite{brauer1952}, the present deflation
procedure requires only the special update
\begin{equation}
\widetilde H
=
H+\delta\lvert u_p\rangle\langle u_p\rvert,
\label{eq:appB_projector_update}
\end{equation}
where $H$ is Hermitian, $\lvert u_p\rangle$ is an exact normalized
eigenvector, and $\delta\in\mathbb{R}$. This case follows directly from
the spectral decomposition and does not require the general
non-Hermitian rank-one theorem.

Let
\begin{equation}
H\lvert u_j\rangle
=
\lambda_j\lvert u_j\rangle,
\qquad
\langle u_j\vert u_k\rangle
=
\delta_{jk},
\label{eq:appB_eigenpairs}
\end{equation}
so that
\begin{equation}
H
=
\sum_j
\lambda_j
\lvert u_j\rangle\langle u_j\rvert.
\label{eq:appB_spectral_decomposition}
\end{equation}
Substitution into Eq.~\eqref{eq:appB_projector_update} gives
\begin{equation}
\widetilde H
=
\sum_{j\neq p}
\lambda_j
\lvert u_j\rangle\langle u_j\rvert
+
(\lambda_p+\delta)
\lvert u_p\rangle\langle u_p\rvert.
\label{eq:appB_projector_shift_identity}
\end{equation}
Hence the eigenvectors are unchanged, while only the selected eigenvalue
is shifted:
\begin{equation}
\lambda_p
\longmapsto
\lambda_p+\delta.
\label{eq:appB_eigenvalue_shift}
\end{equation}
Equivalently,
\begin{equation}
\widetilde H\lvert u_j\rangle
=
\begin{cases}
(\lambda_p+\delta)\lvert u_p\rangle,
& j=p,\\[2mm]
\lambda_j\lvert u_j\rangle,
& j\neq p.
\end{cases}
\label{eq:appB_action_on_eigenbasis}
\end{equation}

The general Brauer update has the form
\begin{equation}
H+\mathbf{u}_p\mathbf{v}^{\dagger},
\label{eq:appB_general_brauer}
\end{equation}
where $\mathbf{u}_p$ is a right eigenvector of $H$. It shifts the
associated eigenvalue by
$\mathbf{v}^{\dagger}\mathbf{u}_p$, but for a general
$\mathbf{v}$ the updated matrix need not be Hermitian, and the theorem
does not imply preservation of all other eigenvectors. The present work
does not rely on that stronger claim. Full eigenbasis preservation follows
here only because Eq.~\eqref{eq:appB_projector_update} uses the orthogonal
projector onto an exact eigenvector of a Hermitian operator.

For exact nondegenerate ground-state deflation, let
$\lambda_0<\lambda_1$ and choose
\begin{equation}
\delta>\lambda_1-\lambda_0.
\label{eq:appB_shift_condition}
\end{equation}
Then $\lambda_0+\delta>\lambda_1$, so the original first excited state,
or first-excited eigenspace if $\lambda_1$ is degenerate, becomes the
lowest eigenspace of
\begin{equation}
H+\delta\lvert u_0\rangle\langle u_0\rvert.
\label{eq:appB_exact_deflation}
\end{equation}
This is the algebraic motivation for the excited-state construction used
in the main text. If the original ground eigenspace is degenerate,
shifting a single ground-state direction does not remove the remaining
ground-state directions; exact deflation would instead require a
projector onto the complete ground-state eigenspace.

In Tx-NQDT, however, the projector is formed from the learned variational
ground state:
\begin{equation}
H_{\rm QA}^{(1)}(s)
=
H_{\rm QA}(s)
+
\delta_0(s)
\frac{
\lvert\Psi_{0,\theta}(s)\rangle
\langle\Psi_{0,\theta}(s)\rvert
}{
\langle\Psi_{0,\theta}(s)
\vert\Psi_{0,\theta}(s)\rangle
}.
\label{eq:appB_variational_deflation}
\end{equation}
Define the normalized learned direction
\begin{equation}
\lvert\phi_{0,\theta}(s)\rangle
=
\frac{
\lvert\Psi_{0,\theta}(s)\rangle
}{
\sqrt{
\langle\Psi_{0,\theta}(s)
\vert\Psi_{0,\theta}(s)\rangle
}
}
=
\sum_j a_j(s)\lvert u_j(s)\rangle,
\label{eq:appB_learned_state_expansion}
\end{equation}
where $\{\lvert u_j(s)\rangle\}$ are the exact instantaneous
eigenvectors of $H_{\rm QA}(s)$. In this basis,
\begin{equation}
\left\langle
u_j(s)
\left|
H_{\rm QA}^{(1)}(s)
\right|
u_k(s)
\right\rangle
=
E_j(s)\delta_{jk}
+
\delta_0(s)a_j(s)a_k(s)^*.
\label{eq:appB_variational_matrix_elements}
\end{equation}
Unless the learned direction lies entirely within a single exact
eigenspace, Eq.~\eqref{eq:appB_variational_matrix_elements} contains
off-diagonal terms that mix the original eigenvectors. Their magnitude
scales with both the learned-state misalignment and the chosen shift
$\delta_0(s)$.

For any trial state $\lvert\chi\rangle$, the shifted Rayleigh quotient is
\begin{equation}
\frac{
\langle\chi|H_{\rm QA}^{(1)}(s)|\chi\rangle
}{
\langle\chi|\chi\rangle
}
=
\frac{
\langle\chi|H_{\rm QA}(s)|\chi\rangle
}{
\langle\chi|\chi\rangle
}
+
\delta_0(s)
\frac{
|\langle\phi_{0,\theta}(s)|\chi\rangle|^2
}{
\langle\chi|\chi\rangle
}.
\label{eq:appB_variational_rayleigh_quotient}
\end{equation}
Thus, the added term penalizes overlap with the learned ground-state
direction. It does not guarantee that the minimizer is the exact first
excited state.

Because $\lvert\Psi_{0,\theta}(s)\rangle$ is generally not an exact
eigenvector, its projector need not commute with $H_{\rm QA}(s)$, and
the remaining eigenvectors are not preserved exactly. In particular, a
condition based only on a learned gap and
$\delta_0(s)>\Delta(s)$ is not sufficient to recover the exact first
excited state: a shift that is too small may leave residual
ground-state contamination, whereas a large shift can amplify mixing
caused by ground-state misalignment. The orthogonality penalty in
Eq.~\eqref{eq:orthogonality_penalty} mitigates this contamination but
does not convert the learned projector into an exact eigenprojector.

Equation~\eqref{eq:appB_variational_deflation} should therefore be
interpreted as a variational overlap penalty that encourages the
subsequent optimization to leave the learned ground-state direction and
target the first-excited-state manifold. It is a practical deflation
surrogate, not an exact eigenpair-preserving transformation for an
approximate neural state. Errors in the learned ground state can
consequently propagate into the reconstructed excited state, gap, and
transition matrix element.

\section{Adiabatic Evolution and First-Order Adiabatic Perturbation Theory}
\label{app:adiabatic_foapt}
\label{app:adiabatic_theorem}
\label{app:foapt}

This appendix derives the adiabatic and first-order perturbative
arguments motivating the spectral-difficulty score and schedule density
defined in Sec.~\ref{sec:foapt_schedule}. It also summarizes the
assumptions under which the resulting local diagnostic is expected to be
quantitatively reliable.

\subsection{Instantaneous-basis formulation}

Let the instantaneous eigenstates and eigenvalues of a time-dependent
Hamiltonian satisfy
\begin{equation}
H(t)\lvert n(t)\rangle
=
E_n(t)\lvert n(t)\rangle ,
\label{eq:app_instantaneous_eigenproblem}
\end{equation}
and let the system evolve according to
\begin{equation}
i\hbar\frac{\partial}{\partial t}\lvert\Psi(t)\rangle
=
H(t)\lvert\Psi(t)\rangle .
\label{eq:app_schrodinger}
\end{equation}
Expanding the state in the instantaneous eigenbasis gives
\begin{equation}
\lvert\Psi(t)\rangle
=
\sum_n
c_n(t)
\exp\left[
-\frac{i}{\hbar}
\int_0^t E_n(t')\,\mathrm{d}t'
\right]
\lvert n(t)\rangle .
\label{eq:app_instantaneous_expansion}
\end{equation}
In the parallel-transport gauge,
$\langle n(t)\vert\dot n(t)\rangle=0$, the coefficients obey
\begin{equation}
\dot c_m(t)
=
-\sum_{n\neq m}
c_n(t)
\exp\left[
\frac{i}{\hbar}
\int_0^t
\bigl(E_m(t')-E_n(t')\bigr)\,\mathrm{d}t'
\right]
\langle m(t)\vert\dot n(t)\rangle .
\label{eq:app_coefficient_evolution}
\end{equation}
Other gauge choices add diagonal geometric-phase terms but do not alter
the transition probabilities.

For nondegenerate instantaneous levels, differentiating
Eq.~\eqref{eq:app_instantaneous_eigenproblem} yields
\begin{equation}
\langle m(t)\vert\dot n(t)\rangle
=
\frac{
\langle m(t)\vert\dot H(t)\vert n(t)\rangle
}{
E_n(t)-E_m(t)
},
\qquad m\neq n .
\label{eq:app_nonadiabatic_coupling}
\end{equation}
A commonly used local adiabatic condition is therefore
\begin{equation}
\eta_{mn}(t)
\equiv
\hbar
\frac{
\left|
\langle m(t)\vert\dot H(t)\vert n(t)\rangle
\right|
}{
\left|E_n(t)-E_m(t)\right|^2
}
\ll 1 .
\label{eq:app_local_adiabatic_condition}
\end{equation}
Small gaps, rapid Hamiltonian variation, and large off-diagonal matrix
elements all increase the local tendency toward diabatic transitions.

For the logical annealing Hamiltonian,
\begin{equation}
H_{\rm QA}(s)
=
A(s)H_0+B(s)H_P,
\qquad
s=s(t),
\label{eq:app_annealing_hamiltonian}
\end{equation}
physical time enters through the monotone schedule $s(t)$, so that
\begin{equation}
\dot H_{\rm QA}(t)
=
\frac{\mathrm{d}H_{\rm QA}}{\mathrm{d}s}
\frac{\mathrm{d}s}{\mathrm{d}t},
\qquad
\frac{\mathrm{d}H_{\rm QA}}{\mathrm{d}s}
=
A'(s)H_0+B'(s)H_P .
\label{eq:app_hamiltonian_derivative}
\end{equation}
Defining
\begin{equation}
\Delta(s)=E_1(s)-E_0(s)
\label{eq:app_gap}
\end{equation}
and
\begin{equation}
M_{01}(s)
=
\left\langle
\Psi_0(s)
\left|
\frac{\mathrm{d}H_{\rm QA}}{\mathrm{d}s}
\right|
\Psi_1(s)
\right\rangle ,
\label{eq:app_transition_matrix_element}
\end{equation}
the ground-to-first-excited local adiabatic parameter becomes
\begin{equation}
\eta_{01}(s)
=
\hbar
\frac{|M_{01}(s)|}{\Delta(s)^2}
\left|
\frac{\mathrm{d}s}{\mathrm{d}t}
\right|.
\label{eq:app_two_level_adiabatic_parameter}
\end{equation}
Equation~\eqref{eq:app_two_level_adiabatic_parameter} motivates slower
evolution where the logical gap is small or the transition matrix element
is large.

\subsection{First-order transition amplitude}

Suppose that the system is initially in the instantaneous ground state,
$\lvert\Psi(0)\rangle=\lvert\Psi_0(s(0))\rangle$. Retaining the leading
ground-to-first-excited contribution in
Eq.~\eqref{eq:app_coefficient_evolution} gives
\begin{align}
c_{0\rightarrow1}^{(1)}(T)
&=
-\int_0^T
\langle\Psi_1(t)\vert\dot\Psi_0(t)\rangle
\exp\left[
\frac{i}{\hbar}
\int_0^t\Delta(t')\,\mathrm{d}t'
\right]\mathrm{d}t
\nonumber\\
&=
\int_0^T
\frac{
\langle\Psi_1(t)\vert\dot H_{\rm QA}(t)\vert\Psi_0(t)\rangle
}{
\Delta(t)
}
\exp\left[
\frac{i}{\hbar}
\int_0^t\Delta(t')\,\mathrm{d}t'
\right]\mathrm{d}t ,
\label{eq:app_first_order_amplitude_time}
\end{align}
up to an overall sign determined by the phase convention. The associated
first-order transition probability is
\begin{equation}
P_{0\rightarrow1}^{(1)}(T)
=
\left|
c_{0\rightarrow1}^{(1)}(T)
\right|^2 .
\label{eq:app_first_order_probability}
\end{equation}

Changing the integration variable from $t$ to $s$ gives
\begin{equation}
c_{0\rightarrow1}^{(1)}(T)
=
\int_0^1
\frac{M_{10}(s)}{\Delta(s)}
\exp\left[
\frac{i}{\hbar}
\int_0^s
\frac{\Delta(u)}{\dot s(u)}\,\mathrm{d}u
\right]\mathrm{d}s ,
\label{eq:app_first_order_amplitude_s}
\end{equation}
where
\begin{equation}
M_{10}(s)
=
\left\langle
\Psi_1(s)
\left|
\frac{\mathrm{d}H_{\rm QA}}{\mathrm{d}s}
\right|
\Psi_0(s)
\right\rangle
=
M_{01}(s)^*
\label{eq:app_M10_M01_relation}
\end{equation}
and
\begin{equation}
\dot s(u)
\equiv
\left.
\frac{\mathrm{d}s}{\mathrm{d}t}
\right|_{t:\,s(t)=u}.
\label{eq:app_schedule_speed}
\end{equation}
The explicit speed factor cancels from the nonoscillatory prefactor in
Eq.~\eqref{eq:app_first_order_amplitude_s}, but the schedule remains
present through the accumulated dynamical phase. The full transition
amplitude can therefore exhibit interference and cancellation between
different anneal regions. The schedule construction used in this work
does not minimize this complete phase-sensitive integral; instead, it
uses the associated local adiabatic condition as a deterministic
feature-construction rule.

\subsection{Local spectral-difficulty interpretation}

Equation~\eqref{eq:app_two_level_adiabatic_parameter} motivates the
regularized spectral-difficulty score $\Lambda(s)$ defined in
Eq.~\eqref{eq:lambda_def}. The regularization parameter
$\epsilon_\Delta$ has the same dimensions as $\Delta^2$ and prevents
numerical divergence near very small learned gaps. The normalized density,
cumulative map, and continuous schedule are defined uniquely in
Eqs.~\eqref{eq:rho_def}--\eqref{eq:adaptive_schedule}. These definitions
allocate a larger fraction of the available annealing time to regions
with greater reconstructed spectral difficulty.

For a piecewise-linear schedule with segment endpoints
$(t_\ell,s_\ell)$, define
\begin{equation}
v_\ell
=
\frac{\Delta s_\ell}{\Delta t_\ell},
\qquad
\overline{\Lambda}_\ell
=
\frac{1}{\Delta s_\ell}
\int_{s_\ell}^{s_{\ell+1}}
\Lambda(s)\,\mathrm{d}s .
\label{eq:app_segment_quantities}
\end{equation}
The local adiabatic parameter scales approximately as
$v_\ell\overline{\Lambda}_\ell$. Local equalization therefore motivates
\begin{equation}
v_\ell
\propto
\frac{1}{\overline{\Lambda}_\ell},
\qquad
\Delta t_\ell
\propto
\Delta s_\ell\overline{\Lambda}_\ell .
\label{eq:app_segment_local_adiabatic_rule}
\end{equation}
Equation~\eqref{eq:app_segment_local_adiabatic_rule} provides motivation
for the normalized-density construction rather than a separate schedule
algorithm. The finite-control-point projection, monotonicity conditions,
and hardware ramp-rate constraint are specified in
Appendix~\ref{app:schedule_projection}.

\subsection{Validity and interpretation}

The derivation above assumes locally isolated instantaneous levels and
uses the ground-to-first-excited transition as the dominant leakage
channel. These assumptions are not uniformly valid near exact
degeneracies, strongly multilevel avoided crossings, or the degenerate
driver and problem endpoints encountered in some TFIM benchmarks. When
$\Lambda(s)$ becomes large, higher-order corrections may also invalidate
a quantitative first-order transition estimate. In such regimes,
$\Lambda(s)$ is used only as a regularized indicator of candidate
slowdown regions.

The score also inherits uncertainty from the learned gap, excited state,
and transition matrix element. Agreement of the variational energies
alone does not guarantee an accurate $M_{01}(s)$ or schedule density.
Reference validation of the gap and transition matrix element is
therefore required wherever exact, sparse-Lanczos, or analytical results
are available.

Finally, the derivation concerns the unembedded closed-system logical
Hamiltonian, not the embedded finite-temperature processor. Accordingly,
$\Lambda(s)$ is neither a microscopic hardware transition rate nor a
processor excitation probability. The physical and statistical
limitations of this distinction are consolidated in
Sec.~\ref{sec:hardware_interpretive_limitations} and
Appendix~\ref{app:dwave_hardware_diagnostics}; the resulting schedules
must ultimately be assessed through empirical hardware performance.

\section{Discrete Hardware Projection of the Adaptive Schedule}
\label{app:schedule_projection}
\label{annealing_schedule}

This appendix describes the finite-control-point projection of the
continuous schedule defined in Sec.~\ref{sec:foapt_schedule}. It does not
redefine the logical FOAPT score, normalized time density, cumulative
map, or continuous schedule. Instead, it takes their discretized values
from Sec.~\ref{sec:foapt_schedule} as inputs and converts the continuous
schedule into a monotone piecewise-linear control-point list
\begin{equation}
\{(t_\ell,s_\ell)\}_{\ell=0}^{K_{\rm cp}-1}
\label{eq:control_points_appendixC}
\end{equation}
that satisfies the hardware control-point and ramp-rate constraints.

The projection changes only the traversal rate along the fixed
annealing path; it does not modify the submitted logical coefficients
$(h,J)$. We write $K=K_{\rm cp}-1$ for the number of piecewise-linear
segments, with
\begin{equation}
0=s_0<s_1<\cdots<s_K=1,
\qquad
0=t_0<t_1<\cdots<t_K=T_{\rm sched}.
\label{eq:appendixC_knot_convention}
\end{equation}

\subsection{Projection inputs and control-point placement}

On the anneal grid $\{s_r\}_{r=0}^{R}$,
Sec.~\ref{sec:foapt_schedule} provides the tabulated logical score
$\Lambda(s_r)$, normalized time density $\rho(s_r)$, and cumulative map
$\tau(s_r)$. The projection does not introduce an alternative estimator
or normalization for these quantities. A monotone interpolation of the
tabulated cumulative map is used to evaluate its inverse.

To obtain $K_{\rm cp}=K+1$ control points, we choose uniformly spaced
normalized-time values
\begin{equation}
\tau_\ell
=
\frac{\ell}{K},
\qquad
\ell=0,\ldots,K,
\label{eq:tau_knots_appendixC}
\end{equation}
and determine the corresponding anneal fractions from
\begin{equation}
s_\ell
=
\tau^{-1}(\tau_\ell).
\label{eq:s_knots_appendixC}
\end{equation}
Equivalently,
\begin{equation}
\int_{s_\ell}^{s_{\ell+1}}
\rho(u)\,\mathrm{d}u
\approx
\frac{1}{K},
\label{eq:equal_mass_appendixC}
\end{equation}
so that the control-point density is greater in regions assigned more
time by the continuous schedule. The reported hardware schedules use
$K_{\rm cp}=12$, corresponding to $K=11$ piecewise-linear segments.

\subsection{Segment timing and ramp-rate enforcement}

For each segment, define
\begin{equation}
\Delta s_\ell
=
s_{\ell+1}-s_\ell,
\qquad
\ell=0,\ldots,K-1.
\label{eq:delta_s_appendixC}
\end{equation}
The hardware ramp-rate constraint is
\begin{equation}
\frac{\Delta s_\ell}{\Delta t_\ell}
\leq
\gamma_{\max},
\qquad
\gamma_{\max}
=
2.0~(\mu{\rm s})^{-1}.
\label{eq:slope_cap_appendixC}
\end{equation}
Hence each segment requires a minimum duration
\begin{equation}
\Delta t_\ell^{\min}
=
\frac{\Delta s_\ell}{\gamma_{\max}},
\label{eq:min_segment_time_appendixC}
\end{equation}
and the minimum feasible total duration is
\begin{equation}
T_{\min}
=
\sum_{\ell=0}^{K-1}\Delta t_\ell^{\min}.
\label{eq:Tmin_appendixC}
\end{equation}

If the requested duration satisfies $T\geq T_{\min}$, the remaining time
\begin{equation}
T_{\rm rem}
=
T-T_{\min}
\label{eq:Trem_appendixC}
\end{equation}
is distributed according to the segment weights
\begin{equation}
W_\ell
=
\int_{s_\ell}^{s_{\ell+1}}
[\Lambda(u)+\epsilon_\Lambda]\,\mathrm{d}u.
\label{eq:segment_weight_appendixC}
\end{equation}
Here $\Lambda(s)$ and $\epsilon_\Lambda$ are defined in
Sec.~\ref{sec:foapt_schedule}. For $\sum_jW_j>0$, the segment durations
are
\begin{equation}
\Delta t_\ell
=
\Delta t_\ell^{\min}
+
T_{\rm rem}
\frac{W_\ell}
{\displaystyle\sum_{j=0}^{K-1}W_j}.
\label{eq:final_dtk_appendixC}
\end{equation}
If all weights are numerically zero, the remaining time is distributed
uniformly:
\begin{equation}
\Delta t_\ell
=
\Delta t_\ell^{\min}
+
\frac{T_{\rm rem}}{K}.
\label{eq:final_dtk_uniform_appendixC}
\end{equation}
Both choices satisfy
\begin{equation}
\sum_{\ell=0}^{K-1}\Delta t_\ell=T,
\qquad
\frac{\Delta s_\ell}{\Delta t_\ell}
\leq
\gamma_{\max}.
\label{eq:feasible_fixed_T_appendixC}
\end{equation}

If $T<T_{\min}$, the requested duration is infeasible. The minimum
segment durations are not shortened, because doing so would violate
Eq.~\eqref{eq:slope_cap_appendixC}. Instead, we set
\begin{equation}
T_{\rm sched}
=
T_{\min},
\qquad
\Delta t_\ell
=
\Delta t_\ell^{\min}.
\label{eq:infeasible_T_appendixC}
\end{equation}
Thus, in general,
\begin{equation}
T_{\rm sched}
=
\max(T,T_{\min}).
\label{eq:Tsched_appendixC}
\end{equation}
The physical-time knots are then obtained cumulatively:
\begin{equation}
t_0=0,
\qquad
t_{\ell+1}
=
t_\ell+\Delta t_\ell,
\qquad
\ell=0,\ldots,K-1,
\label{eq:time_knots_appendixC}
\end{equation}
with $t_K=T_{\rm sched}$.

\subsection{Uploaded piecewise-linear schedule}

The uploaded schedule is obtained by linear interpolation between
adjacent control points:
\begin{equation}
s_{\rm upload}(t)
=
s_\ell
+
\frac{s_{\ell+1}-s_\ell}
{t_{\ell+1}-t_\ell}
(t-t_\ell),
\qquad
t\in[t_\ell,t_{\ell+1}].
\label{eq:piecewise_s_of_t_appendixC}
\end{equation}
The corresponding programmed real-time energy scales are
\begin{equation}
A_{\rm upload}(t)
=
A\!\left(s_{\rm upload}(t)\right),
\qquad
B_{\rm upload}(t)
=
B\!\left(s_{\rm upload}(t)\right).
\label{eq:At_Bt_appendixC}
\end{equation}
These quantities are the programmed compositions of the
processor-specific functions $A(s)$ and $B(s)$ with the uploaded
schedule. They are not measurements of the instantaneous Hamiltonian or
spectrum of the embedded physical system.

The hardware-projection stage can therefore be summarized as
\begin{equation}
\{\Lambda(s_r),\tau(s_r)\}_{r=0}^{R}
\longrightarrow
\{s_\ell\}
\longrightarrow
\{\Delta t_\ell\}
\longrightarrow
\{(t_\ell,s_\ell)\}
\longrightarrow
s_{\rm upload}(t).
\label{eq:appendixC_projection_pipeline}
\end{equation}
The resulting control-point list is monotone, contains at most
$K_{\rm cp}=12$ points, and satisfies the segment ramp-rate bound. If
the requested duration is infeasible, it is increased only to the
shortest feasible value.

\section{Examples Trained with an MLP NQS}
\label{MLP46}

This appendix presents small-system baselines obtained with a
multilayer-perceptron neural quantum state (MLP NQS). We consider
$N=4$ and $N=6$ transverse-field Ising models (TFIMs) and random
Ising Hamiltonians (RHMs) on the same annealing grid used in the main
text. At each anneal point, the MLP estimates the ground- and
first-excited-state energies $E_0(s)$ and $E_1(s)$ using variational
Monte Carlo and the same deflation procedure employed in the Tx-NQDT
calculations. The results are compared with exact diagonalization (ED).
These examples provide transparent supplementary tests of the numerical
workflow rather than a finite-size scaling or state-of-the-art NQS
benchmark.

When the $N=4,6$ MLP results are shown together with the
$N\geq10$ Tx-NQDT results in Table~\ref{tab:rel_errors}, the entries
involve different ansatzes, representative instances, and optimization
settings. Moreover, the tabulated errors are signed and averaged over
the annealing path, so positive and negative deviations can cancel.
Nonmonotonic signed means, such as a smaller value at $N=10$ than at
$N=6$, therefore do not imply that the variational problem becomes
easier as $N$ increases.

\begin{figure}[t]
\centering
\includegraphics[width=\columnwidth]{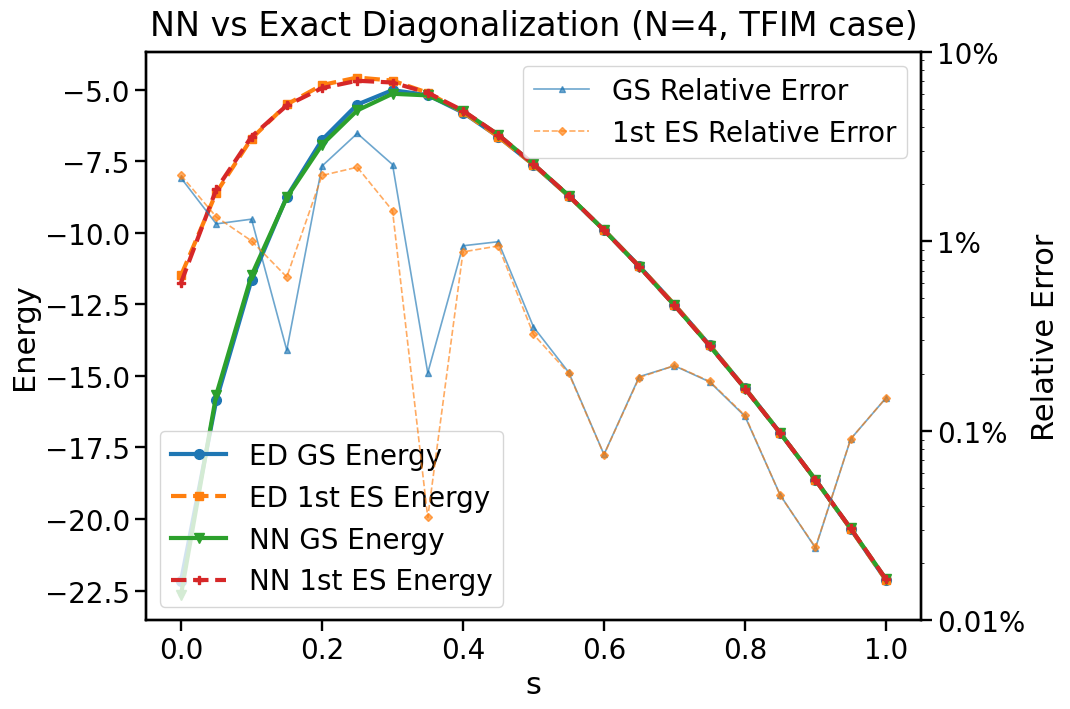}
\caption{
MLP and ED ground- and first-excited-state spectra for the $N=4$ TFIM.
The plotted eigenenergies are reported in frequency-equivalent units,
$E/h$, expressed in GHz.
}
\label{fig:n=4_TFIM}
\end{figure}

\begin{figure}[t]
\centering
\includegraphics[width=\columnwidth]{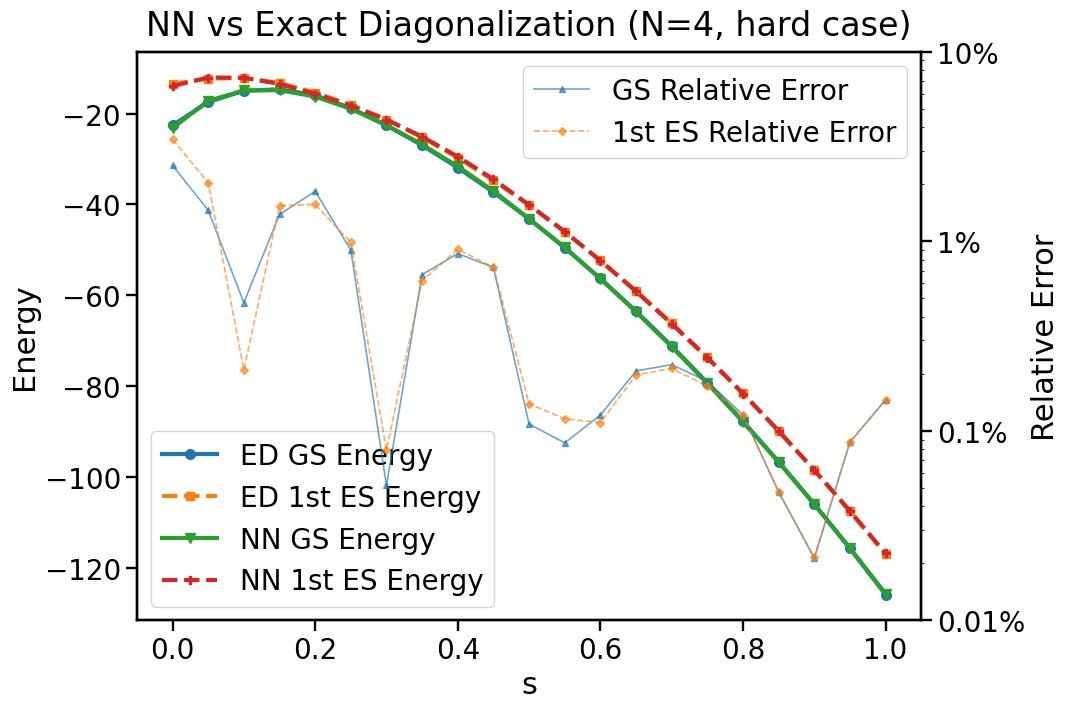}
\caption{
MLP and ED ground- and first-excited-state spectra for a representative
hard $N=4$ random Ising instance. The plotted eigenenergies are reported
in frequency-equivalent units, $E/h$, expressed in GHz.
}
\label{fig:n=4_hard}
\end{figure}

\begin{figure}[t]
\centering
\includegraphics[width=\columnwidth]{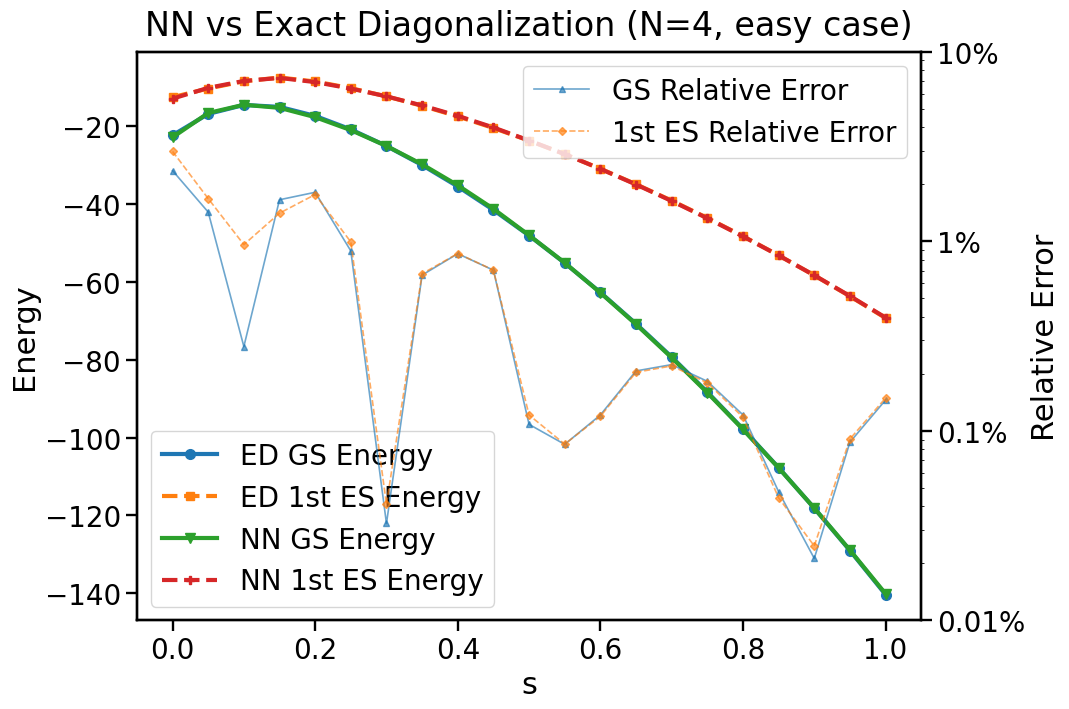}
\caption{
MLP and ED ground- and first-excited-state spectra for a representative
easy $N=4$ random Ising instance. The plotted eigenenergies are reported
in frequency-equivalent units, $E/h$, expressed in GHz.
}
\label{fig:n=4_easy}
\end{figure}

\begin{figure}[t]
\centering
\includegraphics[width=\columnwidth]{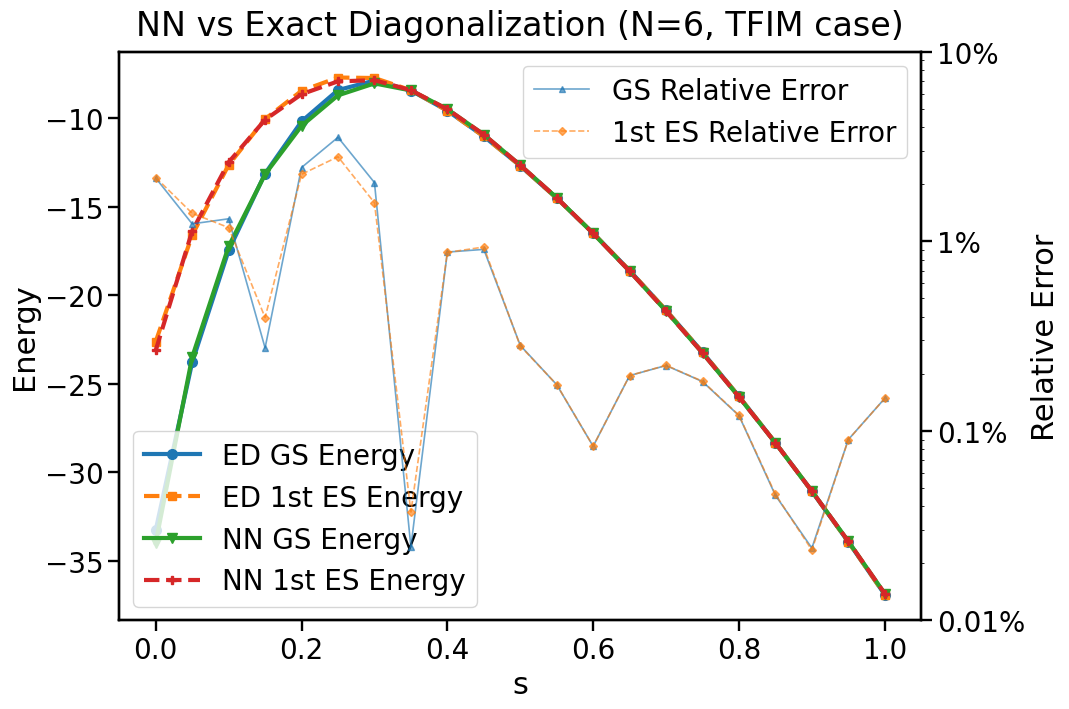}
\caption{
MLP and ED ground- and first-excited-state spectra for the $N=6$ TFIM.
The plotted eigenenergies are reported in frequency-equivalent units,
$E/h$, expressed in GHz.
}
\label{fig:n=6_TFIM}
\end{figure}

Figures~\ref{fig:n=4_TFIM}--\ref{fig:n=6_TFIM} show representative
spectra. Their right axes report the absolute relative energy error
\begin{equation}
\mathrm{RE}_k(s)
=
\frac{
\left|
E_k^{\mathrm{MLP}}(s)-E_k^{\mathrm{ED}}(s)
\right|
}{
\left|E_k^{\mathrm{ED}}(s)\right|+\epsilon_E
},
\qquad
k=0,1,
\label{eq:mlp_relative_error}
\end{equation}
where $\epsilon_E>0$ regularizes the denominator when necessary. The
learned energies track the ED curves closely for these representative
small systems. Excited-state estimates are generally more sensitive near
small gaps, avoided crossings, degeneracies, or regions where the
deflated target is difficult to isolate.

The MLP calculations use the same Hamiltonian convention, annealing
functions $A(s)$ and $B(s)$, Monte Carlo sampling protocol,
runtime-controlled stopping criteria, deflation method, and warm-start
strategy as the corresponding Transformer calculations; only the
wavefunction parameterization is changed. Agreement with ED therefore
tests the estimator, ground- and excited-state optimization, overlap
control, deflation, and anneal-path training independently of the
Transformer backbone.

The principal spectral validation and runtime study use Tx-NQDT at
$N=10,15,$ and $20$. ED is used at $N=10$ and $N=15$, while
low-lying sparse-Lanczos calculations provide reference eigenvalues at
$N=20$. The small-$N$ MLP results are therefore supplementary
implementation tests rather than evidence for Tx-NQDT scalability or
state-of-the-art accuracy. Moreover, agreement of the energies alone
does not certify the wavefunctions, overlaps, transition matrix elements,
nodal structure, gap locations, or FOAPT-derived schedule densities,
which may be more sensitive to variational errors.

\FloatBarrier

\section{Time Complexity Analysis and Comparison with Exact Diagonalization}
\label{app:complexity}

This appendix compares the practical computational costs of Tx-NQDT,
dense exact diagonalization (ED), and sparse low-lying eigensolvers. The
estimates below are implementation-level cost models and do not establish
polynomial worst-case complexity or asymptotic computational advantage
for arbitrary problem instances.

\subsection{Dense and sparse reference calculations}

For $N$ qubits, the Hilbert-space dimension is
\begin{equation}
D=2^N.
\end{equation}
A dense Hamiltonian requires
$\mathcal{O}(D^2)=\mathcal{O}(4^N)$ memory, while full dense
diagonalization has nominal cost
\begin{equation}
T_{\rm dense}
=
\mathcal{O}(D^3)
=
\mathcal{O}(2^{3N}).
\label{eq:dense_ed_cost}
\end{equation}
These costs become prohibitive when the spectrum must be recomputed at
many anneal points.

The transverse-field Ising Hamiltonians considered here are sparse in the
computational basis: the problem Hamiltonian is diagonal, and the driver
connects each basis configuration to its $N$ single-spin-flip
neighbors. A matrix-free Hamiltonian--vector product therefore costs
approximately
\begin{equation}
T_{\rm matvec}
=
\mathcal{O}\!\left[(N+|E|)2^N\right],
\label{eq:sparse_matvec_cost}
\end{equation}
where $|E|$ is the number of nonzero Ising couplings. Thus
$|E|=\mathcal{O}(N)$ for bounded-degree graphs and
$|E|=\Theta(N^2)$ for the dense random instances used here.

If $n_{\rm Krylov}$ iterations are required to obtain the lowest few
eigenpairs at each of $n_s$ anneal points, the corresponding
matrix-free Lanczos cost is approximately
\begin{equation}
T_{\rm Lanczos}
=
\mathcal{O}\!\left[
n_s n_{\rm Krylov}(N+|E|)2^N
\right].
\label{eq:lanczos_cost}
\end{equation}
This remains exponential because each Krylov vector has $2^N$
components, but it is far cheaper than dense full-spectrum
diagonalization.

This distinction is important for the numerical benchmarks. Dense ED is
used at $N=10$ and $N=15$. At $N=20$, repeated dense
full-spectrum diagonalization is not used, but matrix-free Lanczos remains
feasible for the lowest few eigenvalues and is used for reference
calculations. Thus $N=20$ lies beyond convenient repeated dense ED in
the present workflow, not beyond all classical low-energy eigensolvers.

\subsection{Tx-NQDT model-evaluation cost}

Let $L$ denote the number of Transformer blocks, $d$ the token
dimension, and $|E_{\rm att}|$ the number of allowed attention edges.
A graph-masked Transformer forward pass has approximate cost
\begin{equation}
T_{\rm fwd}
=
\mathcal{O}\!\left[
L\left(Nd^2+|E_{\rm att}|d\right)
\right],
\label{eq:fwd-cost}
\end{equation}
up to constants from projections, feed-forward layers, normalization, and
multiple attention heads. For a $k$-hop mask,
$|E_{\rm att}|$ denotes the size of the enlarged attention graph rather
than the original Ising graph.

For bounded-degree sparse graphs,
$|E_{\rm att}|=\mathcal{O}(N)$, so the attention term is approximately
linear in $N$ for fixed $L$ and $d$. For the dense random
Hamiltonians,
$|E_{\rm att}|=\Theta(N^2)$, giving
\begin{equation}
T_{\rm fwd}^{\rm dense}
=
\mathcal{O}\!\left[
L\left(Nd^2+N^2d\right)
\right].
\label{eq:fwd_dense}
\end{equation}
Graph masking therefore provides its largest computational advantage on
sparse or moderately sparse logical graphs; it does not remove the
quadratic graph dependence of dense instances.

\subsection{VMC training cost}

Let $M$ be the number of Monte Carlo samples used in one optimization
epoch. The per-epoch cost includes sampling, wavefunction evaluation, and
local-energy evaluation:
\begin{equation}
T_{\rm epoch}
=
\mathcal{O}\!\left[
M\left(T_{\rm fwd}+T_{\rm loc}\right)
\right].
\label{eq:epoch_cost}
\end{equation}
The diagonal Ising contribution costs
$\mathcal{O}(|E|)$ per configuration. The transverse-field contribution
requires the amplitudes or amplitude ratios of $N$ single-spin-flipped
configurations, so a conservative local-energy cost is
\begin{equation}
T_{\rm loc}
=
\mathcal{O}\!\left(
|E|+NC_{\rm ratio}
\right),
\label{eq:local_energy_cost}
\end{equation}
where $C_{\rm ratio}$ is the effective cost of one amplitude ratio.
Batching, caching, or incremental evaluation can reduce constants but not
the dependence on sample count, graph density, and model complexity.

If $n_{\rm epoch}(s)$ epochs are used at anneal coordinate $s$, then
\begin{equation}
T_{\rm train}(s)
=
n_{\rm epoch}(s)T_{\rm epoch}(s).
\label{eq:train_s_cost}
\end{equation}
The required number of epochs is not fixed by $N$ alone. It depends on
the instance, ansatz expressivity, optimizer, Monte Carlo variance,
sampler mixing, spectral gaps, and excited-state conditioning. Tx-NQDT
avoids explicit storage of a $2^N$-component state vector, but its
runtime can still increase through larger models, longer training, or
larger sample requirements.

The reported implementation uses AdamW rather than stochastic
reconfiguration. A full or approximate natural-gradient method would add
the cost of constructing and solving a quantum-geometric
preconditioning problem, while potentially reducing the number of
optimization epochs. That trade-off is not included in the timings
reported here.

\subsection{Training along the annealing path}

For $n_s$ anneal points, the total ground-state training cost is
\begin{equation}
T_{\rm path}
=
\sum_{r=0}^{n_s-1}
n_{\rm epoch}(s_r)T_{\rm epoch}(s_r).
\label{eq:path_cost_general}
\end{equation}
The reported experiments use $n_s=21$. Parameters learned at $s_r$
initialize training at $s_{r+1}$, exploiting the typically smooth
variation of the low-energy eigenspaces.

If $n_{\rm epoch}^{(0)}$ denotes a representative cold-start cost and
warm starts reduce subsequent training costs by an average factor
$\alpha\in(0,1]$, a heuristic runtime model is
\begin{equation}
T_{\rm path}
\approx
T_{\rm epoch}n_{\rm epoch}^{(0)}
\left[1+\alpha(n_s-1)\right].
\label{eq:path_cost_heuristic}
\end{equation}
This heuristic model summarizes the practical benefit of warm starting.
Warm starts may provide little benefit near narrow avoided crossings or
degeneracies, and they can propagate variational errors between adjacent
grid points. Local grid refinement, additional restarts, or a
parameter-conditioned NQS may therefore increase or reduce the practical
cost depending on the instance.

\subsection{Excited-state and matrix-element costs}

The present work targets the first excited state using the deflation and
overlap-penalty procedure of Sec.~\ref{sec:brauer}. If
$K_{\rm exc}$ excited states are trained separately, a simple practical
cost model is
\begin{equation}
T_{\rm exc}
\approx
\sum_{k=1}^{K_{\rm exc}}
\beta_k T_{\rm path},
\label{eq:excited_cost}
\end{equation}
where $\beta_k$ is the cost of the $k$-th excited-state path relative
to the ground-state path. In this work $K_{\rm exc}=1$.

The value of $\beta_k$ depends on whether the backbone is shared,
frozen, or retrained, and on the cost of overlap and transition-matrix
estimators. Sequential deflation is computationally convenient but can
propagate lower-state errors. Simultaneous multi-state optimization may
improve excited-state and matrix-element accuracy, although it introduces
a more expensive coupled objective.

Once the state estimates are available, evaluating
$E_0(s_r)$, $E_1(s_r)$, and
$M_{01}(s_r)$ requires additional Monte Carlo estimators. Their cost is
normally comparable to or smaller than the training cost, but their
variance can be substantially larger than that of the energies,
particularly for sign-changing excited states.

\subsection{Schedule-construction cost}

After obtaining
\begin{equation}
\left\{
E_0(s_r),E_1(s_r),M_{01}(s_r)
\right\}_{r=0}^{n_s-1},
\end{equation}
the FOAPT-informed score defined in Eq.~\eqref{eq:lambda_def} is
evaluated over the anneal grid in $\mathcal{O}(n_s)$ time. Constructing
the normalized density, cumulative map, inverse schedule, and
$K_{\rm cp}$ hardware control points requires
\begin{equation}
T_{\rm schedule}
=
\mathcal{O}(n_s+K_{\rm cp}),
\label{eq:schedule_cost}
\end{equation}
which is negligible compared with the spectral-learning cost.

\subsection{Comparison and scope of the scaling claims}

The relevant comparison is between the sparse-reference cost
\begin{equation}
T_{\rm Lanczos}
=
\mathcal{O}\!\left[
n_s n_{\rm Krylov}(N+|E|)2^N
\right]
\end{equation}
and the empirical Tx-NQDT cost
\begin{equation}
T_{\rm path}
=
\sum_{r=0}^{n_s-1}
n_{\rm epoch}(s_r)
\,
\mathcal{O}\!\left[
M\left(
T_{\rm fwd}(N,|E_{\rm att}|)
+
T_{\rm loc}(N,|E|)
\right)
\right].
\label{eq:txnqdt_total_cost}
\end{equation}
Lanczos manipulates vectors in the full Hilbert space and therefore
retains an explicit exponential dependence on $N$. Tx-NQDT replaces
these vectors with a parameterized, sample-based wavefunction whose
per-epoch cost is controlled by graph density, model size, sample count,
and local-energy evaluation.

For fixed $L$, $d$, and $M$, the model-evaluation cost can be
approximately linear in $N$ on bounded-degree graphs and at least
quadratic on dense graphs. This statement concerns the cost of one epoch.
The total training cost additionally depends on the model size, number of
samples, autocorrelation time, and number of optimization epochs, each of
which may grow with problem difficulty or system size.

Two practical qualifications are important.

First, the nominal sample count $M$ is not the number of independent
samples when the Markov chain is correlated. If the integrated
autocorrelation time is $\tau_{\rm int}$, the effective sample size is
schematically
\begin{equation}
M_{\rm eff}
\sim
\frac{M}{2\tau_{\rm int}},
\label{eq:effective_sample_complexity}
\end{equation}
up to convention-dependent factors. Slow mixing can therefore increase
the sampling cost substantially, especially near difficult spectral
regions. This motivates effective-sample-size monitoring, improved
proposals, and direct or autoregressive sampling.

Second, for small and intermediate systems, ED or sparse Lanczos can be
both faster and more reliable than training a Transformer. These methods
are therefore retained as validation tools wherever feasible. The role
of Tx-NQDT is not to replace reference eigensolvers in regimes where they
are efficient, but to provide a reusable spectral surrogate and
schedule-generation workflow when repeated dense calculations become
impractical and full Hilbert-space Krylov calculations become increasingly
costly.

Overall, Tx-NQDT does not establish a universal polynomial-time
replacement for exact spectral methods. Its practical advantage is that
it avoids dense Hilbert-space storage, represents the low-energy states
through a trainable sampled ansatz, and reuses the learned spectral
information for gap estimation, transition-matrix-element evaluation, and
FOAPT-informed schedule construction. The measured runtime growth over
the limited range $N\leq20$ should therefore be interpreted as an
implementation-level scaling diagnostic rather than evidence of
asymptotic computational advantage.

\section{Closed-System Duration Scan for Linear Annealing}
\label{app:linear_duration_scan}

To examine the role of total annealing time, we performed a
closed-system duration scan of the linear schedule for a representative
unembedded logical instance. For each duration $T$, the state was
evolved according to the Schr\"odinger equation along the same logical
Hamiltonian path using
\begin{equation}
s_{\rm lin}(t)
=
\frac{t}{T},
\qquad
t\in[0,T].
\label{eq:linear_schedule_duration_scan}
\end{equation}
We then evaluated the final population of the first-excited eigenstate of
the final Hamiltonian,
\begin{equation}
P_{0\to1}(T)
=
\left|
\left\langle
\psi_1(1)
\middle|
\psi(T)
\right\rangle
\right|^2,
\label{eq:p01_duration_scan}
\end{equation}
where $\lvert\psi(T)\rangle$ is the state obtained from the
closed-system evolution and $\lvert\psi_1(1)\rangle$ is the final
first-excited eigenstate.

\begin{figure}[t]
\centering
\includegraphics[width=\columnwidth]{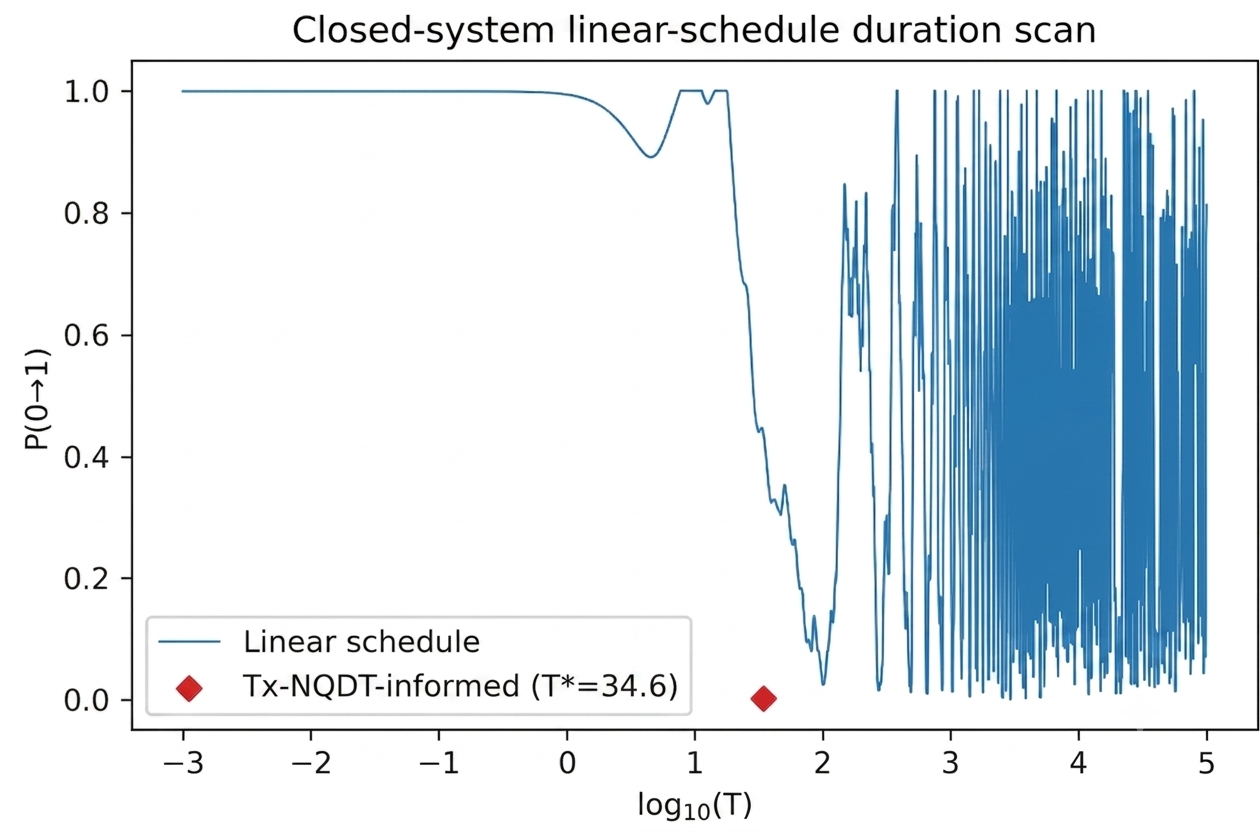}
\caption{
Closed-system duration scan of the linear schedule for a representative
unembedded logical instance. The blue curve shows the final
first-excited-state probability $P_{0\to1}(T)$ versus
$\log_{10}T$. Its nonmonotonic dependence on $T$ results from
phase-sensitive coherent finite-time dynamics. The red marker denotes
the Tx-NQDT-informed nonlinear schedule for the same instance at
$T^\ast=34.6\,\mu s$.
}
\label{fig:linear_duration_scan}
\end{figure}

As shown in Fig.~\ref{fig:linear_duration_scan},
$P_{0\to1}(T)$ is strongly nonmonotonic. Increasing $T$ therefore
does not necessarily reduce the final excitation probability over the
finite-duration range shown. This behavior arises from coherent
interference between nonadiabatic amplitudes accumulated at different
parts of the annealing path, whose relative phases depend on the
dynamical phase accumulated during the evolution. A longer linear
schedule is consequently not automatically a stronger baseline in this
closed-system reference calculation.

The red marker shows the Tx-NQDT-informed nonlinear schedule for the same
logical instance, with total duration $T^\ast=34.6\,\mu s$. It provides a
reference point for comparing the nonlinear result with the
linear-schedule duration scan.

This single-instance closed-system calculation does not establish that
the nonlinear schedule outperforms every linear duration or that any
difference arises solely from schedule shape, and it does not replace a
matched-duration hardware comparison. The D-Wave processor implements minor-embedded, autoscaled,
finite-temperature open-system dynamics
\cite{king2022coherent}, whereas the available hardware data do not
contain $p_{\rm lin}(T_{\rm NQDT})$ for every tested instance. The scan therefore
provides only theoretical context showing that linear-schedule
performance need not vary monotonically with duration. Isolating
schedule-shape effects on hardware requires direct comparison of
$p_{\rm lin}(T_{\rm NQDT})$ and
$p_{\rm NQDT}(T_{\rm NQDT})$ for each instance.

\clearpage

\bibliography{apssamp}

\end{document}

%% file: Intro.tex
\section{Introduction}

Quantum annealing provides a hardware-efficient framework for solving combinatorial optimization problems, particularly quadratic unconstrained binary optimization (QUBO) and equivalent Ising formulations, by evolving from a simple driver Hamiltonian toward a problem Hamiltonian whose low-energy states encode candidate solutions~\cite{kadowaki1998quantum,farhi2000quantum,albash2018adiabatic,rajak2023quantum}. In the ideal adiabatic limit, sufficiently slow evolution allows the system to remain in its instantaneous ground state throughout the anneal, motivating local-adiabatic and perturbative schedule-design strategies~\cite{roland2002quantum,jansen2007bounds,lidar2009adiabatic}.

In practice, however, annealing times are necessarily finite. Nonadiabatic transitions become most likely near regions where the spectral gap between the ground state and low-lying excited states becomes small or where transition matrix elements become large. Consequently, different stages of the anneal have markedly different dynamical importance: evolution should proceed more slowly near spectral bottlenecks while allowing faster evolution elsewhere. Despite significant advances in quantum annealing hardware, practical performance remains strongly influenced by the choice of annealing schedule. This raises the central question addressed in this work: \emph{How can a logical spectral surrogate efficiently reconstruct the low-energy spectral quantities required to generate adaptive annealing schedules under realistic hardware constraints?}

Addressing this question requires access to the instantaneous spectral structure – ground-state energy, low-lying excited states, and energy gaps – along the full time-dependent evolution. However, present-day quantum annealers cannot directly measure these quantities during execution~\cite{quinton2024}. As a result, effective schedule design depends on accurate classical surrogates capable of tracking spectral evolution under time-dependent Hamiltonians.

Exact diagonalization provides direct spectral information but scales exponentially. Sparse eigensolvers and Lanczos methods extend the accessible range, but remain costly when repeated over many anneal points and instances. Quantum Monte Carlo is powerful for many stoquastic systems but can be expensive for repeated low-energy spectral estimation and is limited by sign problems in more general settings~\cite{foulkes2001quantum,ceperley1980ground,troyer2005computational}. Tensor-network methods are highly effective for one-dimensional and low-entanglement systems but can become costly for dense or highly connected problem graphs~\cite{Schollwock2005RMP,orus2014practical}. This motivates flexible variational representations that combine graph structure, transferability along the annealing path, and access to low-lying excited states.

Neural network quantum states (NQS) provide such a framework~\cite{carleo2017,Lange2024}. They have been developed for ground-state learning, symmetry-adapted optimization, autoregressive representations, neural quantum chemistry, and scalable many-body simulation~\cite{carleo2017,choo2018symmetries,sharir2020deep,pfau2024excited}. Transformer-based variational states have shown strong performance for frustrated spin systems, while improved optimization methods have enhanced the trainability of deep NQS~\cite{viteritti2023transformer,viteritti2025transformer,chen2024empowering}. NQS methods have also been extended to real- and imaginary-time dynamics~\cite{schmitt2020quantum,sinibaldi2023unbiasing,sinibaldi2026neuralgalerkin}, transfer learning across Hamiltonian families~\cite{zen2020transfer}, and excited-state calculations through symmetry resolution, state-specific objectives, orthogonality constraints, deflation, and simultaneous multi-state optimization~\cite{choo2018symmetries,entwistle2023electronic,pfau2024excited,hendry2025grassmann}. Foundation NQS approaches further aim to represent entire Hamiltonian families within a parameter-conditioned ansatz~\cite{rende2025foundation}.

Despite these advances, the use of neural quantum states (NQS) as spectrum-aware surrogates for adaptive quantum-annealing schedule synthesis remains comparatively unexplored. This task differs fundamentally from endpoint optimization and direct time evolution, requiring path-resolved estimates of low-lying states, spectral gaps, transition matrix elements, and other spectral quantities along the annealing path. Accordingly, the objective is to reconstruct the logical spectral information required to generate hardware-compatible adaptive annealing schedules rather than directly predict the final optimization outcome.

In this work, we introduce a \emph{Transformer-based Neural Quantum Digital Twin} (Tx-NQDT) for reconstructing low-lying spectral quantities along quantum-annealing paths. Tx-NQDT combines a graph-aware Transformer ansatz with variational Monte Carlo, transfer learning between neighboring anneal points, and a projector-shift deflation strategy to efficiently reconstruct ground- and first-excited-state energies, spectral gaps, and transition matrix elements required for adaptive schedule construction. The reconstructed spectral gaps and transition matrix elements are integrated with first-order adiabatic perturbation theory (FOAPT) to construct adaptive annealing schedules satisfying practical hardware constraints. In the present framework, FOAPT serves as a closed-system spectral diagnostic of the logical annealing Hamiltonian, identifying regions where slower evolution is expected to reduce diabatic transitions and guiding the placement of piecewise-linear control points along the annealing path. Together, these components provide a practical workflow for generating hardware-compatible adaptive annealing schedules from reconstructed logical spectral information.

We validate Tx-NQDT on transverse-field Ising models and random Ising Hamiltonians, equivalently expressible as QUBOs, at problem sizes of $N=10, 15,$ and $20$, demonstrating accurate reconstruction of ground- and first-excited-state energies, spectral gaps, and minimum-gap locations. The reconstructed spectral information is then used to generate adaptive annealing schedules, which are evaluated on a D-Wave quantum annealer. Across $60$ tested instances, the proposed schedules achieve higher empirical ground-state success probabilities than the default $20\,\mu\mathrm{s}$ linear schedule in $44$ cases, illustrating the practical utility of learned logical spectral information for adaptive quantum-annealing schedule design under realistic hardware conditions.

The remainder of this paper is organized as follows. Section~\ref{sec:methodology} presents the Tx-NQDT architecture, variational training, excited-state construction, and FOAPT-guided schedule synthesis. Section~\ref{sec:experiments} reports numerical validation against exact and sparse-reference calculations. Section~\ref{sec:application} describes the D-Wave implementation, hardware constraints, schedule comparisons, and hardware-level limitations. Section~\ref{sec:conclusion} discusses limitations and future extensions.

%% file: method.tex
\section{Methodology}
\label{sec:methodology}

The proposed Tx-NQDT framework consists of four components: (i) formulation of the logical quantum-annealing Hamiltonian, (ii) neural reconstruction of low-lying spectral quantities, (iii) first-excited-state construction via projector-shift deflation, and (iv) FOAPT-guided adaptive schedule generation. Together, these components transform a logical Ising/QUBO instance into a hardware-compatible candidate annealing schedule. The following subsections describe each component in turn.

\subsection{Logical annealing Hamiltonian and schedule notation}
\label{sec:logical_hamiltonian}

A quadratic unconstrained binary optimization problem is specified by a
real symmetric matrix $Q\in\mathbb{R}^{N\times N}$ and the objective
\begin{equation}
\min_{\mathbf{x}\in\{0,1\}^{N}}
f_Q(\mathbf{x})
=
\mathbf{x}^{\top}Q\mathbf{x}.
\label{eq:qubo}
\end{equation}
Using $x_i=(1+z_i)/2$, with $z_i\in\{-1,+1\}$, this is equivalent,
up to an additive constant, to
\begin{equation}
E_{\rm Ising}(\mathbf{z})
=
\frac{1}{2}
\left(
\sum_i h_i z_i
+
\sum_{i>j}J_{ij}z_i z_j
\right).
\label{eq:ising_objective}
\end{equation}
QUBO and Ising formulations encode broad classes of NP-hard optimization
problems
\cite{karp1972reducibility,garey1979computers,
papadimitriou1998combinatorial,barahona1982computational,
lucas2014ising,glover2019qubo,chen2025benchmarking,koch:2025}.

The logical problem and driver Hamiltonians are
\begin{align}
H_{\rm P}
&=
\frac{1}{2}
\left(
\sum_i h_i\hat{\sigma}_z^{(i)}
+
\sum_{i>j}J_{ij}
\hat{\sigma}_z^{(i)}
\hat{\sigma}_z^{(j)}
\right),
\label{eq:problem_hamiltonian}
\\
H_0
&=
-\frac{1}{2}
\sum_i\hat{\sigma}_x^{(i)}.
\label{eq:driver_hamiltonian}
\end{align}
Tx-NQDT analyzes the closed-system logical path
\begin{equation}
H_{\rm QA}(s)
=
A(s)H_0+B(s)H_{\rm P},
\qquad
s\in[0,1],
\label{eq:H_QA_general}
\end{equation}
or explicitly
\begin{equation}
H_{\rm QA}(s)
=
-\frac{A(s)}{2}
\sum_i\hat{\sigma}_x^{(i)}
+
\frac{B(s)}{2}
\left(
\sum_i h_i\hat{\sigma}_z^{(i)}
+
\sum_{i>j}J_{ij}
\hat{\sigma}_z^{(i)}
\hat{\sigma}_z^{(j)}
\right).
\label{eq:H_QA_explicit}
\end{equation}

The dimensionless coordinate $s$ labels the Hamiltonian path, whereas
$t\in[0,T]$ denotes physical time. An annealing schedule is a monotone
map
\begin{equation}
s=s(t),
\qquad
s(0)=0,
\qquad
s(T)=1.
\label{eq:schedule_map}
\end{equation}
The default linear schedule is
\begin{equation}
s_{\rm lin}(t)=\frac{t}{T},
\label{eq:linear_schedule}
\end{equation}
whereas the Tx-NQDT-informed schedule is denoted by
$s_{\rm ad}(t)$ and constructed in
Sec.~\ref{sec:foapt_schedule}. Thus, $A(s)$ and $B(s)$ specify the fixed
annealing path, while $s(t)$ determines its traversal in physical time.

Equation~\eqref{eq:H_QA_explicit} is the unembedded logical
Hamiltonian reconstructed by Tx-NQDT; its distinction from the physical
processor implementation is discussed in
Sec.~\ref{sec:hardware_interpretive_limitations}. The experimental
implementation is described in Sec.~\ref{sec:application} and
Appendix~\ref{app:dwave_implementation}, while unresolved hardware-level
systematics are collected in
Appendix~\ref{app:dwave_hardware_diagnostics}.

We denote the two lowest logical eigenpairs by
\begin{equation}
H_{\rm QA}(s)\lvert\Psi_k(s)\rangle
=
E_k(s)\lvert\Psi_k(s)\rangle,
\qquad
k=0,1,
\end{equation}
with instantaneous gap
\begin{equation}
\Delta(s)
=
E_1(s)-E_0(s).
\label{eq:gap_definition_method}
\end{equation}
Because the Hamiltonians considered here are real in the computational
basis, the reported calculations use real-valued variational
wavefunctions.

\subsection{Graph-aware neural quantum-state surrogate}
\label{sec:nnqsst}

For each instance $(h,J)$ and anneal coordinate $s$, Tx-NQDT
estimates $E_0(s)$, $E_1(s)$, $\Delta(s)$, and
\begin{equation}
M_{01}(s)
=
\left\langle
\Psi_0(s)
\middle|
\frac{\mathrm{d}H_{\rm QA}}{\mathrm{d}s}
\middle|
\Psi_1(s)
\right\rangle.
\label{eq:M01_method}
\end{equation}
Because gaps, overlaps, and transition matrix elements can be more
sensitive to state errors than variational energies, these quantities are
validated individually wherever suitable references are available.

Each spin is represented by a token containing its computational-basis
value $x_i$, local field $h_i$, graph degree
\begin{equation}
d_i
=
\sum_j\mathbf{1}_{\{J_{ij}\neq0\}},
\label{eq:degree_feature}
\end{equation}
and weighted degree
\begin{equation}
w_i
=
\sum_j|J_{ij}|.
\label{eq:weighted_degree_feature}
\end{equation}
The annealing context $(s,A(s),B(s))$ is supplied through a global
token and feature-wise conditioning
\cite{devlin2019bert,dosovitskiy2020vit,perez2018film}.
Laplacian and random-walk positional features encode graph topology, and
attention is masked by the logical coupling graph
\cite{dwivedi2021lspe,chen2022sat}. The mask yields sparse attention for
sparse graphs but becomes effectively dense for the fully connected
random instances.

For the stoquastic ground state, we use
\begin{equation}
\Psi_{0,\theta}(x;s)
=
\exp\!\left[g_{0,\theta}(x;s)\right],
\label{eq:ground_state_head}
\end{equation}
which enforces positive amplitudes. The first-excited-state head is
real-valued and sign-unconstrained,
\begin{equation}
\Psi_{1,\theta}(x;s)
=
\widetilde{g}_{1,\theta}(x;s).
\label{eq:excited_state_head}
\end{equation}
The normalized sampling distribution is
\begin{equation}
p_{k,\theta}(x;s)
=
\frac{
|\Psi_{k,\theta}(x;s)|^2
}{
\sum_{x'}|\Psi_{k,\theta}(x';s)|^2
}.
\label{eq:born_probability}
\end{equation}
For sufficiently small systems, the sums are evaluated exactly.
Otherwise, configurations are generated using
Metropolis--Hastings single-spin-flip updates with acceptance probability
\begin{equation}
a(x\rightarrow x')
=
\min\!\left\{
1,
\frac{|\Psi_\theta(x';s)|^2}
     {|\Psi_\theta(x;s)|^2}
\right\}.
\label{eq:mh_acceptance}
\end{equation}

For fixed $s$, the local and variational energies are
\begin{align}
E_{\rm loc}(x;\theta,s)
&=
\frac{
\langle x|H_{\rm QA}(s)|\Psi_\theta\rangle
}{
\Psi_\theta(x;s)
},
\label{eq:local_energy}
\\
E_\theta(s)
&=
\mathbb{E}_{p_\theta}
\left[
E_{\rm loc}(x;\theta,s)
\right].
\label{eq:vmc_energy}
\end{align}
The reported calculations use the score-function gradient
\begin{equation}
\nabla_\theta E_\theta(s)
\approx
2\,\mathbb{E}_{p_\theta}
\left[
\left(E_{\rm loc}-\bar E_\theta\right)
\nabla_\theta\log|\Psi_\theta|
\right],
\label{eq:vmc_gradient}
\end{equation}
optimized with AdamW. This is \emph{not} a stochastic-reconfiguration
(SR) update. SR instead solves
\begin{equation}
\left(S+\lambda_{\rm SR}I\right)\Delta\theta=-g,
\label{eq:sr_linear_system}
\end{equation}
where $S$ estimates the quantum geometric tensor from the covariance of
logarithmic derivatives
\cite{sorella1998,sorella2007}. AdamW was adopted as a stable,
runtime-controlled optimizer.

Equation~\eqref{eq:vmc_gradient} inherits support and variance
limitations when variational amplitudes vanish or become very small,
particularly for sign-changing excited states. Detailed support,
variance, and mitigation considerations are given in
Appendix~\ref{app:vmc_support_variance}.

Parameters learned at $s_i$ initialize the optimization at
$s_{i+1}=s_i+\Delta s$. This warm start exploits smooth spectral
variation and reduces repeated training cost. Its convergence limitations
are discussed in Sec.~\ref{sec:conclusion}.

\subsection{First-excited-state construction}
\label{sec:brauer}

Excited-state NQS methods include symmetry-resolved and state-specific
objectives, orthogonality constraints, and simultaneous multi-state
formulations
\cite{choo2018symmetries,entwistle2023electronic,
pfau2024excited,hendry2025grassmann}. We use sequential projector-shift
deflation to obtain the first excited state required by the two-level
FOAPT diagnostic. 

For a Hermitian operator
\begin{equation}
H
=
\sum_j\lambda_j
\lvert u_j\rangle\langle u_j\rvert,
\label{eq:spectral_decomposition}
\end{equation}
the exact projector update
\begin{equation}
\widetilde H
=
H+\delta
\lvert u_p\rangle\langle u_p\rvert
\label{eq:projector_shift}
\end{equation}
shifts only $\lambda_p$ to $\lambda_p+\delta$, leaving all eigenvectors
and the remaining eigenvalues unchanged. This result follows directly
from Eq.~\eqref{eq:spectral_decomposition}. A complete derivation of the
projector-shift identity, together with its relation to Brauer-type
rank-one eigenvalue shifts and the limitations arising when the
projector is constructed from an approximate variational state, is given
in Appendix~\ref{brauer_proof}.

Using the learned ground state, we define
\begin{equation}
H_{\rm QA}^{(1)}(s)
=
H_{\rm QA}(s)
+
\delta_0(s)
\frac{
\lvert\Psi_{0,\theta}(s)\rangle
\langle\Psi_{0,\theta}(s)\rvert
}{
\langle\Psi_{0,\theta}(s)
\vert\Psi_{0,\theta}(s)\rangle
}.
\label{eq:deflated_hamiltonian}
\end{equation}
For a nondegenerate exact ground state, the original first excited state
becomes the lowest state of the shifted Hamiltonian when
\begin{equation}
\delta_0(s)>\Delta(s).
\label{eq:deflation_shift_condition}
\end{equation}
For an approximate neural ground state, however, the learned projector
need not commute with $H_{\rm QA}(s)$, and the remaining eigenvectors
are not preserved exactly. Equation~\eqref{eq:deflated_hamiltonian} is
therefore a variational deflation surrogate that suppresses the learned
ground-state direction rather than an exact eigenpair-preserving
transformation.

Residual contamination is controlled through
\begin{equation}
\mathcal{L}_{\perp}(s)
=
\lambda_{\perp}
\frac{
|\langle\Psi_0(s)|\Psi_1(s)\rangle|^2
}{
\langle\Psi_0|\Psi_0\rangle
\langle\Psi_1|\Psi_1\rangle
+\epsilon_{\rm norm}
}.
\label{eq:orthogonality_penalty}
\end{equation}
For small systems, the overlap and norms are evaluated exactly.
Otherwise, they are estimated by importance sampling from
\begin{equation}
q(x)
=
\frac{1}{2}p_{0,\theta}(x;s)
+
\frac{1}{2}p_{1,\theta}(x;s),
\label{eq:overlap_mixture_distribution}
\end{equation}
using
\begin{equation}
\langle\Psi_0|\Psi_1\rangle
=
\mathbb{E}_{x\sim q}
\left[
\frac{
\Psi_0^*(x;s)\Psi_1(x;s)
}{
q(x)
}
\right].
\label{eq:overlap_estimator}
\end{equation}
The mixture distribution reduces support mismatch by sampling
configurations important to either state.

A shared Transformer backbone with state-specific heads represents the
two states:
\begin{equation}
\Psi_k(x;s)
=
\mathcal H_k\!\left(
\mathcal T_\phi
(x;h,J,s,A(s),B(s))
\right),
\qquad
k=0,1.
\label{eq:state_heads}
\end{equation}
The ground state is optimized first and then used in the deflated
excited-state objective. Training is warm-started across the two states
and between neighboring anneal points.

\subsection{FOAPT-informed schedule construction}
\label{sec:foapt_schedule}
\label{nas}

Tx-NQDT converts the reconstructed logical spectrum into a continuous
candidate schedule through
\begin{equation}
E_0(s),\,E_1(s),\,M_{01}(s)
\longrightarrow
\Lambda(s)
\longrightarrow
\rho(s)
\longrightarrow
\tau(s)
\longrightarrow
s_{\rm ad}(t).
\label{eq:schedule_pipeline_main}
\end{equation}
This construction is motivated by first-order adiabatic perturbation
theory for an isolated system and is used as a closed-system
feature-construction rule. The detailed derivation, validity conditions,
and limitations of the first-order approximation are given in
Appendix~\ref{app:foapt}.

Because
\begin{equation}
\frac{\mathrm{d}H_{\rm QA}}{\mathrm{d}s}
=
A'(s)H_0+B'(s)H_{\rm P},
\label{eq:dHds}
\end{equation}
the leading two-level adiabatic parameter contains
\begin{equation}
\frac{
\left|
\left\langle
\Psi_0
\middle|
\frac{\mathrm{d}H_{\rm QA}}{\mathrm{d}t}
\middle|
\Psi_1
\right\rangle
\right|
}{
\Delta(s)^2
}
=
\frac{|M_{01}(s)|}{\Delta(s)^2}
\left|
\frac{\mathrm{d}s}{\mathrm{d}t}
\right|.
\label{eq:foapt_parameter}
\end{equation}
We define the regularized logical difficulty score
\begin{equation}
\Lambda(s)
=
\frac{|M_{01}(s)|}
{\Delta(s)^2+\epsilon_\Delta},
\label{eq:lambda_def}
\end{equation}
where $\epsilon_\Delta>0$ prevents numerical instability near very
small learned gaps. Large $\Lambda(s)$ identifies regions with small
gaps, strong ground--excited-state coupling, or both. The quantitative
limitations of this first-order diagnostic are discussed in
Sec.~\ref{sec:conclusion}.

The normalized time density and cumulative map are
\begin{align}
\rho(s)
&=
\frac{
\Lambda(s)+\epsilon_\Lambda
}{
\displaystyle
\int_0^1
[\Lambda(u)+\epsilon_\Lambda]\,\mathrm{d}u
},
\label{eq:rho_def}
\\
\tau(s)
&=
\int_0^s\rho(u)\,\mathrm{d}u,
\label{eq:tau_def}
\end{align}
where $\epsilon_\Lambda>0$ ensures a strictly positive baseline time
density. For a requested duration $T$, the continuous candidate schedule
is
\begin{equation}
s_{\rm ad}(t)
=
\tau^{-1}\!\left(\frac{t}{T}\right),
\qquad
t\in[0,T].
\label{eq:adaptive_schedule}
\end{equation}
Equivalently,
\begin{equation}
\frac{\mathrm{d}t}{\mathrm{d}s}
=
T\rho(s),
\qquad
\frac{\mathrm{d}s_{\rm ad}}{\mathrm{d}t}
=
\frac{1}{T\rho(s)}.
\label{eq:schedule_derivatives}
\end{equation}
Thus, more physical time is allocated to regions of greater reconstructed
logical spectral difficulty.

All quantities entering this construction are evaluated for the
unembedded logical Hamiltonian. Their hardware interpretation is
discussed in Sec.~\ref{sec:hardware_interpretive_limitations} and
Appendix~\ref{app:dwave_hardware_diagnostics}.

The integrals are evaluated on the discrete anneal grid using the
trapezoidal rule. The continuous schedule is then projected onto at most
$K_{\rm cp}=12$ monotone control points with maximum segment ramp rate
$\gamma_{\max}=2.0~(\mu{\rm s})^{-1}$. The complete control-point placement, timing, and
piecewise-linear interpolation procedure is given only in
Appendix~\ref{app:schedule_projection}.

Figure~\ref{fig:txnqdt_pipeline} summarizes the overall Tx-NQDT workflow. Starting from a logical Ising/QUBO instance, the graph-aware neural quantum state reconstructs the low-lying spectral quantities required for FOAPT-guided schedule construction. The resulting candidate schedules are projected onto hardware-compatible control points. In the following sections, we first validate the reconstructed spectral quantities through comparisons with exact or sparse-reference solutions and then evaluate the resulting schedules empirically on a D-Wave quantum annealer.

\begin{figure*}[!t]
\centering
\includegraphics[width=0.8\textwidth]{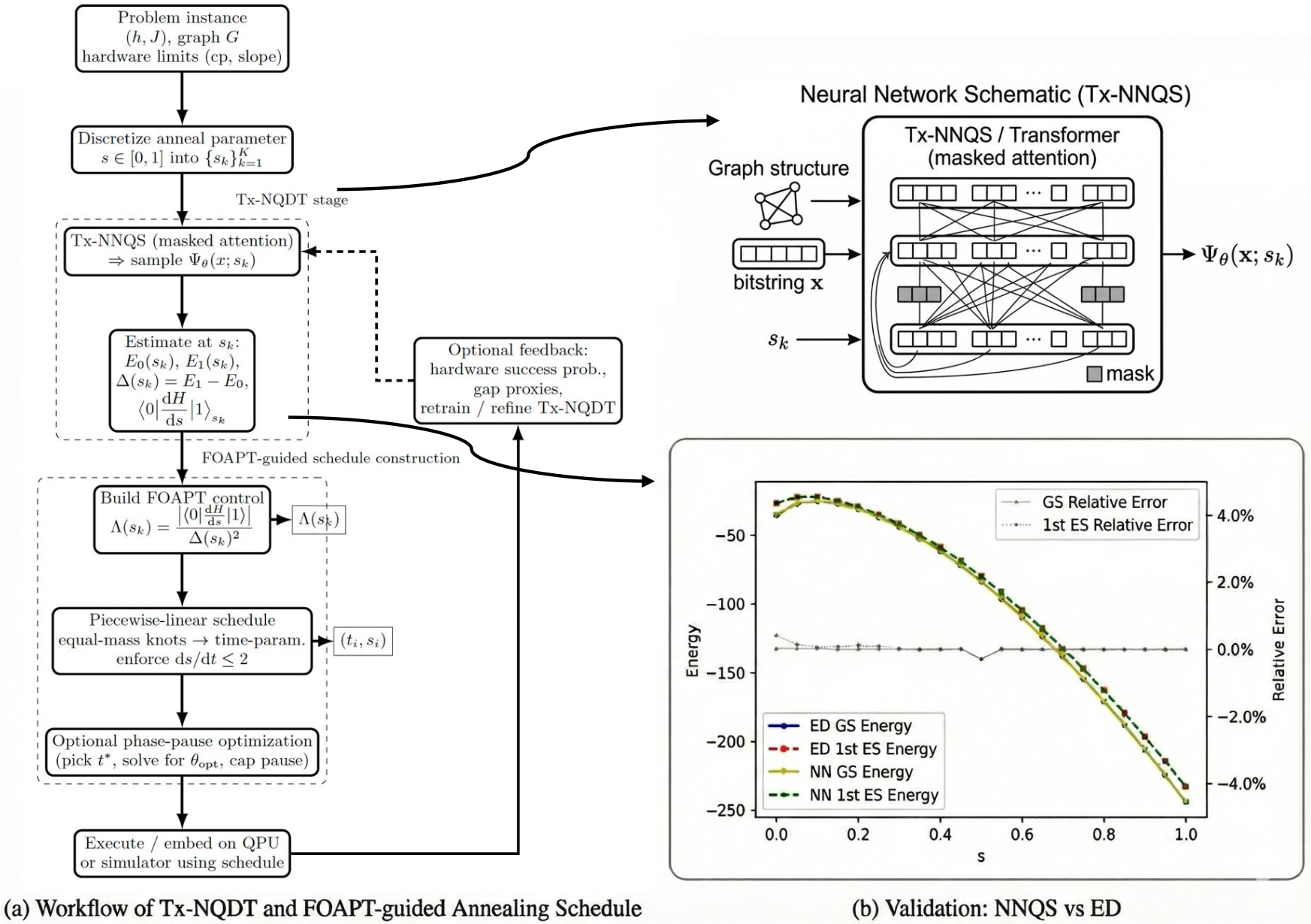}
\caption{
Overview of the Tx-NQDT workflow. Panel~(a) illustrates the complete pipeline, from a logical Ising/QUBO instance through neural spectral reconstruction and FOAPT-guided schedule generation to hardware evaluation. Panel~(b) shows the graph-aware Transformer neural quantum-state architecture. The validation plot corresponds to a representative $N=10$ dense random-Ising benchmark instance.
}
\label{fig:txnqdt_pipeline}
\end{figure*}

%% file: sim.tex
\section{Simulation Experiments}
\label{sec:experiments}

This section evaluates Tx-NQDT as a closed-system spectral surrogate for adaptive quantum-annealing schedule design. The objectives are fourfold: (i) to assess the accuracy of the reconstructed low-lying spectra against exact-diagonalization (ED), sparse-Lanczos, or exact analytical references where available; (ii) to quantify errors in the ground- and first-excited-state energies, spectral gaps, and minimum-gap locations that are central to schedule construction; (iii) to examine the robustness of the proposed method across both structured and disordered Hamiltonians; and (iv) to characterize the practical computational cost of the current implementation over the tested problem sizes.

The principal random-Ising and hardware benchmarks consider problem sizes of $N=10$, $15$, and $20$. Exact diagonalization is used as the reference for $N=10$ and $15$, while sparse-Lanczos calculations provide low-lying spectral references for $N=20$. To further examine scalability in a structured setting, the spectral validation is extended to the open one-dimensional transverse-field Ising model (TFIM) at $N=38$, using the exact Jordan--Wigner solution without constructing the full $2^{38}$-dimensional Hilbert-space Hamiltonian. Auxiliary results for $N=4$ and $N=6$, obtained using the MLP baseline described in Appendix~\ref{MLP46}, provide additional small-system validation.

\subsection{Benchmark models}

\paragraph*{Transverse-field Ising model.}
We benchmark the method on the open one-dimensional ferromagnetic
TFIM, whose low-energy structure provides a
controlled test of spectral reconstruction
\cite{kadowaki1998quantum,santoro2006optimization}. We set
\begin{equation}
h_i=0,
\qquad
J_{i,i+1}=-J_{\rm F},
\qquad
J_{\rm F}>0,
\qquad
J_{ij}=0
\quad\text{otherwise}.
\label{eq:tfim_parameters}
\end{equation}
With the convention of Eq.~\eqref{eq:problem_hamiltonian}, the negative
coupling is ferromagnetic. The absence of longitudinal fields gives a
global spin-flip symmetry and a degenerate ferromagnetic endpoint at
$s=1$; the TFIM is therefore used as a spectral benchmark rather than as
a generic optimization instance. We test whether Tx-NQDT reproduces the two lowest energies, the gap
profile, and the location of the dominant spectral bottleneck. ED is used
at $N=10$ and $N=15$, sparse matrix-free Lanczos provides the two
lowest reference energies at $N=20$, and the exact Jordan--Wigner
solution provides the reference spectrum at $N=38$.

\paragraph*{Random Ising Hamiltonian models.}
To test less structured spectra, we generate dense random Ising
Hamiltonians for $N\in\{10,15,20\}$. The fields $h_i$ and couplings
$J_{ij}$, $i<j$, are sampled independently from the uniform distribution
on $[-5,5]$. The logical graph is therefore complete with probability
one,
\begin{equation}
|E|=\frac{N(N-1)}{2}=\Theta(N^2),
\label{eq:dense_graph_edges}
\end{equation}
so graph-masked attention is effectively dense. ED references are used at
$N=10$ and $N=15$, whereas sparse Lanczos is used at $N=20$ when
reference eigenvalues are required.

Random instances are classified using the logical minimum gap
\begin{equation}
\Delta_{\min}
=
\min_{s\in[0,1]}
\left[E_1(s)-E_0(s)\right].
\label{eq:delta_min_rhm}
\end{equation}
Relatively large-gap instances are labeled \emph{easy}, and small-gap
instances are labeled \emph{hard}. These labels refer exclusively to the logical minimum gap; their hardware
performance is evaluated independently in Sec.~\ref{sec:application}.

\subsection{Training and reference protocol}
\label{sec:training}

The annealing coordinate is evaluated on the uniform grid
\begin{equation}
s_j=0.05j,
\qquad
j=0,\ldots,20.
\label{eq:anneal_grid}
\end{equation}
Local refinement may be used diagnostically near a rapidly varying gap,
but all results reported below use the 21-point grid unless stated
otherwise. At every grid point, Tx-NQDT approximates the ground and first
excited states of the logical Hamiltonian
\begin{equation}
H_{\rm QA}(s)
=
-\frac{A(s)}{2}\sum_i\hat{\sigma}_x^{(i)}
+
\frac{B(s)}{2}
\left(
\sum_i h_i\hat{\sigma}_z^{(i)}
+
\sum_{i>j}J_{ij}
\hat{\sigma}_z^{(i)}
\hat{\sigma}_z^{(j)}
\right).
\label{eq:training_hamiltonian}
\end{equation}
The neural and reference calculations use identical values of
$A(s)$, $B(s)$, $h$, and $J$.

The implementation uses Metropolis--Hastings VMC with single-spin-flip
updates and the ordinary score-function energy gradient described in
Sec.~\ref{sec:nnqsst}. Optimization is performed with AdamW using learning
rate $3\times10^{-4}$, cosine decay after a short warm-up, weight decay
$10^{-4}$, and gradient clipping at norm $1$. The reported calculations
use AdamW rather than stochastic reconfiguration (SR).

The estimator inherits the support and variance limitations summarized
in Sec.~\ref{sec:nnqsst} and analyzed in
Appendix~\ref{app:vmc_support_variance}, particularly for sign-changing
excited states.

Training is terminated using recent energy fluctuations, Markov-chain
effective sample size, gradient norm, and a maximum number of optimization
steps. This is a runtime-control rule, which provides a common runtime
budget. Finite sampling and incomplete optimization may therefore produce
localized errors in the reconstructed spectra.

After the ground state is trained, the first excited state is obtained
using the deflation and overlap-penalty procedure of
Sec.~\ref{sec:brauer}. Parameters at $s_j$ initialize training at
$s_{j+1}$, providing a warm start along the annealing path and improving
runtime and continuity.

At $N=10$ and $N=15$, the learned TFIM spectra are compared with ED
for the same Hamiltonian. At $N=20$, sparse Lanczos is used for the
lowest eigenvalues, while at $N=38$ the open-chain TFIM is compared
with the exact Jordan--Wigner reference. The dense random Ising
benchmarks remain restricted to $N=10,15,$ and $20$, with ED used at
$N=10,15$ and sparse Lanczos at $N=20$. These comparisons assess
whether the runtime-controlled surrogate supplies useful
schedule-generation features.

\subsection{Spectral accuracy and practical scaling}
\label{sec:spectral_accuracy_scaling}

\begin{figure}[t]
\centering
\includegraphics[width=0.6\columnwidth]{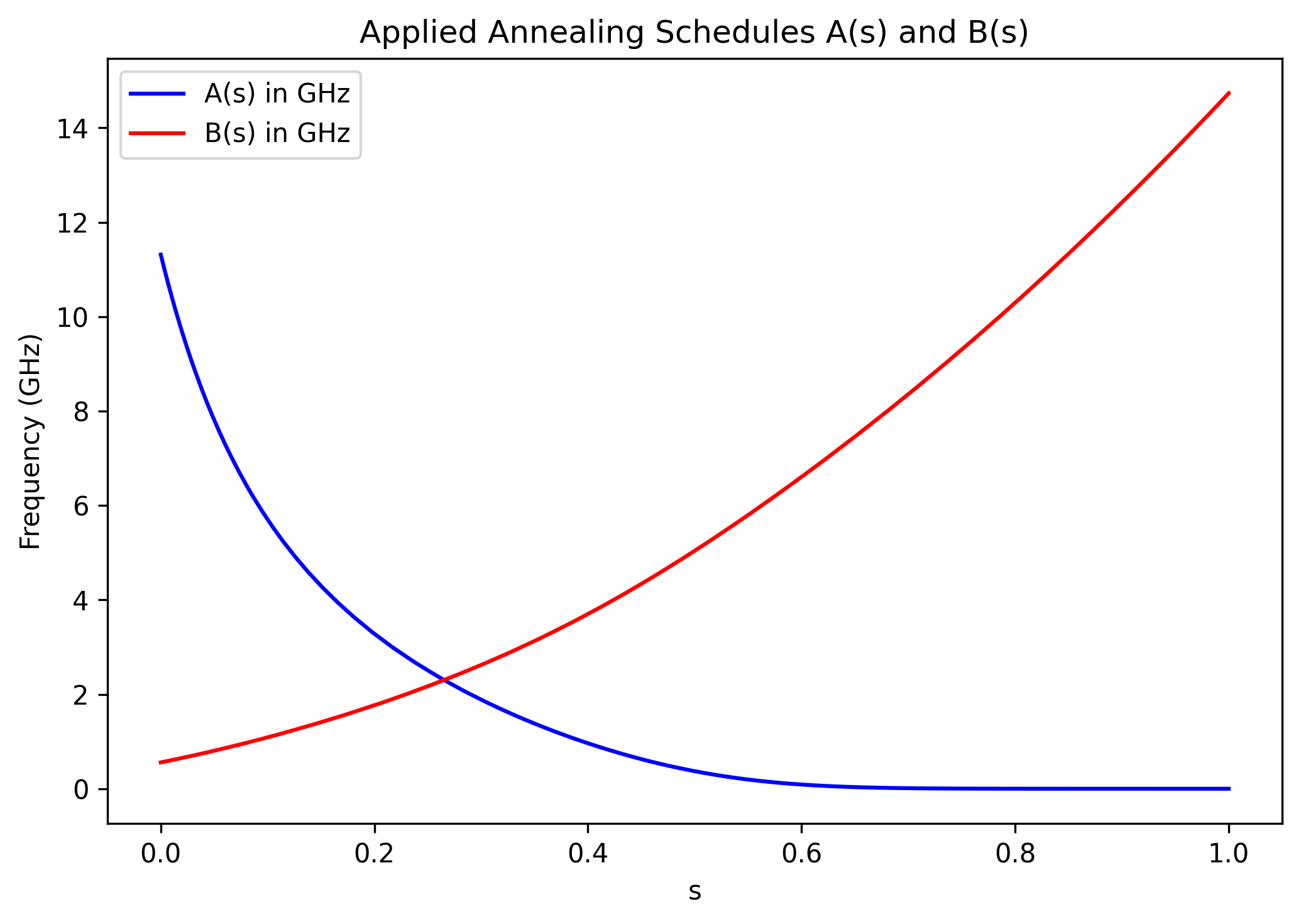}
\caption{
Annealing coefficients $A(s)$ and $B(s)$ defining the logical Hamiltonian
path used in the spectral-reconstruction experiments. Following the
D-Wave convention, the plotted coefficients are frequency-equivalent
energy scales, $A(s)/h$ and $B(s)/h$, expressed in GHz, where $h$ is
Planck's constant.
}
\label{fig:applied_A_B}
\end{figure}

We first assess the accuracy of the reconstructed low-lying spectral quantities and then examine the practical computational scaling of the current implementation. All experiments follow the same annealing path defined by the D-Wave annealing coefficients $A(s)$ and $B(s)$ shown in Fig.~\ref{fig:applied_A_B}.

\begin{figure}[p!]
\centering

\begin{subfigure}[t]{0.47\textwidth}
\centering
\includegraphics[width=\textwidth,height=0.285\textheight,keepaspectratio]{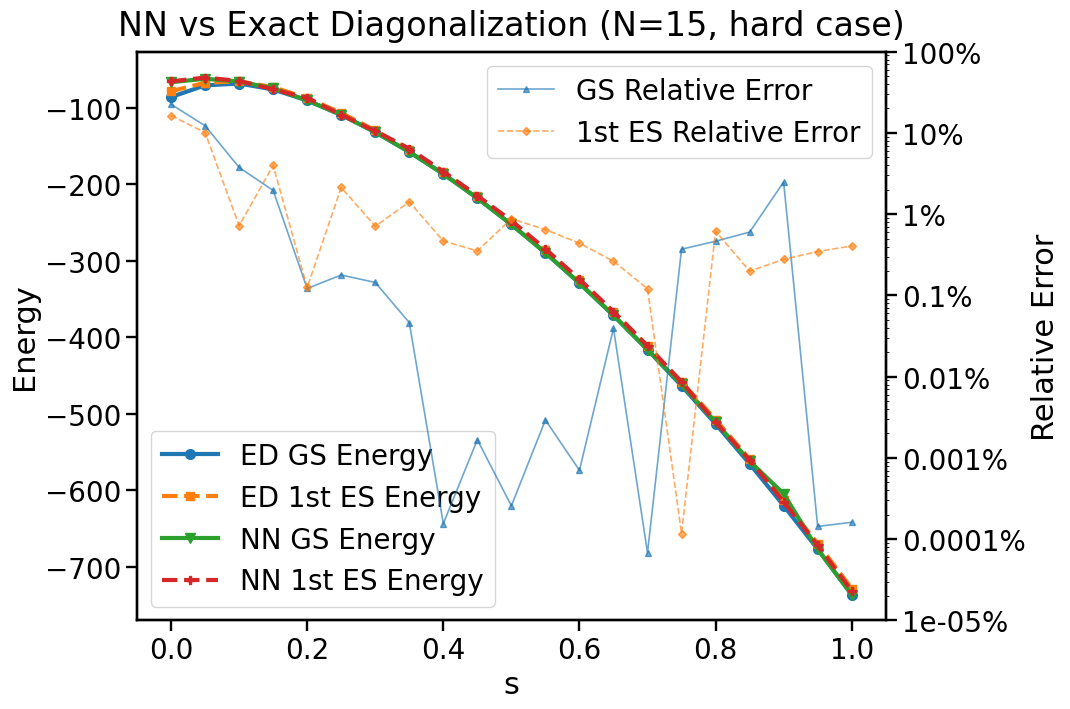}
\caption{$N=15$ hard case}
\label{fig:n15_hard}
\end{subfigure}
\hfill
\begin{subfigure}[t]{0.47\textwidth}
\centering
\includegraphics[width=\textwidth,height=0.285\textheight,keepaspectratio]{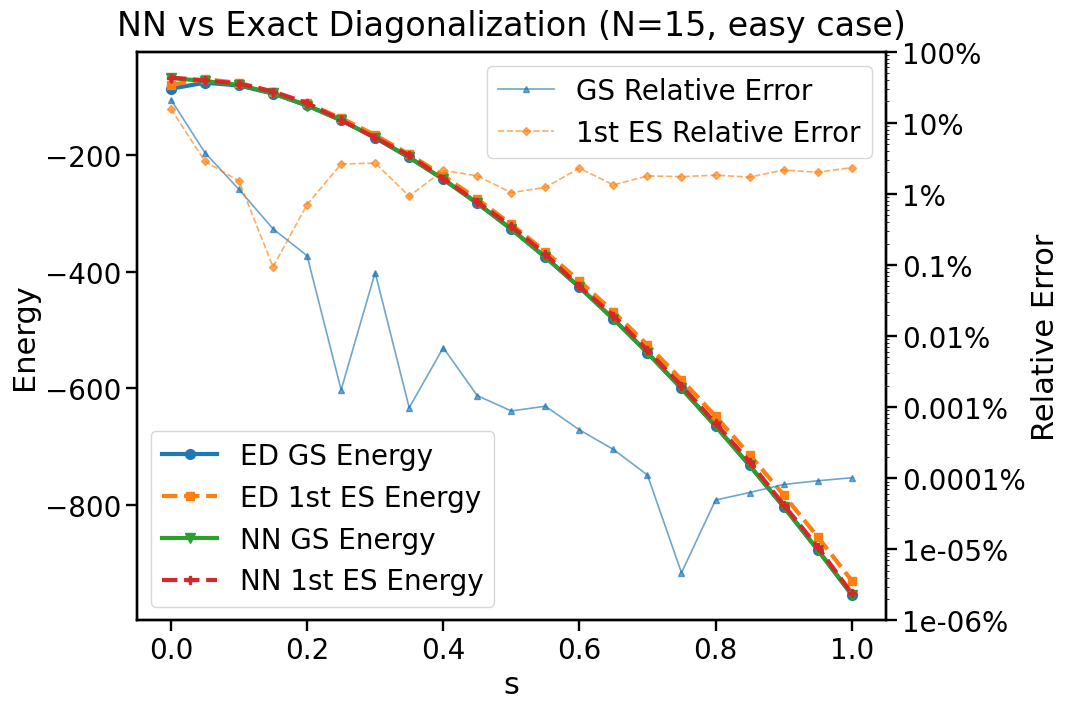}
\caption{$N=15$ easy case}
\label{fig:n15_easy}
\end{subfigure}

\vspace{0.6em}

\begin{subfigure}[t]{0.47\textwidth}
\centering
\includegraphics[width=\textwidth,height=0.285\textheight,keepaspectratio]{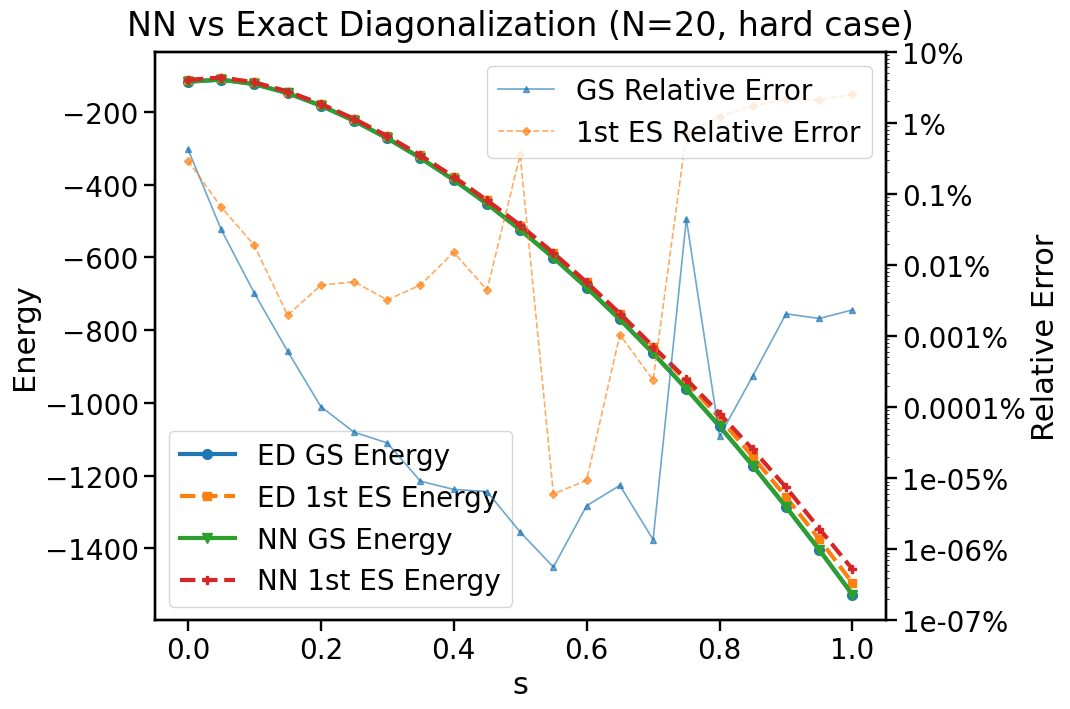}
\caption{$N=20$ hard case}
\label{fig:n20_hard}
\end{subfigure}
\hfill
\begin{subfigure}[t]{0.47\textwidth}
\centering
\includegraphics[width=\textwidth,height=0.285\textheight,keepaspectratio]{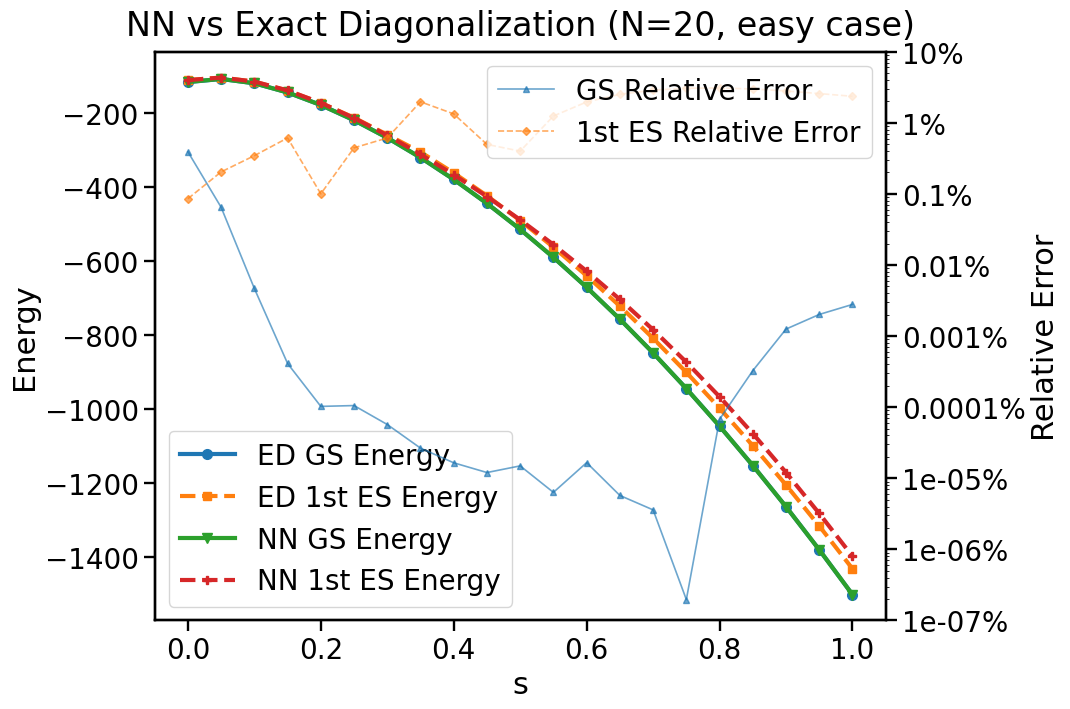}
\caption{$N=20$ easy case}
\label{fig:n20_easy}
\end{subfigure}
\caption{
Representative low-lying spectra for $N=15$ and $N=20$. The $N=15$
results are compared with ED. Dense full-spectrum diagonalization is not
used at $N=20$; sparse-Lanczos references are reported where available.
The plotted eigenenergies are reported in frequency-equivalent units,
$E/h$, expressed in GHz.
}
\label{fig:nn_spectra_1520}
\end{figure}

Figure~\ref{fig:nn_spectra_1520} compares representative spectral reconstructions for both easy and hard problem instances. Exact diagonalization provides the reference spectra for $N=15$, while sparse-Lanczos calculations are used for $N=20$. Across these representative examples, Tx-NQDT accurately captures the evolution of the two lowest energy levels and the locations of the minimum-gap regions. Complementary small-system baseline results at $N=4$ and $N=6$ are
provided in Appendix~\ref{MLP46}.
A quantitative analysis of the reconstruction errors is presented below.

\clearpage
\FloatBarrier

We quantify the energy, gap, and minimum-gap-location errors using
\begin{align}
\epsilon_{E_k}(s)
&=
\left|
E_k^{\rm NQDT}(s)-E_k^{\rm ref}(s)
\right|,
\qquad
k=0,1,
\label{eq:energy_abs_error}
\\
r_{E_k}(s)
&=
\frac{
E_k^{\rm NQDT}(s)-E_k^{\rm ref}(s)
}{
|E_k^{\rm ref}(s)|+\epsilon_E
},
\qquad
k=0,1,
\label{eq:energy_rel_error}
\\
\epsilon_\Delta(s)
&=
\left|
\Delta^{\rm NQDT}(s)-\Delta^{\rm ref}(s)
\right|,
\label{eq:gap_abs_error}
\\
\epsilon_{s_{\min}}
&=
\left|
s_{\min}^{\rm NQDT}-s_{\min}^{\rm ref}
\right|.
\label{eq:smin_error}
\end{align}
Here ``ref'' denotes ED or sparse Lanczos, as appropriate. The gap,
minimum-gap location, and transition matrix element are more directly
relevant to FOAPT-informed schedule construction than the individual
energy errors alone.

\begin{table*}[t]
\centering
\scriptsize
\setlength{\tabcolsep}{2.6pt}
\renewcommand{\arraystretch}{1.08}
\begin{tabular}{lll|cc|cc|cc|rrr}
\hline
&  &  &
\multicolumn{2}{c|}{Early $s$} &
\multicolumn{2}{c|}{Mid $s$} &
\multicolumn{2}{c|}{Late $s$} &
\multicolumn{3}{c}{All $s$ statistics} \\
$N$ & case & level &
0.00 & 0.20 &
0.40 & 0.60 &
0.80 & 1.00 &
Mean & Max & Min \\
\hline
4 & TFIM & GS &
-0.022 & -0.025 &
0.009 & $-7.5{\times}10^{-4}$ &
-0.001 & 0.001 &
-0.003 & 0.013 & -0.037 \\

4 & TFIM & ES &
-0.022 & -0.022 &
0.009 & $-7.5{\times}10^{-4}$ &
-0.001 & 0.001 &
-0.002 & 0.013 & -0.025 \\

4 & easy & GS &
-0.024 & -0.018 &
0.009 & -0.001 &
-0.001 & 0.001 &
-0.002 & 0.014 & -0.024 \\

4 & easy & ES &
-0.030 & -0.018 &
0.009 & -0.001 &
-0.001 & 0.001 &
-0.001 & 0.017 & -0.030 \\

4 & hard & GS &
-0.025 & -0.018 &
0.009 & -0.001 &
-0.001 & 0.001 &
-0.003 & 0.009 & -0.025 \\

4 & hard & ES &
-0.034 & -0.016 &
0.009 & -0.001 &
-0.001 & 0.001 &
-0.002 & 0.020 & -0.034 \\
\hline
6 & TFIM & GS &
-0.022 & -0.024 &
0.009 & $-8.3{\times}10^{-4}$ &
-0.001 & 0.001 &
-0.003 & 0.013 & -0.035 \\

6 & TFIM & ES &
-0.022 & -0.023 &
0.009 & $-8.3{\times}10^{-4}$ &
-0.001 & 0.001 &
-0.002 & 0.014 & -0.028 \\

6 & easy & GS &
-0.024 & -0.018 &
0.009 & -0.001 &
-0.001 & 0.001 &
-0.002 & 0.014 & -0.024 \\

6 & easy & ES &
-0.026 & -0.017 &
0.009 & -0.001 &
-0.001 & 0.001 &
$-4.8{\times}10^{-4}$ & 0.016 & -0.026 \\

6 & hard & GS &
-0.026 & -0.018 &
0.009 & -0.001 &
-0.001 & 0.001 &
-0.002 & 0.015 & -0.026 \\

6 & hard & ES &
-0.031 & -0.017 &
0.009 & -0.001 &
-0.001 & 0.001 &
-0.002 & 0.018 & -0.031 \\
\hline
10 & easy & GS &
$4.8{\times}10^{-4}$ & $1.0{\times}10^{-5}$ &
$7.2{\times}10^{-6}$ & $2.2{\times}10^{-5}$ &
$1.1{\times}10^{-6}$ & $1.9{\times}10^{-6}$ &
$6.8{\times}10^{-5}$ & $5.4{\times}10^{-4}$ & $9.3{\times}10^{-7}$ \\

10 & easy & ES &
-0.147 & -0.045 &
-0.029 & $2.3{\times}10^{-6}$ &
$7.9{\times}10^{-6}$ & $1.7{\times}10^{-5}$ &
-0.023 & 0.065 & -0.147 \\

10 & hard & GS &
0.019 & 0.013 &
0.005 & 0.005 &
0.005 & 0.005 &
0.010 & 0.043 & 0.003 \\

10 & hard & ES &
-0.138 & -0.018 &
0.008 & 0.008 &
0.008 & $2.0{\times}10^{-6}$ &
-0.007 & 0.010 & -0.138 \\
\hline
15 & easy & GS &
0.212 & -0.001 &
$7.1{\times}10^{-5}$ & $-3.5{\times}10^{-6}$ &
$1.3{\times}10^{-6}$ & $1.5{\times}10^{-6}$ &
0.013 & 0.212 & -0.001 \\

15 & easy & ES &
0.157 & 0.007 &
-0.021 & -0.023 &
-0.018 & -0.023 &
-0.008 & 0.157 & -0.029 \\

15 & hard & GS &
0.225 & 0.001 &
$3.7{\times}10^{-6}$ & $-5.7{\times}10^{-6}$ &
0.005 & $2.1{\times}10^{-6}$ &
0.021 & 0.225 & $-2.8{\times}10^{-5}$ \\

15 & hard & ES &
0.163 & 0.001 &
0.005 & 0.004 &
-0.006 & -0.004 &
0.011 & 0.163 & -0.040 \\
\hline
20 & easy & GS &
0.004 & $1.0{\times}10^{-6}$ &
$1.7{\times}10^{-7}$ & $1.7{\times}10^{-7}$ &
$6.8{\times}10^{-7}$ & $2.8{\times}10^{-5}$ &
$2.2{\times}10^{-4}$ & 0.004 & $2.0{\times}10^{-9}$ \\

20 & easy & ES &
$8.5{\times}10^{-4}$ & 0.001 &
-0.013 & 0.020 &
0.031 & 0.024 &
0.010 & 0.031 & -0.020 \\

20 & hard & GS &
0.004 & $1.0{\times}10^{-6}$ &
$7.0{\times}10^{-8}$ & $4.1{\times}10^{-8}$ &
$3.9{\times}10^{-7}$ & $2.3{\times}10^{-5}$ &
$2.0{\times}10^{-4}$ & 0.004 & $-4.4{\times}10^{-4}$ \\

20 & hard & ES &
0.003 & $5.2{\times}10^{-5}$ &
$1.5{\times}10^{-4}$ & $9.4{\times}10^{-8}$ &
0.012 & 0.025 &
0.005 & 0.025 & $6.0{\times}10^{-8}$ \\
\hline
\end{tabular}
\caption{
Signed relative energy errors
$r_{E_k}(s)=[E_{k,\mathrm{NN}}(s)-E_{k,\mathrm{ref}}(s)]/[
|E_{k,\mathrm{ref}}(s)|+\epsilon_{E}]$
for the ground state (GS) and first excited state (ES). Errors are
evaluated at 21 anneal points; selected values and full-grid statistics
are shown. ED is used where feasible and sparse Lanczos at $N=20$ where
indicated. Values below three-decimal resolution are written in scientific
notation. The $N=4,6$ entries use the MLP baseline, whereas the
$N\geq10$ entries use Tx-NQDT on different representative instances.
}
\label{tab:rel_errors}
\end{table*}

\begin{table*}[t]
\centering
\scriptsize
\setlength{\tabcolsep}{2.6pt}
\renewcommand{\arraystretch}{1.08}
\begin{tabular}{lll|cc|cc|cc|rrr}
\hline
&  &  &
\multicolumn{2}{c|}{Early $s$} &
\multicolumn{2}{c|}{Mid $s$} &
\multicolumn{2}{c|}{Late $s$} &
\multicolumn{3}{c}{All $s$ statistics} \\
$N$ & case & level &
0.00 & 0.20 &
0.40 & 0.60 &
0.80 & 1.00 &
Mean & Max & Min \\
\hline
10 & TFIM & GS &
$3.6{\times}10^{-13}$ & $4.6{\times}10^{-5}$ &
$1.1{\times}10^{-5}$ & $5.8{\times}10^{-7}$ &
$8.4{\times}10^{-5}$ & $2.0{\times}10^{-4}$ &
$2.7{\times}10^{-5}$ & $2.0{\times}10^{-4}$ & $-2.8{\times}10^{-4}$ \\

10 & TFIM & ES &
$6.7{\times}10^{-5}$ & $1.0{\times}10^{-4}$ &
$1.8{\times}10^{-5}$ & $8.7{\times}10^{-6}$ &
0.002 & 0.006 &
$6.4{\times}10^{-4}$ & 0.006 & $4.9{\times}10^{-6}$ \\
\hline
15 & TFIM & GS &
$6.5{\times}10^{-12}$ & 0.031 &
$5.5{\times}10^{-6}$ & $2.7{\times}10^{-4}$ &
0.001 & -0.002 &
0.007 & 0.055 & -0.002 \\

15 & TFIM & ES &
$1.8{\times}10^{-4}$ & $3.4{\times}10^{-4}$ &
$5.5{\times}10^{-6}$ & $2.7{\times}10^{-4}$ &
0.001 & -0.002 &
0.001 & 0.019 & -0.002 \\
\hline
20 & TFIM & GS &
$9.4{\times}10^{-13}$ & 0.002 &
$3.5{\times}10^{-5}$ & $4.0{\times}10^{-5}$ &
0.003 & $3.3{\times}10^{-4}$ &
0.002 & 0.035 & $9.4{\times}10^{-13}$ \\

20 & TFIM & ES &
$2.0{\times}10^{-5}$ & 0.015 &
$3.5{\times}10^{-5}$ & $4.0{\times}10^{-5}$ &
0.003 & $3.3{\times}10^{-4}$ &
0.002 & 0.015 & $7.7{\times}10^{-7}$ \\
\hline
38 & TFIM & GS &
$4.6{\times}10^{-9}$ & 0.006 &
$9.0{\times}10^{-5}$ & $-1.1{\times}10^{-6}$ &
$8.9{\times}10^{-6}$ & $-4.2{\times}10^{-4}$ &
0.001 & 0.013 & $-4.2{\times}10^{-4}$ \\

38 & TFIM & ES &
-0.056 & 0.001 &
$9.0{\times}10^{-5}$ & $-1.1{\times}10^{-6}$ &
$2.0{\times}10^{-5}$ & $-4.2{\times}10^{-4}$ &
-0.008 & 0.006 & -0.056 \\
\hline
\end{tabular}
\caption{
Signed relative energy errors for the TFIM benchmarks on the 21-point
anneal grid. ED references are used at $N=10$ and $N=15$, the
lowest sparse-Lanczos eigenvalues are used at $N=20$, and the exact
Jordan--Wigner solution is used at $N=38$. Values that would
round to $0.000$ are reported in scientific notation.
}
\label{tab:tfim_rel_errors}
\end{table*}

\begin{figure*}[t]
\centering

\begin{subfigure}[t]{0.45\textwidth}
    \centering
    \includegraphics[width=\linewidth]{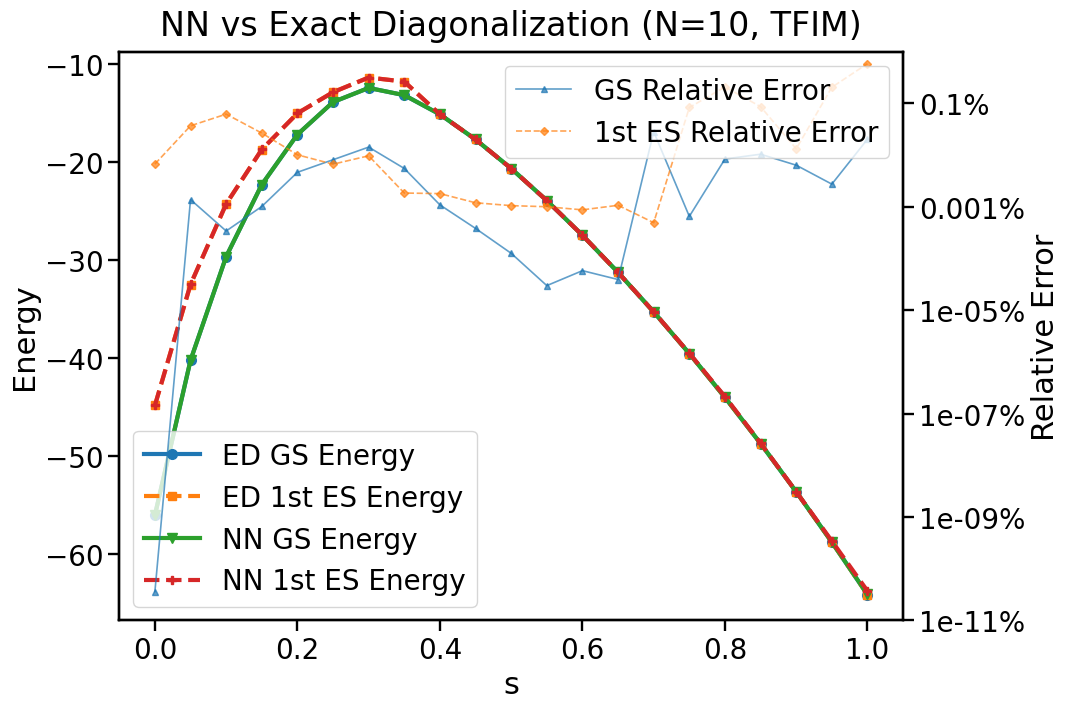}
    \caption{$N=10$ TFIM.}
    \label{fig:tfim_n10_spectrum}
\end{subfigure}
\hfill
\begin{subfigure}[t]{0.45\textwidth}
    \centering
    \includegraphics[width=\linewidth]{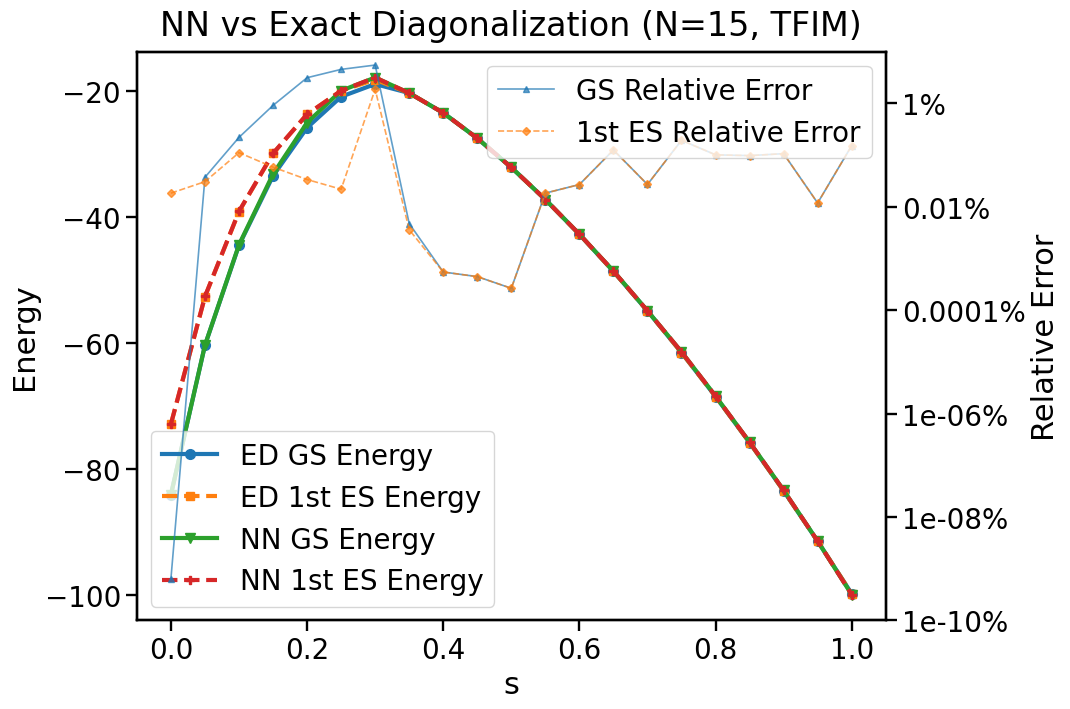}
    \caption{$N=15$ TFIM.}
    \label{fig:tfim_n15_spectrum}
\end{subfigure}
\hfill
\begin{subfigure}[t]{0.45\textwidth}
    \centering
    \includegraphics[width=\linewidth]{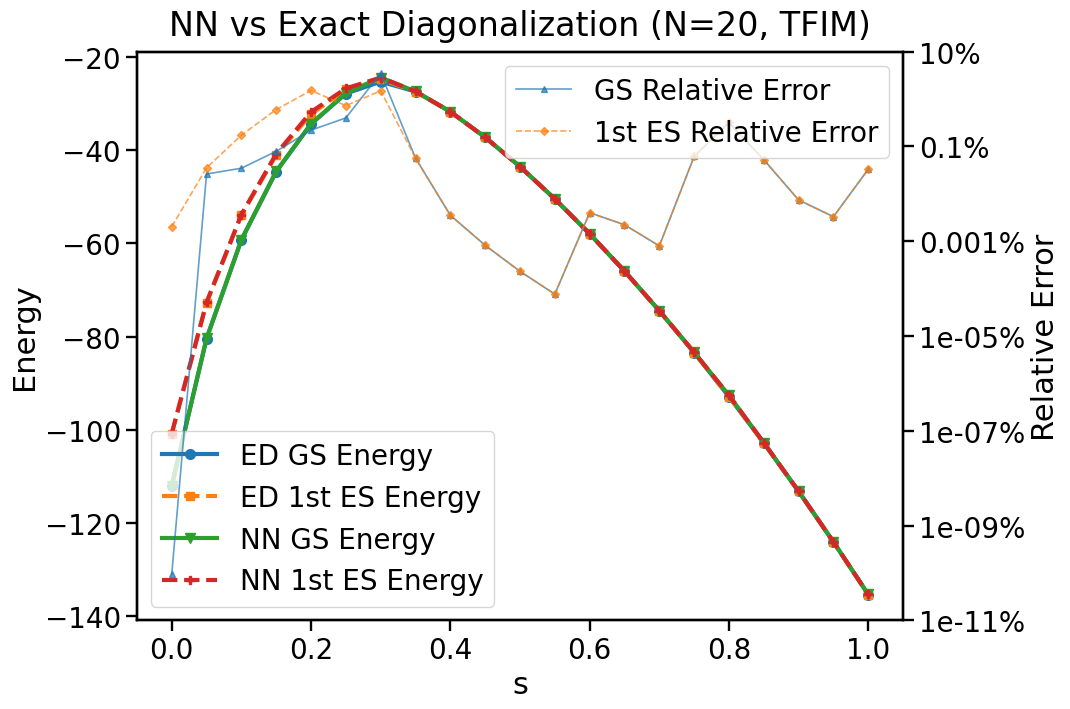}
    \caption{$N=20$ TFIM.}
    \label{fig:tfim_n20_spectrum}
\end{subfigure}
\hfill
\begin{subfigure}[t]{0.45\textwidth}
    \centering
    \includegraphics[width=\linewidth]{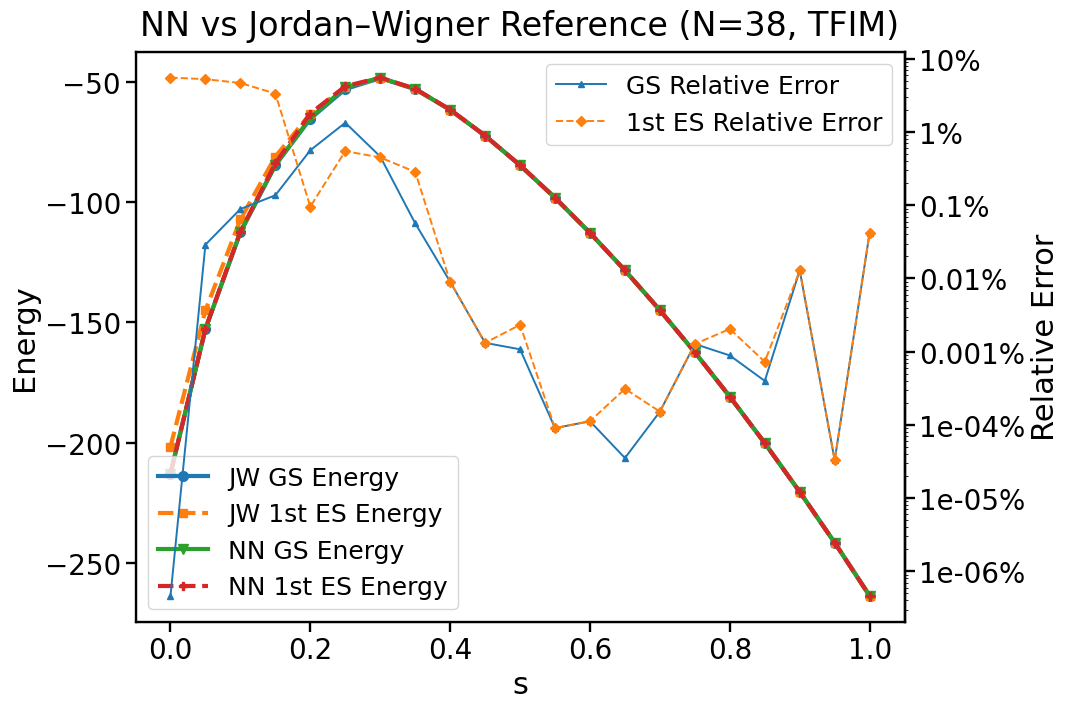}
    \caption{$N=38$ TFIM.}
    \label{fig:tfim_n38_spectrum}
\end{subfigure}

\caption{
Tx-NQDT and reference low-energy spectra for the open one-dimensional
TFIM at $N=10,15,20,$ and $38$. Exact diagonalization is used for
$N=10$ and $N=15$, sparse matrix-free Lanczos for $N=20$, and the
exact Jordan--Wigner solution for $N=38$. In each panel, the left axis
shows the ground- and first-excited-state energies, while the right axis
shows the corresponding absolute relative errors. The eigenenergies on
the left axis are reported in frequency-equivalent units, $E/h$,
expressed in GHz. The errors are plotted on a logarithmic scale for
$N=10,15,$ and $20$. The Jordan--Wigner transformation provides the
$N=38$ spectral reference without constructing the
$2^{38}$-dimensional spin-basis Hamiltonian.
}
\label{fig:tfim_nn_vs_ed}
\end{figure*}

Table~\ref{tab:rel_errors} reports signed relative energy errors for
representative benchmark instances. Values that would round to zero are reported in scientific notation. The
structured TFIM comparison is summarized in
Table~\ref{tab:tfim_rel_errors} and
Fig.~\ref{fig:tfim_nn_vs_ed}. The $N=38$ TFIM extends the spectral
validation beyond the problem sizes used in the random-instance and
hardware benchmarks while retaining the same 21-point anneal grid and
logical Hamiltonian path.

Across all benchmark problems, the reconstruction errors are
nonuniform and concentrated at particular anneal points, with the
largest deviations generally occurring in the driver-dominated region
and for the first excited state. These include the $O(10^{-1})$ maxima
for some $N=15$ instances in Table~\ref{tab:rel_errors} and the
approximately $5.6\%$ first-excited-state error at $s=0$ for the
$N=38$ TFIM. For the $N=38$ TFIM, the reconstruction errors remain
small over most of the annealing path, apart from a few localized
increases. The methodological origins of these localized deviations and
their implications for wavefunction reconstruction and schedule
generation are discussed in
Sec.~\ref{sec:conclusion}.

Table~\ref{tab:rel_errors} summarizes representative benchmark
instances spanning different problem classes and neural ansatzes,
whereas Table~\ref{tab:tfim_rel_errors} focuses exclusively on the TFIM
to provide a controlled comparison across system sizes from $N=10$ to
$38$. Together, these results characterize the spectral accuracy of
Tx-NQDT across the tested benchmark problems rather than establishing
finite-size convergence behavior.

\begin{figure}[t!]
\centering
\includegraphics[width=\columnwidth]{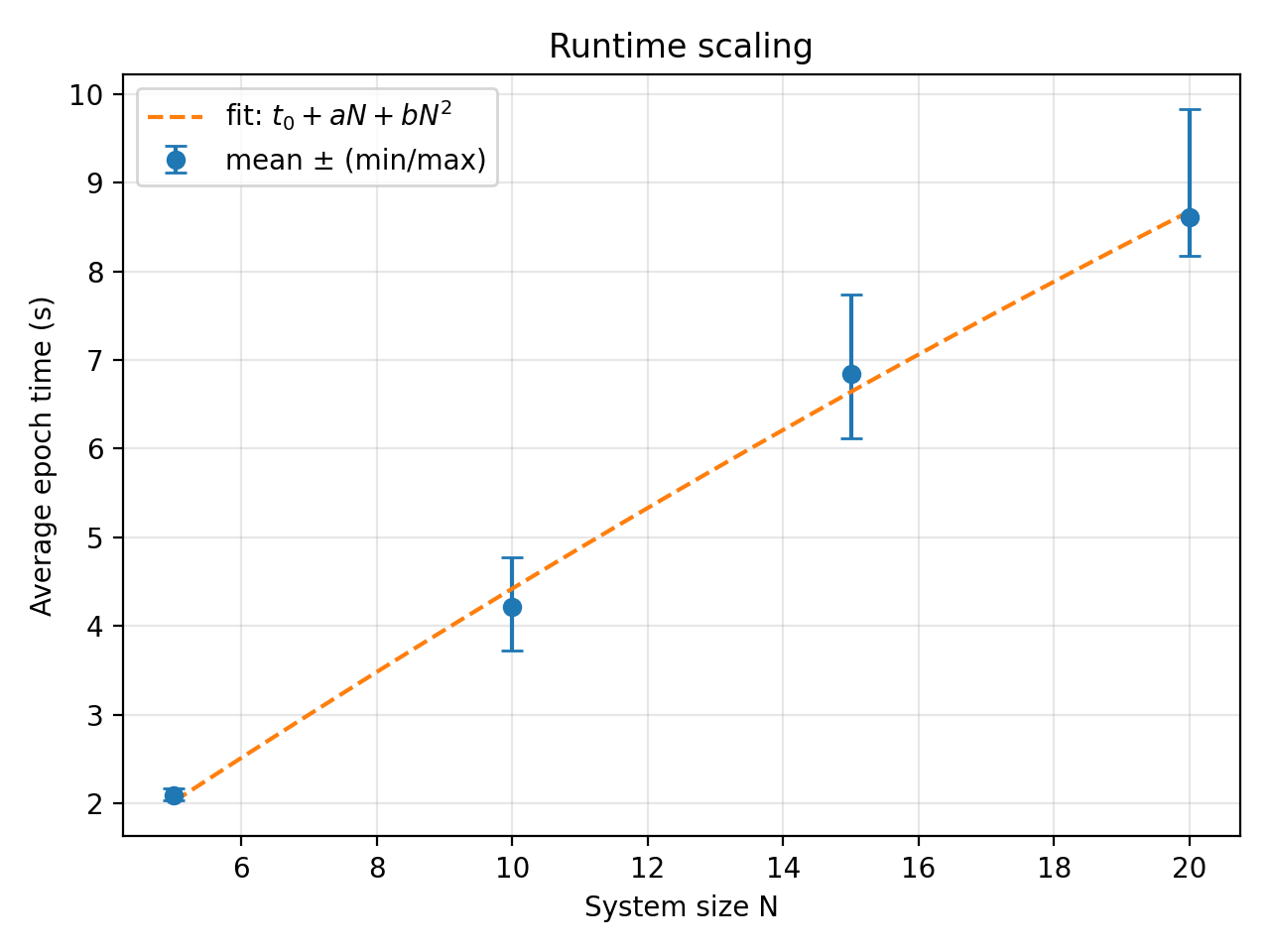}
\caption{
Mean epoch wall-clock time on an Apple M2, averaged over epochs 6--25.
Error bars indicate the minimum and maximum times in the averaging window.
The dashed curve is the fit
$t_{\rm epoch}(N)=t_0+aN+bN^2$.
}
\label{fig:avg-epoch-time}
\end{figure}

Figure~\ref{fig:avg-epoch-time} reports the measured ground-state
training time per epoch at a fixed anneal point. Timings are averaged over
epochs 6--25 to reduce warm-up effects, and the error bars show the
minimum and maximum values in the same window. Measurements use one
representative dense random Ising Hamiltonian models (RHM) instance per system size on an Apple M2. The
mean epoch time increases from approximately $2.1\,\mathrm{s}$ at
$N=6$ to $8.6\,\mathrm{s}$ at $N=20$ and is fitted over this limited
range by
\begin{equation}
t_{\rm epoch}(N)
=
t_0+aN+bN^2,
\label{eq:epoch_time_fit}
\end{equation}
with $R^2=0.996$. The observed quadratic component is consistent with
dense attention and dense Ising-energy evaluation, both of which require
$\mathcal{O}(N^2)$ pairwise operations. The measurements cover
$N\leq20$, one Apple M2 computing platform, and one representative dense
RHM instance per size, so the fit characterizes the implementation over
the measured range. The $N=38$ TFIM spectral result is not included in
this timing analysis. A detailed operation-count analysis of the
model-evaluation, Monte Carlo, anneal-path training, excited-state, and
schedule-construction costs is provided in Appendix~\ref{app:complexity}.

Overall, the reference comparisons support the use of Tx-NQDT as a
runtime-controlled source of path-resolved estimates of
$\Delta(s)$, $s_{\min}$, and $M_{01}(s)$ for the downstream
FOAPT-informed schedule workflow. The hardware utility of the resulting
schedule features is evaluated empirically in
Sec.~\ref{sec:application}.

%% file: dwave.tex
\section{D-Wave Optimization Experiments}
\label{sec:application}

This section evaluates whether the logical spectral information
reconstructed by Tx-NQDT can be translated into practical performance
improvements on quantum annealing hardware. The generated candidate
schedules are executed on a D-Wave Advantage processor and compared with
the default linear annealing schedule on identical logical QUBO
instances. The objective is to assess the practical benefit of
spectrum-informed schedule construction under realistic hardware
constraints.

For each instance, Tx-NQDT supplies the logical gap
$\Delta(s)=E_1(s)-E_0(s)$ and transition matrix element $M_{01}(s)$.
The FOAPT-informed score, normalized time-allocation density, and
continuous schedule are defined in Sec.~\ref{sec:foapt_schedule}. The
continuous schedule is projected onto at most $12$ monotone
piecewise-linear control points, with a maximum segment ramp rate of
$2.0~(\mu{\rm s})^{-1}$, using the procedure in
Appendix~\ref{annealing_schedule}. 

The logical-instance construction, easy--hard classification, Pegasus
minor embedding, autoscaling and calibration settings, and hardware
submission procedure are documented in
Appendix~\ref{app:dwave_implementation}. Hardware diagnostics and
unresolved experimental systematics are collected separately in
Appendix~\ref{app:dwave_hardware_diagnostics}.

\subsection{Experimental protocol and baselines}

The hardware benchmark contains $20$ independently generated dense
random Ising instances at each of $N=10,15,$ and $20$, giving $60$
logical problems in total. At each size, $10$ instances are classified
as easy and $10$ as hard according to the Tx-NQDT estimate of the
logical minimum gap. Because the dense logical graphs are not native
Pegasus subgraphs, the submitted problems require minor embedding. Both
schedules for a given logical instance use the same submitted
coefficients and programming convention.

For each instance, we compare the Tx-NQDT-informed schedule with the
default linear schedule of duration $20\,\mu{\rm s}$. Each schedule is
executed with $1000$ anneals per instance, and the experiment is repeated
$10$ times. The empirical ground-state success probability is the
frequency of returned logical samples having the exact minimum classical
energy after unembedding.

The default $20\,\mu{\rm s}$ linear anneal is used as a
\emph{low-overhead, no-tuning baseline}. It requires no
instance-specific schedule scan, pause optimization, reverse-annealing
initialization, anneal-offset calibration, user-level shimming, or
additional QPU optimization budget. It is not claimed to be the strongest
available annealing baseline.

Because the Tx-NQDT-informed schedules generally have
instance-dependent durations $T_{\rm NQDT}$, the present comparison does
not isolate schedule shape from total-duration effects. A complete
shape-controlled comparison would require
\begin{equation}
p_{\rm lin}(20\,\mu{\rm s}),
\qquad
p_{\rm lin}(T_{\rm NQDT}),
\qquad
p_{\rm NQDT}(T_{\rm NQDT}),
\label{eq:matched_duration_baselines}
\end{equation}
where the last two schedules have equal duration. Complete
matched-duration linear QPU runs were unavailable because the QPU
allocation had been exhausted. Appendix~\ref{app:linear_duration_scan}
provides a closed-system duration scan showing that linear-schedule
excitation probabilities need not vary monotonically with duration, but
this does not replace a matched-duration comparison on the embedded,
autoscaled, finite-temperature processor.

The endpoint neural state defines a separate classical heuristic that
samples $|\Psi_\theta(x;1)|^2$ and evaluates the exact final Ising
energies. It was not included in the present QPU schedule comparison.
Stronger future baselines should include this endpoint-NQS sampler,
matched-duration linear schedules, tuned pauses, reverse annealing,
anneal offsets, gauge and embedding optimization, calibration-aware
controls, and established classical heuristics.

\subsection{Empirical schedule performance}

\begin{figure}[!t]
\centering
\includegraphics[width=0.48\textwidth]{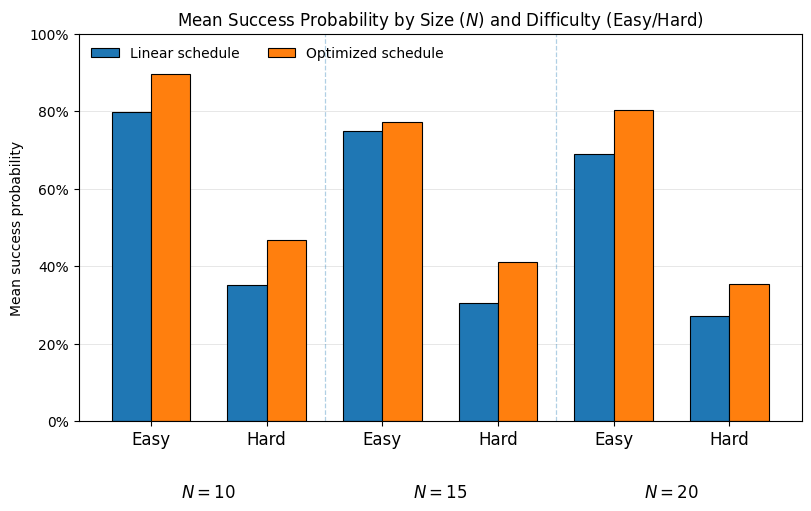}
\caption{
Mean empirical ground-state success probabilities for the default
$20\,\mu{\rm s}$ linear schedule and Tx-NQDT-informed schedules over
the easy and hard subsets at $N=10,15,$ and $20$.
}
\label{fig:success_bar_means}
\end{figure}

Figure~\ref{fig:success_bar_means} compares the mean empirical success
probabilities of the default and Tx-NQDT-informed schedules. Across the
tested ensemble, the Tx-NQDT schedules improve performance on many
instances, with especially visible gains in the hard subsets.

Table~\ref{tab:schedule_success_summary} reports the detailed aggregate
statistics. Tx-NQDT-informed schedules outperform the default schedule on
$16/20$ instances at $N=10$, $15/20$ at $N=15$, and $13/20$ at
$N=20$, giving $44/60$ wins overall.

\begin{table*}[!t]
\centering
\small
\setlength{\tabcolsep}{4.5pt}
\begin{tabular}{c c c c c c c c c}
\toprule
$N$ & subset & wins (NQDT$>$lin) &
mean $p_{\rm lin}$ & median $p_{\rm lin}$ &
mean $p_{\rm NQDT}$ & median $p_{\rm NQDT}$ &
mean $\Delta p$ & median $\Delta p$ \\
\midrule
10 & easy & $8/10$ & 79.9\% & 84.6\% & 89.7\% & 91.1\% & +9.8\% & +6.5\% \\
10 & hard & $8/10$ & 35.1\% & 37.6\% & 46.8\% & 50.2\% & +11.7\% & +12.7\% \\
\midrule
15 & easy & $8/10$ & 75.0\% & 72.4\% & 77.2\% & 79.1\% & +2.2\% & +6.8\% \\
15 & hard & $7/10$ & 30.4\% & 30.4\% & 41.2\% & 48.0\% & +10.8\% & +17.6\% \\
\midrule
20 & easy & $7/10$ & 68.9\% & 75.0\% & 80.2\% & 76.1\% & +11.3\% & +1.1\% \\
20 & hard & $6/10$ & 27.1\% & 30.1\% & 35.5\% & 39.4\% & +8.5\% & +9.3\% \\
\bottomrule
\end{tabular}
\caption{
Empirical ground-state success probabilities for the default
$20\,\mu{\rm s}$ linear schedule and Tx-NQDT-informed schedules. Each
size contains $10$ easy and $10$ hard instances. Here
$\Delta p=p_{\rm NQDT}-p_{\rm lin}$.
}
\label{tab:schedule_success_summary}
\end{table*}

\begin{figure*}[t]
\centering
\begin{subfigure}[t]{0.32\textwidth}
    \centering
    \includegraphics[width=\textwidth]{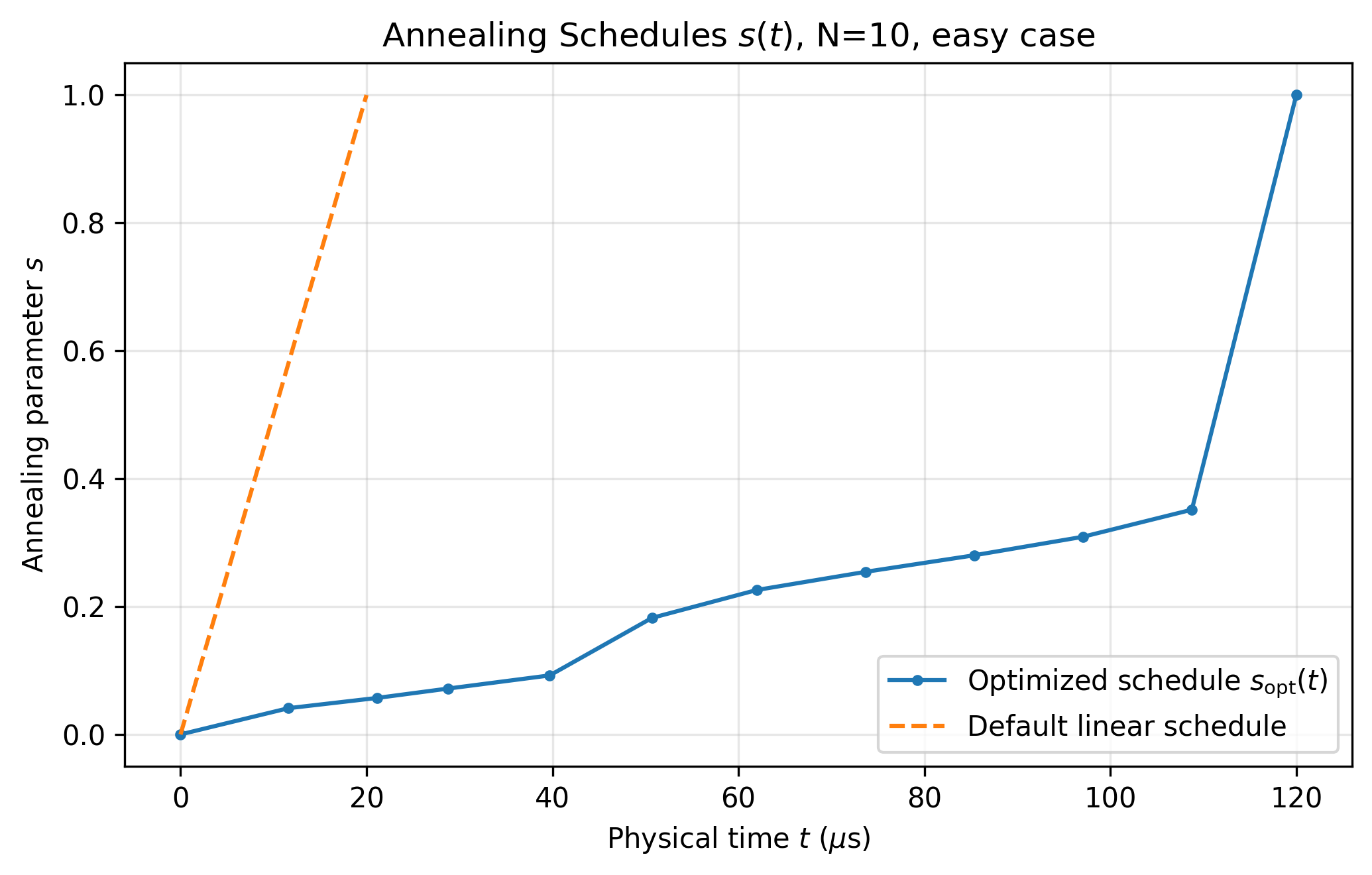}
    \caption{$N=10$ easy}
    \label{fig:anneal_N10_easy}
\end{subfigure}\hfill
\begin{subfigure}[t]{0.32\textwidth}
    \centering
    \includegraphics[width=\textwidth]{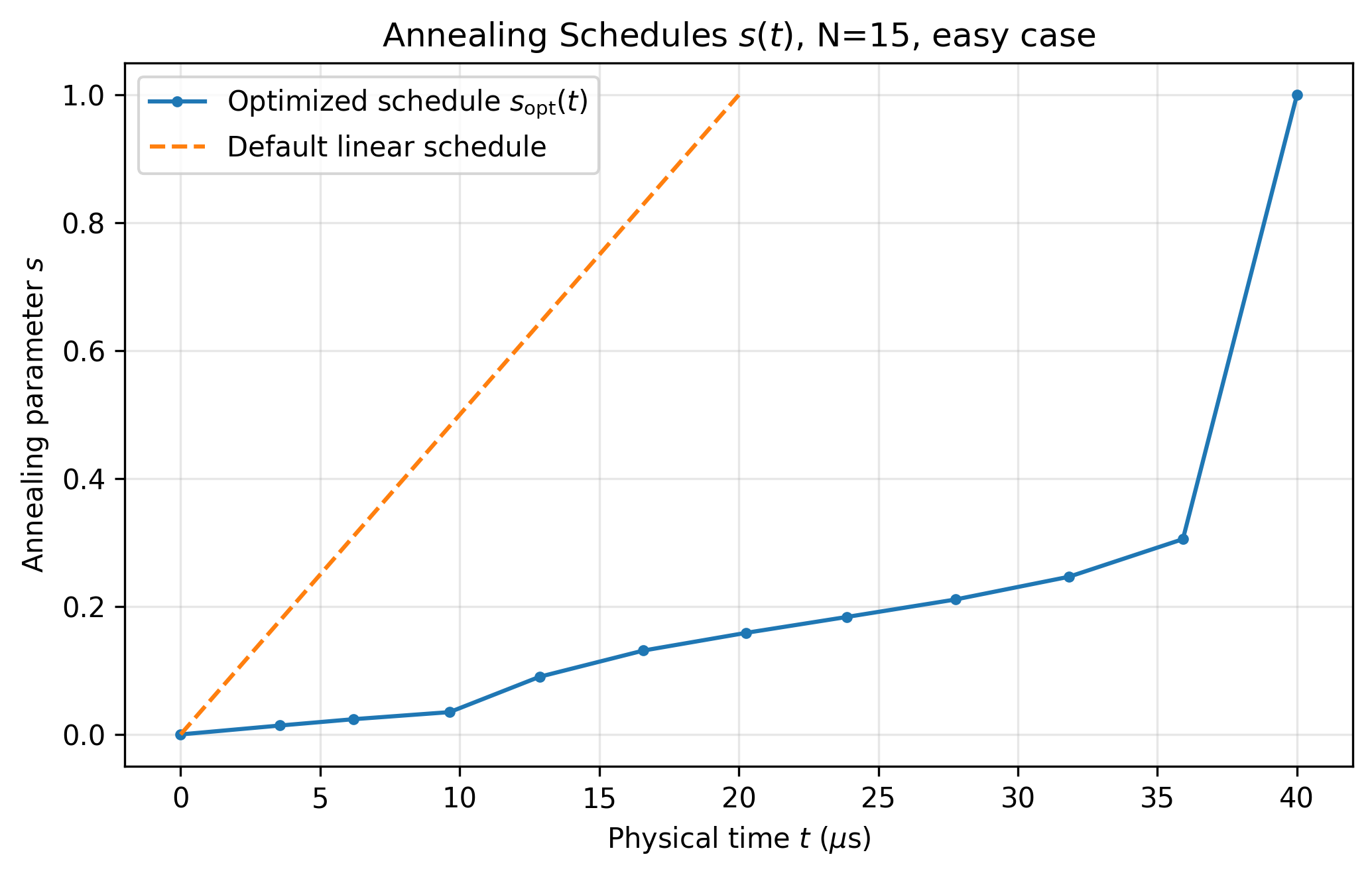}
    \caption{$N=15$ easy}
    \label{fig:anneal_N15_easy}
\end{subfigure}\hfill
\begin{subfigure}[t]{0.32\textwidth}
    \centering
    \includegraphics[width=\textwidth]{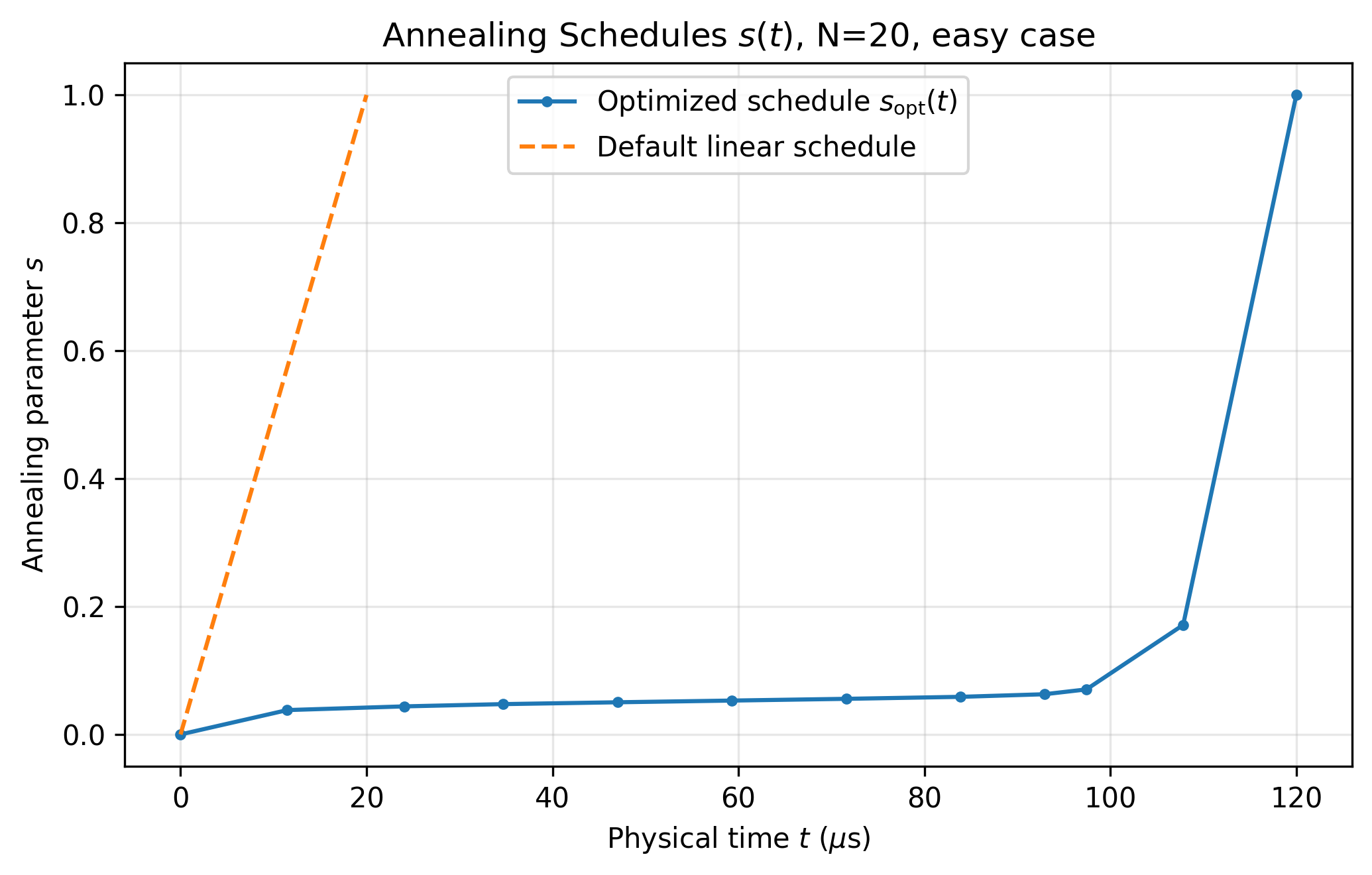}
    \caption{$N=20$ easy}
    \label{fig:anneal_N20_easy}
\end{subfigure}

\vspace{0.6em}

\begin{subfigure}[t]{0.32\textwidth}
    \centering
    \includegraphics[width=\textwidth]{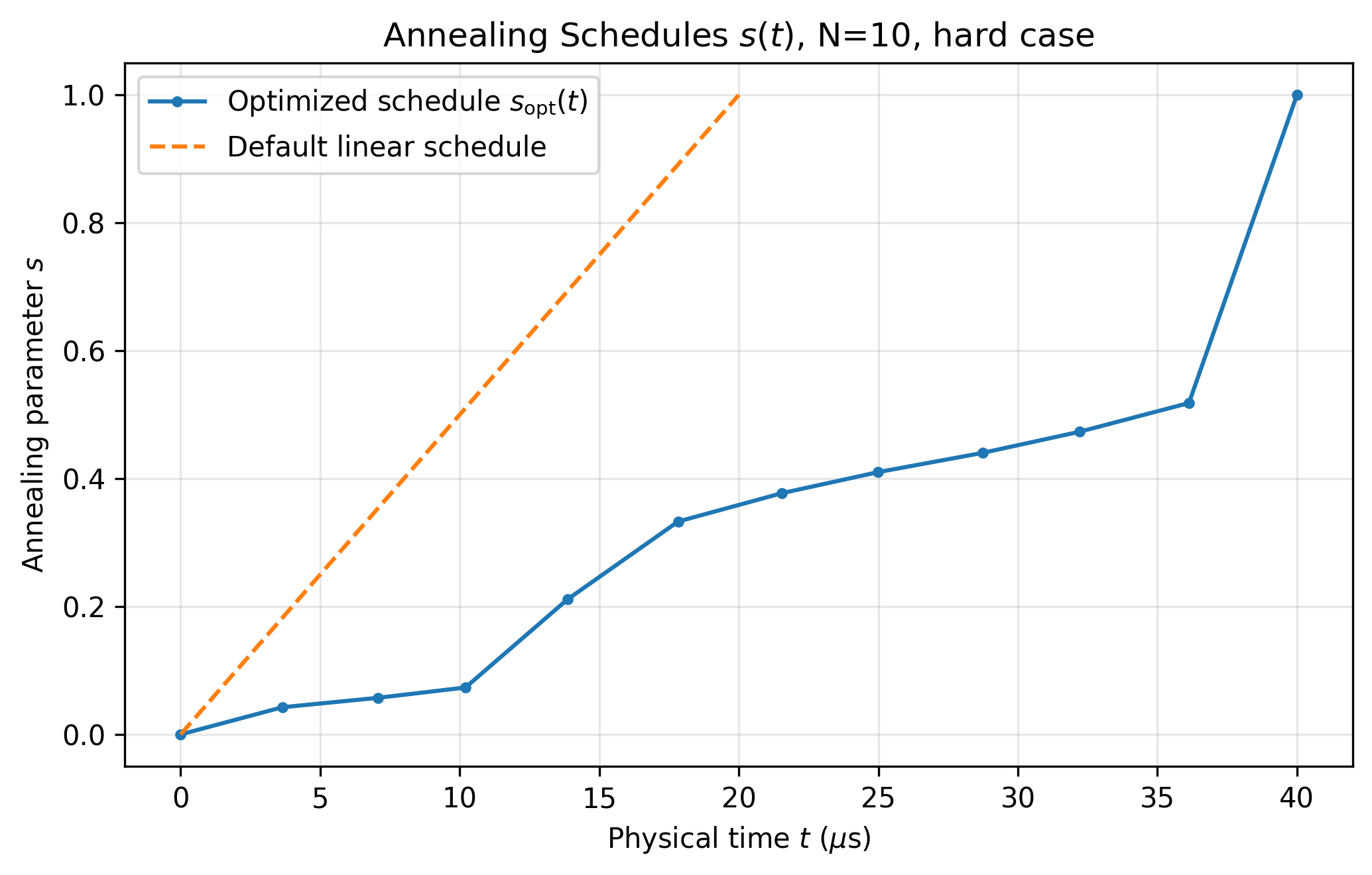}
    \caption{$N=10$ hard}
    \label{fig:anneal_N10_hard}
\end{subfigure}\hfill
\begin{subfigure}[t]{0.32\textwidth}
    \centering
    \includegraphics[width=\textwidth]{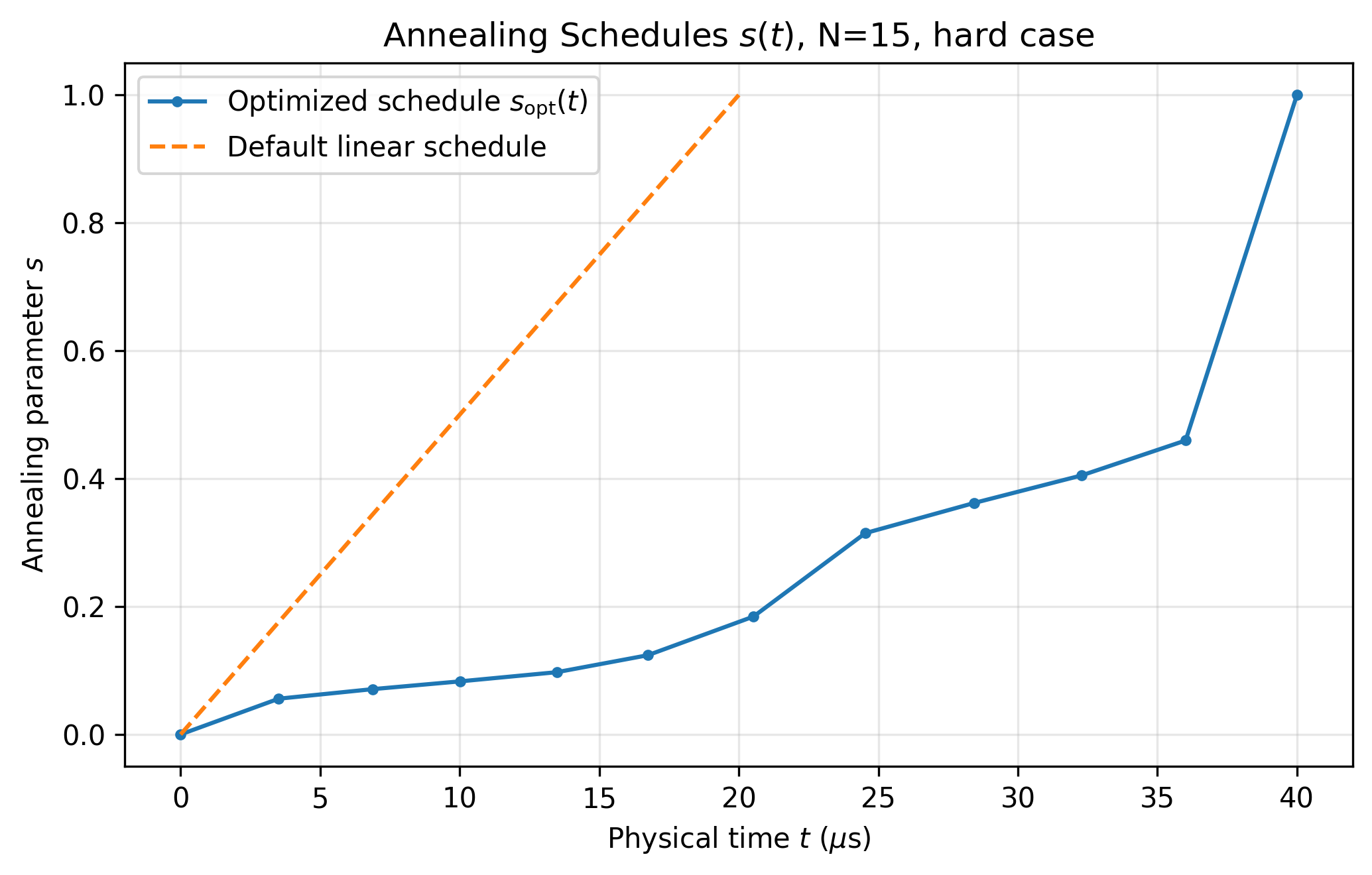}
    \caption{$N=15$ hard}
    \label{fig:anneal_N15_hard}
\end{subfigure}\hfill
\begin{subfigure}[t]{0.32\textwidth}
    \centering
    \includegraphics[width=\textwidth]{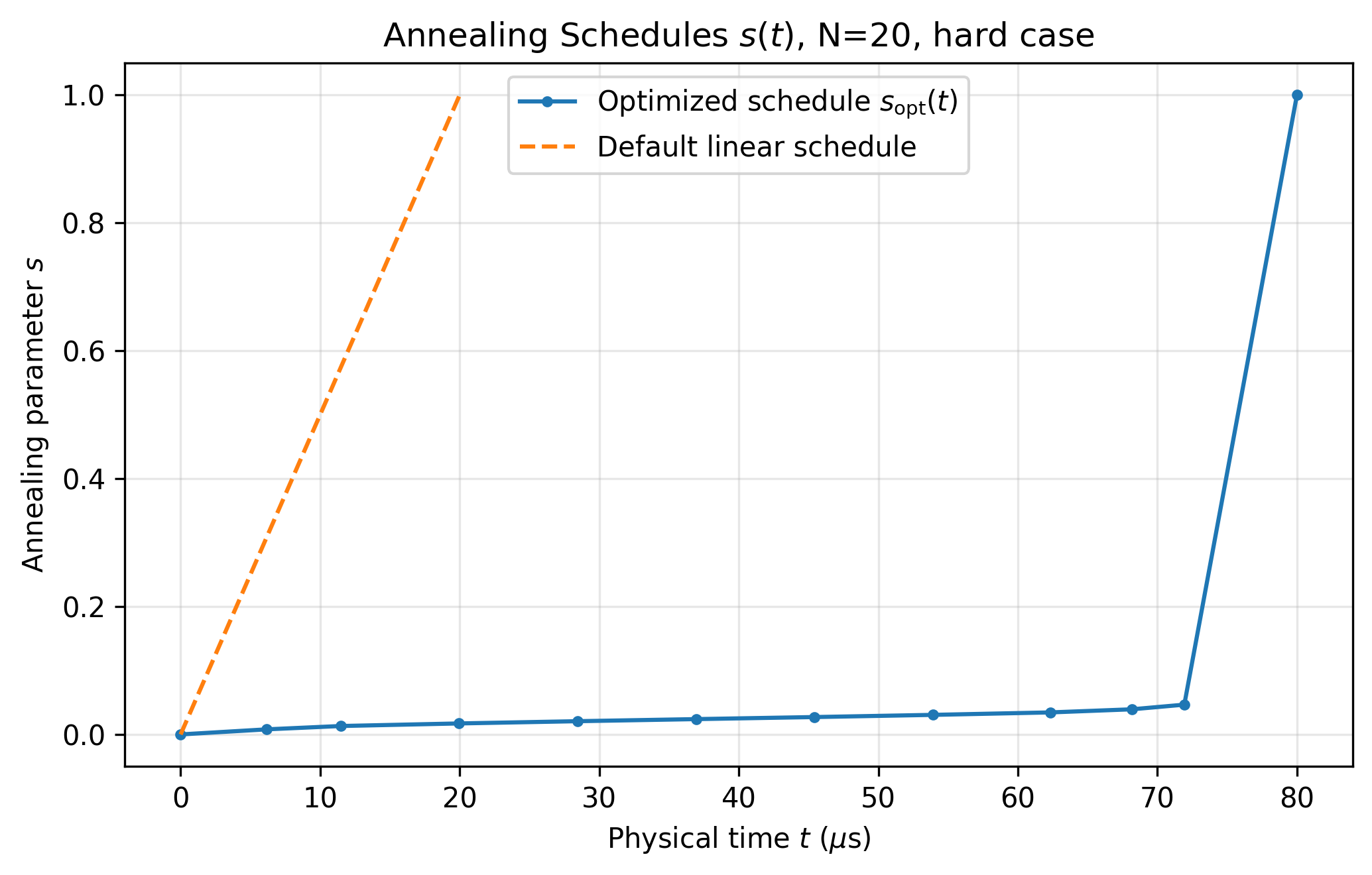}
    \caption{$N=20$ hard}
    \label{fig:anneal_N20_hard}
\end{subfigure}
\caption{
Representative Tx-NQDT-informed anneal-fraction schedules $s(t)$ and the
default $20\,\mu{\rm s}$ linear schedule. The candidates are projected
onto the hardware control-point and ramp-rate constraints described in
Appendix~\ref{annealing_schedule}. Their total durations are
generally instance dependent. The corresponding programmed energy-scale
trajectories are shown in Fig.~\ref{fig:hardware_AB_schedules}.
}
\label{fig:optimized-schedule}
\end{figure*}

Representative uploaded schedules are shown in
Fig.~\ref{fig:optimized-schedule}. They commonly exhibit a slow--fast
profile, allocating additional time near the logical FOAPT bottleneck and
accelerating after the logical gap reopens.

\begin{figure*}[t]
\centering
\begin{subfigure}[t]{0.32\textwidth}
    \centering
    \includegraphics[width=\textwidth]{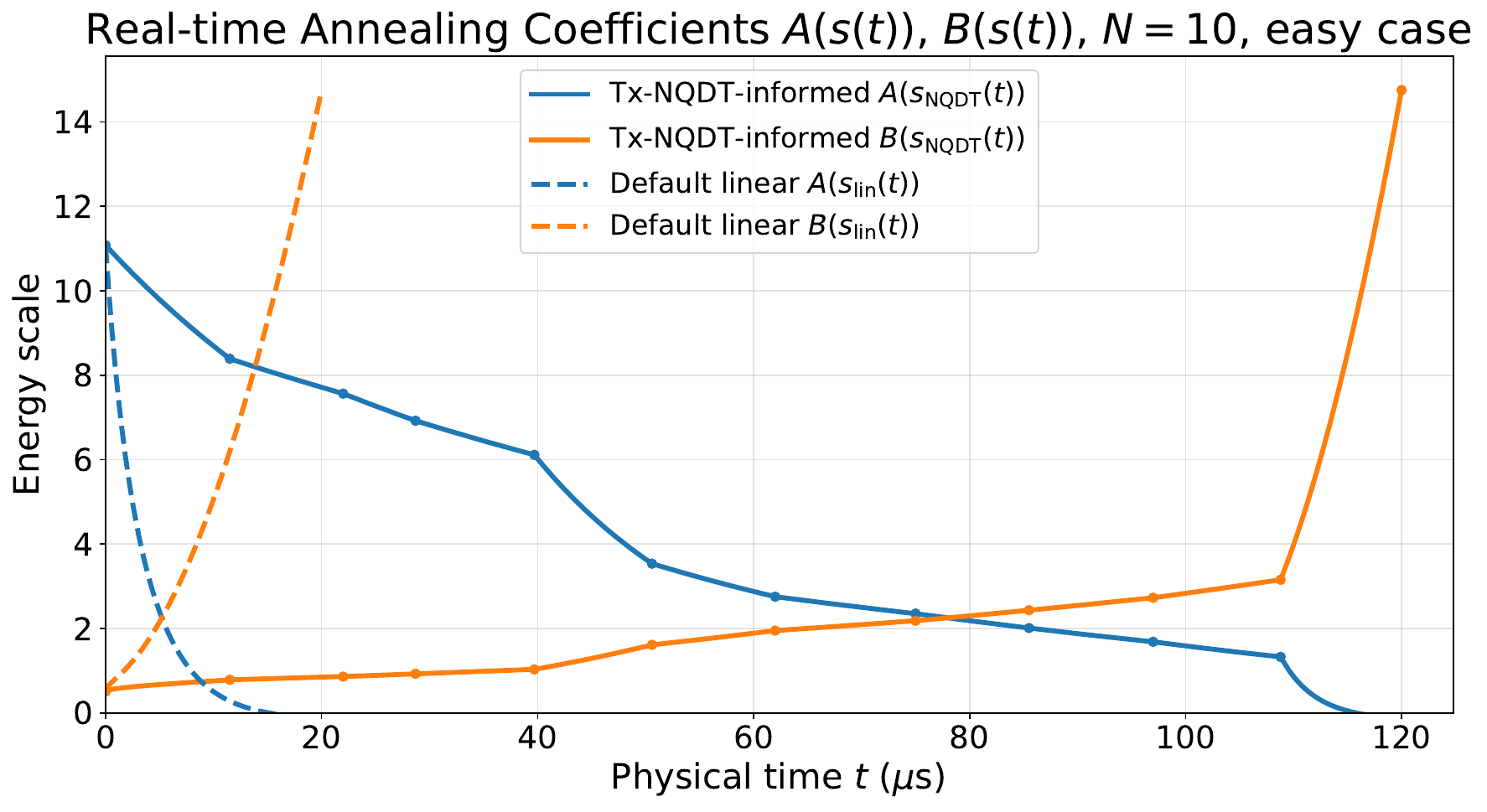}
    \caption{$N=10$ easy}
    \label{fig:AB_schedule_N10_easy}
\end{subfigure}\hfill
\begin{subfigure}[t]{0.32\textwidth}
    \centering
    \includegraphics[width=\textwidth]{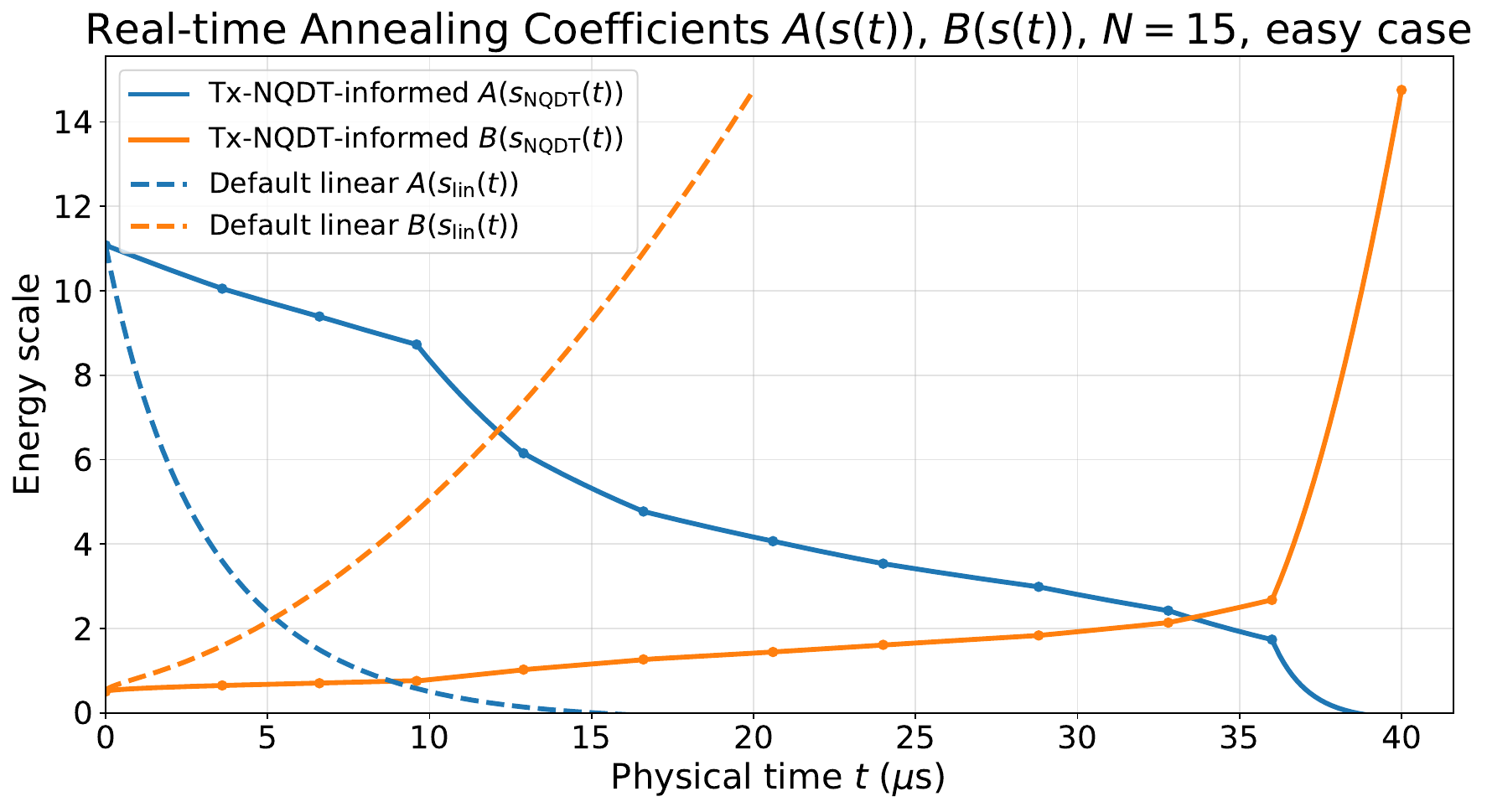}
    \caption{$N=15$ easy}
    \label{fig:AB_schedule_N15_easy}
\end{subfigure}\hfill
\begin{subfigure}[t]{0.32\textwidth}
    \centering
    \includegraphics[width=\textwidth]{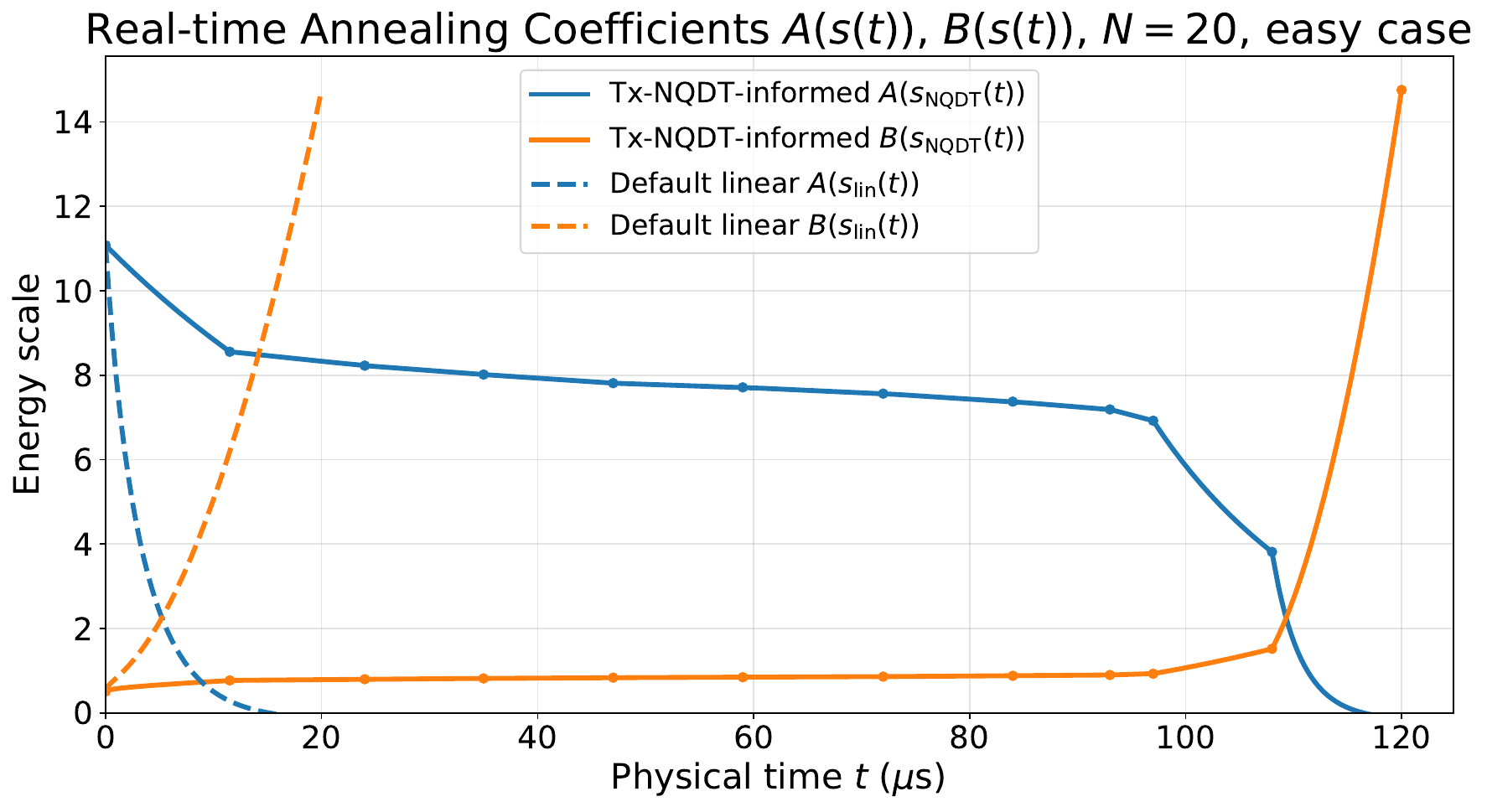}
    \caption{$N=20$ easy}
    \label{fig:AB_schedule_N20_easy}
\end{subfigure}

\vspace{0.6em}

\begin{subfigure}[t]{0.32\textwidth}
    \centering
    \includegraphics[width=\textwidth]{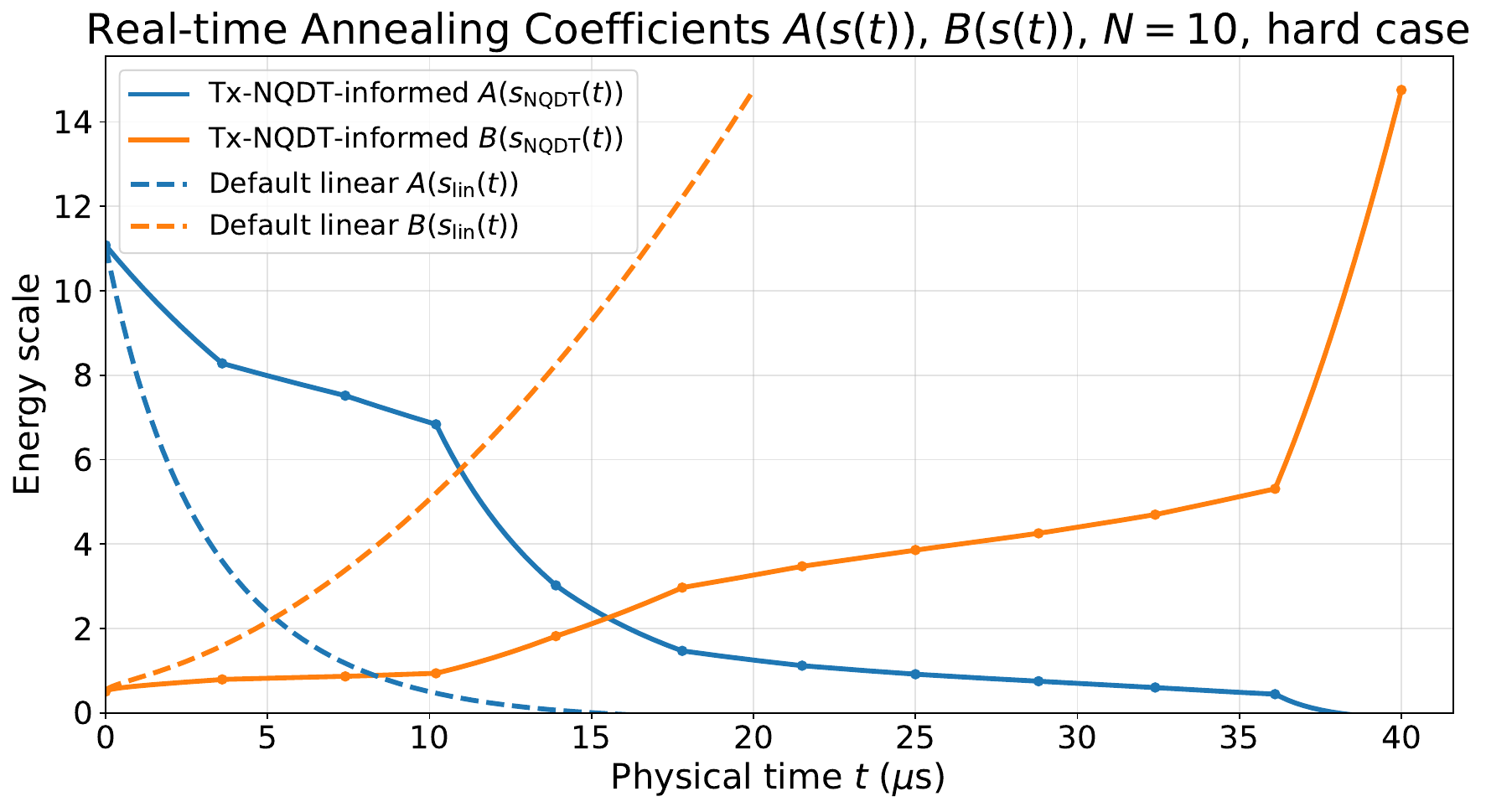}
    \caption{$N=10$ hard}
    \label{fig:AB_schedule_N10_hard}
\end{subfigure}\hfill
\begin{subfigure}[t]{0.32\textwidth}
    \centering
    \includegraphics[width=\textwidth]{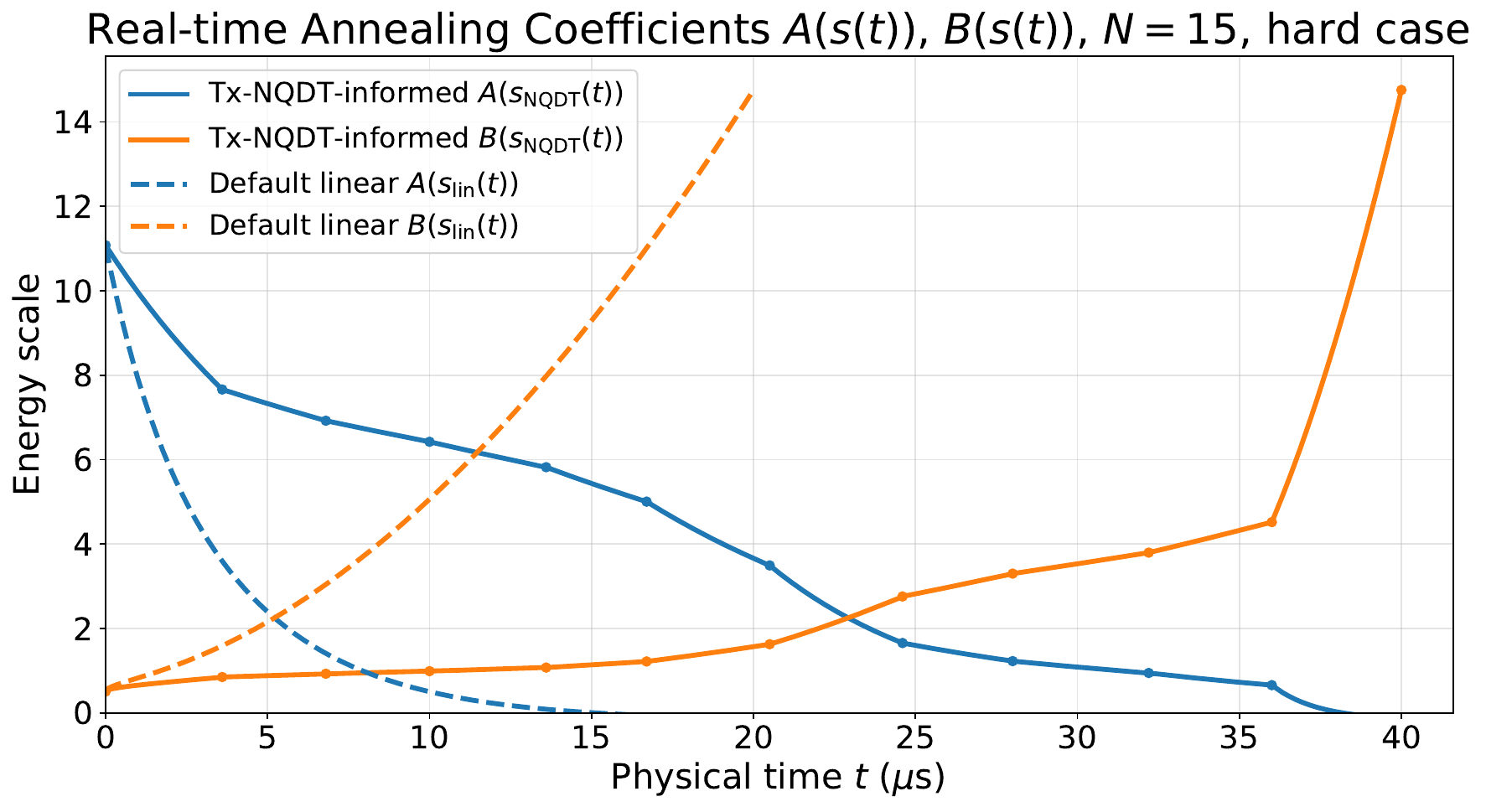}
    \caption{$N=15$ hard}
    \label{fig:AB_schedule_N15_hard}
\end{subfigure}\hfill
\begin{subfigure}[t]{0.32\textwidth}
    \centering
    \includegraphics[width=\textwidth]{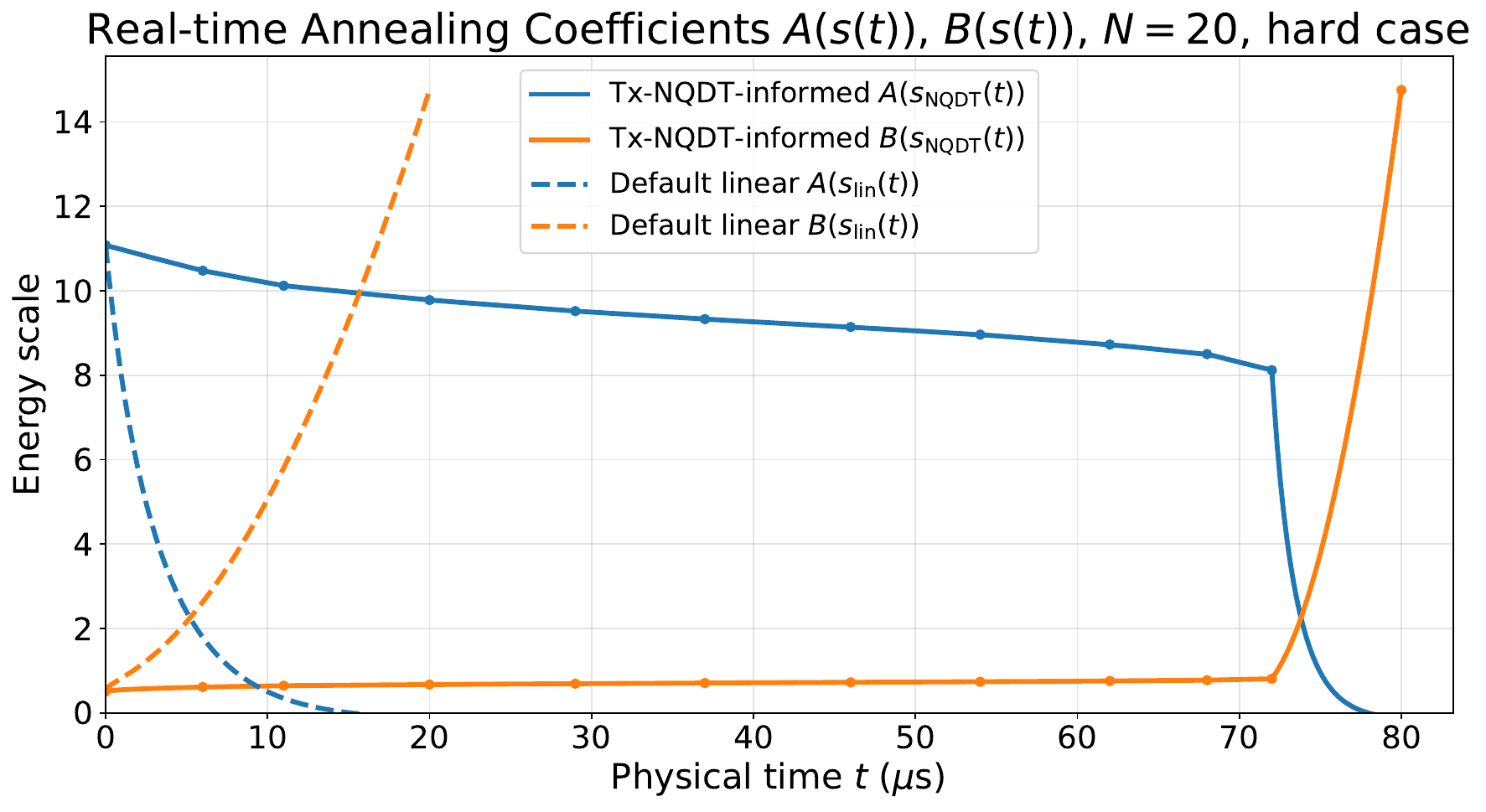}
    \caption{$N=20$ hard}
    \label{fig:AB_schedule_N20_hard}
\end{subfigure}

\caption{
Programmed real-time annealing coefficients for the schedules in
Fig.~\ref{fig:optimized-schedule}. Solid curves show
$A(s_{\rm upload}(t))$ and $B(s_{\rm upload}(t))$ for the Tx-NQDT-informed schedules; dashed curves show the default $20\,\mu{\rm s}$ linear
trajectories. Following the D-Wave convention, the vertical-axis values
are the frequency-equivalent energy scales $A/h$ and $B/h$, expressed
in GHz.
}
\label{fig:hardware_AB_schedules}
\end{figure*}

Figure~\ref{fig:hardware_AB_schedules} shows the corresponding programmed
real-time energy scales
\begin{equation}
A_{\rm exp}(t)
=
A(s_{\rm upload}(t)),
\qquad
B_{\rm exp}(t)
=
B(s_{\rm upload}(t)),
\label{eq:programmed_AB}
\end{equation}
together with the default trajectories
\begin{equation}
A_{\rm lin}(t)
=
A\!\left(\frac{t}{20\,\mu{\rm s}}\right),
\qquad
B_{\rm lin}(t)
=
B\!\left(\frac{t}{20\,\mu{\rm s}}\right).
\label{eq:linear_AB}
\end{equation}
Thus Fig.~\ref{fig:applied_A_B} shows the fixed functions $A(s)$ and
$B(s)$, Fig.~\ref{fig:optimized-schedule} shows the uploaded map $s(t)$,
and Fig.~\ref{fig:hardware_AB_schedules} shows their composition in
physical time. Because $A(s)$ and $B(s)$ are nonlinear functions, these
curves are not simple rescalings of $s(t)$. They represent programmed
control trajectories rather than measurements of the instantaneous
embedded Hamiltonian.

\subsection{Interpretive limitations of the hardware results}
\label{sec:hardware_interpretive_limitations}

The hardware experiments evaluate whether the logical spectral
information reconstructed by Tx-NQDT can be translated into practical
improvements in quantum annealing performance. They should not be
interpreted as a direct validation of the underlying physical annealing
dynamics. Tx-NQDT reconstructs the spectrum of the unembedded logical
Hamiltonian, whereas the QPU executes a minor-embedded, autoscaled,
finite-temperature open-system implementation. Experiments on a
programmable quantum annealer have demonstrated coherent
Kibble--Zurek behavior consistent with a closed-system model for
sufficiently fast anneals, followed by anti-Kibble--Zurek scaling in a
crossover to an open quantum regime for slower anneals
\cite{king2022coherent}. Universal quantum-critical scaling has also
been used to benchmark coherence and noise limitations in gate-based
digital quantum simulations \cite{miessen2024benchmarking}; the latter
study provides related methodological context but is not direct evidence
about D-Wave annealing dynamics. Consequently, the FOAPT-derived
schedule serves as a logical schedule-design heuristic rather than a
direct estimate of the embedded physical spectrum or transition
dynamics. Moreover, the results of Ref.~\cite{king2022coherent} do not
determine the coherent-to-open-system crossover or freeze-out location
for the dense, minor-embedded instances considered here.

Several hardware effects—including minor embedding, autoscaling,
thermal relaxation, analog-control errors, chain-break resolution,
unembedding, and readout—may modify the relationship between the
reconstructed logical spectrum and the executed physical dynamics.
Furthermore, because the instance-, embedding-, and schedule-dependent
freeze-out point was not measured, the programmed slowdown regions
cannot be directly related to the dynamically active portion of the
annealing process.

The experiments used identical submitted logical coefficients and
\texttt{auto\_scale=True} for each paired comparison together with the
standard D-Wave calibration. Nevertheless, complete embedding,
autoscaling, calibration, and ordered readout metadata were not retained
uniformly, preventing retrospective reconstruction of the calibrated
physical Hamiltonian and a consistent analysis of temporal readout
correlations. Consequently, the reported success probabilities and
summary statistics should be interpreted as descriptive results for the
tested experimental ensemble. Additional hardware diagnostics are
provided in Appendix~\ref{app:dwave_hardware_diagnostics}. Within this
experimental scope, the results demonstrate the practical feasibility of
converting learned logical spectral information into
hardware-executable candidate schedules.

%% file: conclusion.tex
\section{Discussion and Conclusions}
\label{sec:conclusion}

This work establishes Tx-NQDT as a spectrum-aware neural surrogate for
reconstructing low-energy properties of time-dependent transverse-field
Ising Hamiltonians and converting them into hardware-compatible
candidate annealing schedules. Across structured TFIM and random
Ising benchmarks, Tx-NQDT accurately reproduces the broad
evolution of the two lowest energy levels, with validation provided by
exact diagonalization, sparse matrix-free Lanczos, and the exact
Jordan--Wigner solution where applicable. Together with the hardware
experiments, these results demonstrate that learned logical spectral
features can be translated into practical, instance-specific annealing
schedules, establishing the feasibility of spectrum-aware neural
surrogates for automated schedule design.

The numerical reconstruction accuracy is nevertheless nonuniform, with
the largest localized errors occurring for the first excited state near
the driver-dominated regime and for several $N=15$ benchmark instances.
These deviations are consistent with finite Monte Carlo sampling,
computational-basis estimator variance, excited-state degeneracy and
sign structure, sequential projector-shift deflation, and the
runtime-limited AdamW optimization used in the present implementation.
Warm-start optimization substantially reduces repeated optimization
costs but does not guarantee uniform convergence, and may propagate
errors between adjacent anneal points, particularly near narrow avoided
crossings or near-degenerate regions.

Although energy reconstruction provides a convenient validation metric,
energy agreement alone does not guarantee accurate wavefunctions or the
schedule-relevant quantities required by FOAPT. Future validation should
therefore emphasize spectral gaps, minimum-gap locations, transition
matrix elements, state overlaps or fidelities, and bottleneck-window
accuracy. Simultaneous multi-state optimization, globally
parameter-conditioned or foundation neural quantum states
\cite{pfau2024excited,hendry2025grassmann,rende2025foundation},
together with adaptive annealing-grid refinement, independent restarts
near avoided crossings, improved sampling, longer optimization, and
scalable stochastic-reconfiguration or natural-gradient methods, offer
promising directions for improving reconstruction accuracy while
reducing error propagation.

The FOAPT score provides a closed-system spectral diagnostic of the
logical Hamiltonian that identifies candidate slowdown regions using the
reconstructed spectral gap and transition matrix element. It is not an
excitation probability or a microscopic transition-rate model of the
quantum annealer, and its quantitative reliability decreases when the
first-order adiabatic approximation becomes inaccurate. Accordingly,
FOAPT should be interpreted as a physically motivated guide for schedule
construction rather than a complete description of processor dynamics.
The present framework therefore combines logical spectral
reconstruction with empirical hardware evaluation, while leaving the
integration of open-system dynamics, embedding effects, and
finite-temperature evolution into the surrogate as an important
direction for future research.

The hardware results should be interpreted within the scope of the
baselines and statistics available in the present study. The default
$20\,\mu{\rm s}$ linear anneal is a low-overhead, no-tuning reference,
whereas the Tx-NQDT-informed schedules generally have
instance-dependent durations. Because matched-duration linear QPU runs
were unavailable, the comparison does not isolate the effect of
nonlinear schedule shape from that of total annealing duration. The
endpoint neural state at $s=1$ also defines a separate classical
heuristic, obtained by sampling
$\lvert\Psi_\theta(x;1)\rvert^2$ and evaluating the corresponding
classical Ising energies, but this endpoint-NQS baseline was not included
in the reported experiments. Accordingly, the present results do not
establish that the QPU-assisted workflow outperforms endpoint-NQS
sampling or other optimized classical and hardware controls. Finally,
the reported win counts, means, and medians are descriptive statistics
for the tested ensemble of $60$ instances; they should not be
interpreted as evidence of distribution-wide superiority.

In summary, Tx-NQDT provides a practical route from learned logical
spectral structure to executable annealing controls. Its principal
conceptual contribution is the separation of logical spectral
reconstruction from hardware execution: the neural surrogate supplies
closed-system spectral diagnostics for schedule generation, while the
quantum annealer empirically evaluates the resulting
hardware-compatible schedules. More broadly, the recurrence of similar
slow--fast profiles across diverse benchmark instances suggests that
aspects of logical spectral information may be compressible into
low-dimensional schedule families, opening opportunities for
transferable schedule representations and scalable schedule
optimization. Together with improved state-level reconstruction,
open-system-aware surrogates, and larger-scale hardware validation,
Tx-NQDT provides a promising foundation for automated, spectrum-aware
quantum annealing schedule design.

\section*{Data and Code Availability}

The data and code supporting the findings of this study are available from the corresponding author upon reasonable request.